\documentclass[onecolumn,showpacs,prc,superscriptaddress,amsmath,amssymb,preprintnumbers]{revtex4-1}

\newcommand{ \be }{\begin{eqnarray}}
\newcommand{ \ee }{\end{eqnarray}}

\usepackage{dcolumn} 
\usepackage{bm} 
\usepackage{fullpage}

\usepackage{graphicx}
\usepackage{color}
\definecolor{dgreen}{cmyk}{1.,0.,1.,0.4} 
\definecolor{orange}{cmyk}{0.,0.353,1.,0.} 

\usepackage[pdftex,bookmarks]{hyperref}

\usepackage{lineno}

\begin{document}


%
\title{Higher order Symmetric Cumulants}
\author{Cindy Mordasini} 
\affiliation{Physik Department, Technische Universit\"{a}t M\"{u}nchen, Munich, Germany}
\author{Ante Bilandzic} 
\affiliation{Physik Department, Technische Universit\"{a}t M\"{u}nchen, Munich, Germany}
\affiliation{Excellence Cluster Universe, Technische Universit\"{a}t M\"{u}nchen, Munich, Germany}
\author{Deniz Karako\c{c}} 
\affiliation{Physik Department, Technische Universit\"{a}t M\"{u}nchen, Munich, Germany}
\author{Seyed Farid Taghavi} 
\affiliation{Physik Department, Technische Universit\"{a}t M\"{u}nchen, Munich, Germany}
\date{\today}

\begin{abstract}
We present the generalization of recently introduced observables for the studies of correlated fluctuations of different anisotropic flow amplitudes, dubbed Symmetric Cumulants. We introduce a new set of higher order observables and outline a unique way how the genuine multi-harmonic correlations can be extracted from multi-particle azimuthal correlators. We argue that correlations among flow amplitudes can be studied reliably with the general mathematical formalism of cumulants only if that formalism is applied directly on the flow amplitudes. We have tested all the desired properties of new observables with the carefully designed Toy Monte Carlo studies. By using the realistic iEBE-VISHNU model, we have demonstrated that their measurements are feasible and we have provided the predictions for the centrality dependence in Pb--Pb collisions at LHC energies. A separate study was performed for their values in the coordinate space. The new observables contain information which is inaccessible to individual flow amplitudes and correlated fluctuations of only two flow amplitudes, and therefore they provide further and independent constraints for the initial conditions and the properties of Quark-Gluon Plasma in high-energy nuclear collisions. 
\end{abstract}


\maketitle













\section{Introduction}
\label{s:Introduction}

In the explorations of the phase diagram of strong nuclear force, some of the most intriguing questions are associated with the phase dubbed Quark-Gluon Plasma (QGP), in which under large temperatures and/or baryon densities quarks and gluons are deconfined. Properties of this extreme state of matter have been investigated in high-energy physics in the past 20 years with the plethora of different observables and across different collision systems, at SPS, RHIC and LHC experiments. One of the most important programs in these studies were the analyses of anisotropic flow phenomenon~\cite{Ollitrault:1992bk,Voloshin:2008dg}, which primarily have been carried out with the two- and multi-particle correlation techniques. The anisotropic flow measurements proved to be particularly successful in the studies of transport properties of QGP. For instance, they were used to constrain the ratio of shear viscosity to entropy density ($\eta/s$) of QGP to the very small values and therefore helped to establish a perfect liquid paradigm about the QGP properties~\cite{Braun-Munzinger:2015hba}.


Traditionally, anisotropic flow is quantified with the amplitudes $v_n$ and symmetry planes $\Psi_n$ in the Fourier series parametrization of anisotropic particle emission as a function of azimuthal angle $\varphi$ in the plane transverse to the beam direction~\cite{Voloshin:1994mz}:
\begin{equation}
f(\varphi) = \frac{1}{2\pi}\left[1+2\sum_{n=1}^\infty v_n\cos[n(\varphi-\Psi_n)]\right]\,.
\label{eq:FourierSeries_vn_psin}
\end{equation}
Correlation techniques have been utilized in the past to provide estimates for the average values of various moments of individual flow amplitudes, $\left<v_n^k\right>$ ($k > 1$), where each moment by definition carries a piece of independent information about the event-by-event fluctuations, i.e. the stochastic nature, of anisotropic flow. Flow fluctuations are unavoidable in ultrarelativistic collisions as they originate from the non-negligible fluctuations of positions of the nucleons inside the colliding nuclei, as well as from quantum fluctuations of the quark and gluon fields inside those nucleons. As a consequence, even the nuclear collisions with the same impact parameter (the vector connecting centers of two colliding nuclei) can generate different anisotropic flow event-by-event. Therefore, a great deal of additional insights about QGP properties can be extracted from flow fluctuations. It was demonstrated recently that more detailed information of QGP properties, like temperature dependence of $\eta/s$, cannot be constrained with the measurements of individual flow amplitudes due to their insensitivity. Instead, a set of more sensitive flow observables, focusing on correlations of fluctuations of two different flow amplitudes $v_n$ and $v_m$, have been proposed~\cite{Niemi:2012aj,Bilandzic:2013kga}. In this paper, we generalize this idea and introduce a new set of independent flow observables, which quantify the correlations of flow fluctuations involving more than two flow amplitudes.


By using solely the orthogonality relations of trigonometric functions, one can from Eq.~(\ref{eq:FourierSeries_vn_psin}) derive that $v_n = \left<\cos[n(\varphi\!-\!\Psi_n)]\right>$, where the average $\left<\cdots\right>$ goes over all azimuthal angles $\varphi$ of particles reconstructed in an event. However, this result has little importance in the measurements of flow amplitudes $v_n$, since symmetry planes $\Psi_n$ cannot be measured reliably in each event. Since azimuthal angles $\varphi$, on the other hand, can be measured with the great precision, we can estimate instead the flow amplitudes $v_n$ and the symmetry planes $\Psi_n$ by using the correlation techniques~\cite{Wang:1991qh,Jiang:1992bw}. The cornerstone of this approach is the following analytic result derived recently~\cite{Bhalerao:2011yg}
\begin{equation}
\left<e^{i(n_1\varphi_1+\cdots+n_k\varphi_k)}\right> = v_{n_1}\cdots v_{n_k}e^{i(n_1\Psi_{n_1}+\cdots+n_k\Psi_{n_k})}\,,
\label{eq:generalResult}
\end{equation}
where the true value of the average $\left<\cdots\right>$ can be estimated experimentally by averaging out all distinct tuples of $k$ different azimuthal angles $\varphi$ reconstructed in the same event. With the advent of a generic framework for flow analyses with multi-particle azimuthal correlations~\cite{Bilandzic:2013kga}, the LHS of Eq.~(\ref{eq:generalResult}) can be evaluated fast and exactly for any number of particles $k$, and for any choice of non-zero integers for harmonics $n_1,\ldots, n_k$. This new technology was successfully used in the first measurements of Symmetric Cumulants (SC) in~\cite{ALICE:2016kpq} and can be used also for their generalized version, which we present in this paper.


Before elaborating further on all the technical points which are relevant for the generalization of SC observables, we briefly review the existing studies of correlated fluctuations of different flow amplitudes, with the special focus on the ones performed with SC, both by theorists and experimentalists. We traced back the first theoretical study in which correlations of amplitude fluctuations of two different flow harmonics were used to extract the independent information about the QGP properties to~\cite{Niemi:2012aj}. The observables utilized in that work are available only when it is feasible to estimate flow amplitudes event-by-event, which is the case only in the theoretical studies. Alternatively, experimentalists use the correlation techniques to estimate the correlated fluctuations of amplitudes of two different flow harmonics, via the SC observables, in the way it was first proposed in the Sec.~IVC of~\cite{Bilandzic:2013kga}. Complementary, such correlated fluctuations can be also probed with the Event Shape Engineering (ESE) technique~\cite{Schukraft:2012ah}. Even though the SC are relatively novel observables, a lot of theoretical studies utilizing them have been already performed: the standard centrality dependence with state-of-the-art models was obtained in \cite{Bhalerao:2014xra,Zhou:2015eya,Gardim:2016nrr,Zhao:2017yhj,Eskola:2017imo,Alba:2017hhe,Gardim:2017ruc,Giacalone:2018wpp,Bernhard:2018hnz,Moreland:2018gsh,Schenke:2019ruo,Moreland:2019szz,Wen:2020amn}; the relation between SC and symmetry plane correlations was studied in~\cite{Giacalone:2016afq}; the more differential studies (including transverse momentum and pseudorapidity dependence, or using subevents) were performed in~\cite{Zhu:2016puf,Ke:2016jrd,Nasim:2016rfv,Jia:2017hbm}; the use of SC to constrain the details of energy density fluctuations in heavy-ion collisions was recently discussed in~\cite{Bhalerao:2019fzp}; extensive coordinate space study was reported in~\cite{Broniowski:2016pvx}; the study in the collisions of smaller systems was carried out in~\cite{Dusling:2017dqg,Dusling:2017aot,Albacete:2017ajt,Nie:2018xog,Sievert:2019zjr,Rybczynski:2019adt,Zhao:2020pty}. The complementary theoretical studies of $v_n-v_m$ correlations, without using SC, has been performed in~\cite{Qian:2016pau,Qian:2017ier}. The nontrivial technical details about the implementation of multi-particle cumulants, with the special mentioning of SC, was recently rediscussed from scratch in~\cite{DiFrancesco:2016srj}. How SC can emerge in the broader context of flow fluctuation studies was briefly mentioned in~\cite{Mehrabpour:2018kjs}.

Measurements of correlations between the amplitude fluctuations of two different flow harmonics in heavy-ion collisions with the ESE technique was first reported by ATLAS in~\cite{Jia:2014jca,Aad:2015lwa,Radhakrishnan:2016tsc}. On the other hand, analogous measurements by using SC observables were first reported by ALICE in~\cite{Zhou:2015slf,ALICE:2016kpq,Kim:2017rfn}. After these initial studies, the measurements of SC have been successfully extended to different collision systems and energies in~\cite{Guilbaud:2017alf,Sirunyan:2017uyl,STAR:2018fpo,Aaboud:2018syf,Gajdosova:2018pqo,Acharya:2019vdf,Aaboud:2019sma,Sirunyan:2019sef}. The detailed differential analyses and the extension to correlations of subleading flow harmonics were published in~\cite{Acharya:2017gsw}. Feasibility of measurements of SC in the future experiments was recently addressed in~\cite{Citron:2018lsq}.

 
The rest of the paper is organized as follows. In Sec.~\ref{s:Generalization-To-Higher-Orders} the generalization of SC observables is motivated and presented. In Sec.~\ref{s:Predictions-from-realistic-Monte-Carlo-studies} the detailed Monte Carlo studies for the higher order SC are presented, and first predictions provided for heavy-ion collisions at LHC energies. We have summarized our findings in Sec.~\ref{s:Summary}. In a series of appendices, we have placed all technical steps which were omitted in the main part of the paper.

\section{Generalization to higher orders}
\label{s:Generalization-To-Higher-Orders}

The possibility of higher order SC was first mentioned in Table 1 of~\cite{Jia:2014jca}, while the first systematic studies of their properties were performed independently in Sec.~3.2 of~\cite{Deniz:2017}. In this section, we elaborate on the generalization of SC observables for the concrete case of three flow amplitudes, but our results and conclusions can be straightforwardly generalized to any number of amplitudes by following the detailed list of requirements presented in Appendix~\ref{a:List-of-requirments}. The few final solutions for generalized SC involving more than three amplitudes we have outlined in Appendix~\ref{a:Definitions-for-higher-order-Symmetric-Cumulants}. We have cross-checked all our statements about higher order SC with the set of carefully designed Toy Monte Carlo studies in Appendix~\ref{a:Checking-the-other-requirements-for-the-Symmetric-Cumulants}. In Appendix~\ref{a:Comment-on-Parseval-theorem} we have outlined the argument that there are no built-in mathematical correlations in SC observables which might originate as a consequence of Parseval theorem. In all our derivations, we use as the starting point the general mathematical formalism of cumulants published in~\cite{Kubo}.

\subsection{Physics motivation}
\label{ss:Physics-motivation}
We now briefly motivate this generalization with the following simple observation: In mid-central heavy-ion collisions the correlated fluctuations of even amplitudes $v_2, v_4, v_6, \ldots$ can solely originate from the fluctuations of magnitude in the ellipsoidal shape~\cite{Kolb:2003zi}. This sort of fluctuations will contribute only to SC(2,4), SC(2,4,6), etc., but not to SC(2,3,4). The genuine correlation among three amplitudes, $v_2$, $v_3$ and $v_4$ can develop only due to fluctuations in the elliptical shape itself. The non-vanishing result for SC(2,3,4) implies that there are additional sources of fluctuations in the system which couple all three amplitudes $v_2, v_3, v_4$. In this sense, SC(2,3,4) can separate the following two sources of fluctuations: a) fluctuations in the shape of ellipsoidal; b) magnitude fluctuations of the persistent ellipsoidal shape. On the other hand, SC(2,4) cannot disentangle these two different sources of fluctuations. This argument is supported by concrete calculus in the simple mathematical model presented in Appendix~\ref{a:ellipse-like-distributions}.

\subsection{Technical details}
\label{ss:Technical-details}
Before going to the generalization of SC, we recall here a few important points about the use of multi-particle azimuthal correlations in flow analyses. In general, these correlations are susceptible to the non-trivial collective phenomena (e.g. anisotropic flow), but also to the phenomena involving only a few particles (e.g. momentum conservation, jet fragmentation, resonance decays, etc.). The latter, called nonflow, is considered as systematic bias in flow analyses. One way to separate the two contributions is to compute the genuine multi-particle correlations or cumulants.

In the general mathematical formalism of cumulants~\cite{Kubo}, for any set of $k$ ($k>1$) stochastic observables, $X_1,\ldots,X_k$, there exists a unique term, the $k$-particle cumulant, sensitive only to the genuine $k$-particle correlation among all the observables. This mathematical formalism has been introduced for anisotropic flow analyses in~\cite{Borghini:2000sa,Borghini:2001vi}, and further improved and generalized in~\cite{Bilandzic:2010jr,Bilandzic:2013kga}. As one has the freedom to choose the experimental observables of interest (e.g. azimuthal angles, flow amplitudes $v_n$, multiplicities of particles in a given momentum bin~\cite{DiFrancesco:2016srj}, etc.), different variants of cumulants can appear in practice. In flow analyses, the traditional choice is to identify $X_i \equiv e^{in_{i}\varphi_i}$, with $\varphi_i$ the azimuthal angles of reconstructed particles and $n_i$ the non-zero flow harmonics, as the observable of interest. This specific choice allows the identification of the averages in the cumulant expansion (e.g. Eq.~(2.8) in~\cite{Kubo}) with the single-event averages of azimuthal correlators (Eq.~(\ref{eq:generalResult})). If the detector has a uniform azimuthal acceptance, only the isotropic correlators with $\sum_{i} n_i$ = 0 are not averaged to zero in the generalization of the single-event averages $\left<\cdots\right>$ to all-event averages $\left<\left<\ldots\right>\right>$. Finally, in the resulting expression one groups together all terms which differ only by the re-labelling of azimuthal angles $\varphi_i$~\cite{Borghini:2000sa,Borghini:2001vi}.

For instance, the flow amplitude $v_n$ estimated with four-particle cumulant, $v_{n}\{4\}$, can be obtained by identifying $X_1 \equiv e^{in\varphi_1}$, $X_2 \equiv e^{in\varphi_2}$, $X_3 \equiv e^{-in\varphi_3}$ and $X_4 \equiv e^{-in\varphi_4}$. Following the steps described above, one obtains
\begin{eqnarray}
\left<\left<\cos[n(\varphi_1\!+\!\varphi_2\!-\!\varphi_3-\!\varphi_4)]\right>\right>_c &=& \left<\left<\cos[n(\varphi_1\!+\!\varphi_2\!-\!\varphi_3-\!\varphi_4)]\right>\right>\nonumber\\
&&{}-2\left<\left<\cos[n(\varphi_1\!-\!\varphi_2)]\right>\right>^2
\label{eq:4p_cumulant}\,.
\end{eqnarray}
A generalization of this idea for the case of non-identical harmonics leads to the definition of new observables dubbed SC~\cite{Bilandzic:2013kga}. With the more general choice $X_1 \equiv e^{in\varphi_1}$, $X_2 \equiv e^{im\varphi_2}$, $X_3 \equiv e^{-in\varphi_3}$, and $X_4 \equiv e^{-im\varphi_4}$, where $n$ and $m$ are two different positive integers, the general formalism of four-particle cumulants from \cite{Kubo} translates now into:
\begin{eqnarray}
\left<\left<\cos(m\varphi_1\!+\!n\varphi_2\!-\!m\varphi_3\!-\!n\varphi_4)\right>\right>_c &=& \left<\left<\cos(m\varphi_1\!+\!n\varphi_2\!-\!m\varphi_3\!-\!n\varphi_4)\right>\right>\nonumber\\
&&{}-\left<\left<\cos[m(\varphi_1\!-\!\varphi_2)]\right>\right>\left<\left<\cos[n(\varphi_1\!-\!\varphi_2)]\right>\right>
\label{eq:4p_sc_cumulant}\,.
\end{eqnarray}
It follows that, in both Eqs.~(\ref{eq:4p_cumulant}) and (\ref{eq:4p_sc_cumulant}) above, the final expressions for the cumulants depend only on the flow amplitudes, since by using Eq.~(\ref{eq:generalResult}) we obtain immediately:
\begin{eqnarray}
\left<\left<\cos[n(\varphi_1\!+\!\varphi_2\!-\!\varphi_3-\!\varphi_4)]\right>\right>_c &=& \left<v_n^4\right> - 2 \left<v_n^2\right>^2 \equiv - v_n\{4\}^4\nonumber\,,\\
\left<\left<\cos(m\varphi_1\!+\!n\varphi_2\!-\!m\varphi_3\!-\!n\varphi_4)\right>\right>_c &=& \left<v_{m}^2v_{n}^2\right>-\left<v_{m}^2\right>\left<v_{n}^2\right> \equiv \mathrm{SC}(m,n)\,.
\label{eq:cumulantsInTermsOfHarmonics}
\end{eqnarray}
The success of observables $v_n\{4\}$ and SC$(m,n)$ lies in the fact that they suppress much better the unwanted nonflow correlations, than for instance the starting azimuthal correlators $\left<\left<\cos[n(\varphi_1\!+\!\varphi_2\!-\!\varphi_3-\!\varphi_4)]\right>\right>$ and $\left<\left<\cos(m\varphi_1\!+\!n\varphi_2\!-\!m\varphi_3\!-\!n\varphi_4)\right>\right>$, and therefore they are much more reliable estimators of anisotropic flow properties. We will elaborate more on systematic biases due to nonflow later in the paper.
%


After these general considerations, to which we refer from this point onward as traditional or standard cumulant expansion in flow analyses, we proceed with the generalization of SC. As a concrete example, we provide the generalization of SC observables for the case of three flow amplitudes, and discuss in detail all consequences. As opposed to the traditional cumulant expansion in Eqs.~(\ref{eq:4p_cumulant}) and (\ref{eq:4p_sc_cumulant}), we now directly define SC$(k,l,m)$ by using the general cumulant expansion for three different observables (Eq.~(2.8) in~\cite{Kubo}), in which as fundamental observables, instead of azimuthal angles, we choose directly the amplitudes of flow harmonics. This difference in the choice of fundamental observables on which the cumulant expansion is performed is the landmark of our new approach. It follows immediately:
\begin{equation}
\mathrm{SC}(k,l,m) \equiv \left<v_k^2v_l^2v_m^2\right> 
- \left<v_k^2v_l^2\right>\left<v_m^2\right>
- \left<v_k^2v_m^2\right>\left<v_l^2\right>
- \left<v_l^2v_m^2\right>\left<v_k^2\right>
+ 2 \left<v_k^2\right>\left<v_l^2\right>\left<v_m^2\right>\,.
\label{SC(k,l,m)_flowHarmonics}
\end{equation}
In theoretical studies, in which flow amplitudes can be computed on an event-by-event basis, this expression suffices. In Appendix~\ref{a:List-of-requirments} we demonstrate that SC$(k,l,m)$ satisfies the fundamental cumulant properties, while the remaining requirements needed for generalization are cross-checked in Appendix~\ref{a:Checking-the-other-requirements-for-the-Symmetric-Cumulants}.

In an experimental analysis, it is impossible to estimate reliably flow amplitudes event-by-event, and all-event averages of azimuthal correlators need to be used to estimate them. This connection is provided by the analytic result in Eq.~(\ref{eq:generalResult}). Solely by using that result, it follows immediately that the SC$(k,l,m)$ observable defined in Eq.~(\ref{SC(k,l,m)_flowHarmonics}) can be estimated experimentally with:
\begin{eqnarray}
\mathrm{SC}(k,l,m) &=&\left<\left<\cos[k\varphi_1\!+\!l\varphi_2\!+\!m\varphi_3\!-\!k\varphi_4\!-\!l\varphi_5\!-\!m\varphi_6]\right>\right>\nonumber\\
&-&\left<\left<\cos[k\varphi_1\!+\!l\varphi_2\!-\!k\varphi_3\!-\!l\varphi_4]\right>\right>\left<\left<\cos[m(\varphi_5\!-\!\varphi_6)]\right>\right>\nonumber\\
&-&\left<\left<\cos[k\varphi_1\!+\!m\varphi_2\!-\!k\varphi_5\!-\!m\varphi_6]\right>\right>\left<\left<\cos[l(\varphi_3\!-\!\varphi_4)]\right>\right>\nonumber\\
&-&\left<\left<\cos[l\varphi_3\!+\!m\varphi_4\!-\!l\varphi_5\!-\!m\varphi_6]\right>\right>\left<\left<\cos[k(\varphi_1\!-\!\varphi_2)]\right>\right>\nonumber\\
&+&2\left<\left<\cos[k(\varphi_1\!-\!\varphi_2)]\right>\right>\left<\left<\cos[l(\varphi_3\!-\!\varphi_4)]\right>\right>\left<\left<\cos[m(\varphi_5\!-\!\varphi_6)]\right>\right>\,.
\label{eq:3pSC}
\end{eqnarray}
Since each harmonic in each azimuthal correlator above appears with an equal number of positive and negative signs, any dependence on the symmetry planes $\Psi_n$ is canceled out by definition in each of the ten correlators (Requirement~4 in Appendix~\ref{a:List-of-requirments}, and proof therein). All correlators in Eq.~(\ref{eq:3pSC}) are invariant under the shift $\varphi_i \rightarrow \varphi_i + \alpha$ of all azimuthal angles, where $\alpha$ is arbitrary, which proves their isotropy (Requirement~5 in Appendix~\ref{a:List-of-requirments}). 

We have arrived at the final result in Eq.~(\ref{eq:3pSC}) by following our new approach in which the cumulant expansion has been performed directly in Eq.~(\ref{SC(k,l,m)_flowHarmonics}) on flow amplitudes. While we have started with the cumulant expansion of order 3 in Eq.~(\ref{SC(k,l,m)_flowHarmonics}), the final expression in Eq.~(\ref{eq:3pSC}) depends on 6 azimuthal angles. We now make a comparison with the traditional approach in which the cumulant expansion is performed directly on 6 azimuthal angles $\varphi$, and discuss the differences.


\subsection{Difference between old and new cumulant expansion in flow analyses}
\label{ss:Difference-between-old-and-new-cumulant-expansion-in-flow-analyses}

We now scrutinize the cumulant expansion for one concrete example of six-particle azimuthal correlators, by following the traditional procedure, in which the starting observables are azimuthal angles. We start first with the analytic expression (derived solely using Eq.~(\ref{eq:generalResult})):
\begin{equation}
\left<\cos[n(3\varphi_1\!+\!2\varphi_2\!+\!\varphi_3-\!3\varphi_4-\!2\varphi_5-\!\varphi_6)]\right> = v_{n}^2v_{2n}^2v_{3n}^2\,.
\label{eq:second6pExample}
\end{equation}
The corresponding cumulant in the traditional approach is:
\begin{eqnarray}
\left<\cos[n(3\varphi_1\!+\!2\varphi_2\!+\!\varphi_3-\!3\varphi_4-\!2\varphi_5-\!\varphi_6)]\right>_c &=& \left<\left<\cos[n(3\varphi_1\!+\!2\varphi_2\!+\!\varphi_3-\!3\varphi_4-\!2\varphi_5-\!\varphi_6)]\right>\right> \\
&-& \left<\left<\cos[n(2\varphi_1\!+\!\varphi_2\!-\!2\varphi_3-\!\varphi_4)]\right>\right>
\left<\left<\cos[3n(\varphi_1\!-\!\varphi_2\!)]\right>\right>\nonumber\\
&-& \left<\left<\cos[n(3\varphi_1\!+\!\varphi_2\!-\!3\varphi_3-\!\varphi_4)]\right>\right>
\left<\left<\cos[2n(\varphi_1\!-\!\varphi_2\!)]\right>\right>\nonumber\\
&-& \left<\left<\cos[n(3\varphi_1\!+\!2\varphi_2\!-\!3\varphi_3-\!2\varphi_4)]\right>\right>
\left<\left<\cos[n(\varphi_1\!-\!\varphi_2\!)]\right>\right>\nonumber\\
&-& \left<\left<\cos[n(3\varphi_1\!-\!2\varphi_2\!-\!\varphi_3)]\right>\right>^2
-\left<\left<\sin[n(3\varphi_1\!-\!2\varphi_2\!-\!\varphi_3)]\right>\right>^2\nonumber\\
&+&2 \left<\left<\cos[3n(\varphi_1\!-\!\varphi_2\!)]\right>\right>
\left<\left<\cos[2n(\varphi_1\!-\!\varphi_2\!)]\right>\right>\left<\left<\cos[n(\varphi_1\!-\!\varphi_2\!)]\right>\right>\nonumber\,.
\label{eq:six3n2n1n3n2n1n}
\end{eqnarray}
In terms of flow amplitudes, this expression evaluates into:
\begin{eqnarray}
\left<v_{n}^2v_{2n}^2v_{3n}^2\right>_c &=& \left<v_{n}^2v_{2n}^2v_{3n}^2\right> 
- \left<v_{n}^2v_{2n}^2\right>\left<v_{3n}^2\right> 
- \left<v_{n}^2v_{3n}^2\right>\left<v_{2n}^2\right>
- \left<v_{2n}^2v_{3n}^2\right>\left<v_{n}^2\right>\nonumber\\ 
&-& \left<v_{n}v_{2n}v_{3n}\cos[n(3\Psi_{3n}\!-\!2\Psi_{2n}\!-\!\Psi_{n})]\right>^2
-\left<v_{n}v_{2n}v_{3n}\sin[n(3\Psi_{3n}\!-\!2\Psi_{2n}\!-\!\Psi_{n})]\right>^2\nonumber\\
&+& 2\left<v_{n}^2\right>\left<v_{2n}^2\right>\left<v_{3n}^2\right>\,.
\label{eq:six3n2n1n3n2n1n_inTermsOfFlow}
\end{eqnarray}
The result in Eq.~(\ref{eq:six3n2n1n3n2n1n_inTermsOfFlow}) is not a valid cumulant of flow amplitudes $v_{n}$, $v_{2n}$ and $v_{3n}$, since it has the extra terms in the 2nd line which depend also on the symmetry planes. Because of this contribution, this expression does not reduce to 0 for three independent flow amplitudes $v_{n}$, $v_{2n}$ and $v_{3n}$, which violates the elementary Theorem~1 on multivariate cumulants from~\cite{Kubo}.

Based on this concrete example, we conclude that one cannot start in general to calculate cumulants for one set of observables (e.g. azimuthal angles $\varphi_1,\varphi_2,\ldots$), and then interpret the final results to be a cumulant of some other observables (e.g. $v_n$'s and/or $\Psi_n$'s). At best one can state that the cumulants of multi-variate distribution of azimuthal angles $P(\varphi_1,\varphi_2,\ldots)$ can be parameterized with $v_n$'s and $\Psi_n$'s (since they are related via Fourier series expansion in Eq.~(\ref{eq:FourierSeries_vn_psin}) and analytic expression in Eq.~(\ref{eq:generalResult}), but one cannot state that the final results are direct cumulants of $v_n$'s and/or $\Psi_n$'s. In particular, the cumulants of flow degrees of freedom $v_n$'s and $\Psi_n$'s can be obtained only from the underlying p.d.f. $P(v_1,v_2,\ldots,\Psi_1,\Psi_2,\ldots)$ which governs the stochastic nature of $v_n$'s and $\Psi_n$'s observables, and not from $P(\varphi_1,\varphi_2,\ldots)$, which governs the stochastic nature of particle azimuthal angles.

In our new approach, we use azimuthal angles merely to estimate flow amplitudes via the mathematical identity in Eq.~(\ref{eq:generalResult}), but only after the cumulant expansion has been already performed directly on flow amplitudes $v_m,v_n,\ldots$. We do not perform the cumulant expansion on azimuthal angles directly, as it was done in flow analyses using cumulants so far (e.g. in Sec.~II of~\cite{Borghini:2001vi}). This subtle difference is one of the main points in our paper. This is a necessary change if one wants to estimate reliably genuine correlations of flow amplitudes, with the measured multi-particle azimuthal correlators.

This change also implies different expressions for cumulants of just one flow amplitude. For instance, the variance (second cumulant) of $v_n^2$ is
$\left<v_n^4\right>-\left<v_n^2\right>^2$, which is not the same as the usual 4-particle cumulant which gives $\left<v_n^4\right>-2\left<v_n^2\right>^2$. 

We now support our conclusions further by presenting the Monte Carlo studies.

\subsection{Comparison of old and new cumulant expansion with Monte Carlo studies}
\label{a:Monte-Carlo-studies-Appendix-A}

To compare results for cumulants calculated by treating azimuthal angles as fundamental observables (tagged `old' in this section) and calculated with our new approach by treating flow amplitudes as fundamental observables (tagged `new'), we use as a concrete example the respective results for the higher order SC built out of the flow amplitudes $v_1, v_2$ and $v_3$. In the traditional approach, we use result in Eq.~(\ref{eq:six3n2n1n3n2n1n_inTermsOfFlow}) by setting $n=1$, and in the new approach we use Eq.~(\ref{SC(k,l,m)_flowHarmonics}) by setting $k=1, l=2, m=3$. Despite this specific choice, the outcome of this study is more general given the generic nature of Eq.~(\ref{eq:six3n2n1n3n2n1n_inTermsOfFlow}), which covers all cases when the order of one harmonic is the sum of orders of two remaining harmonics.

\begin{figure}
	\begin{tabular}{c c}
		\includegraphics[scale=0.43]{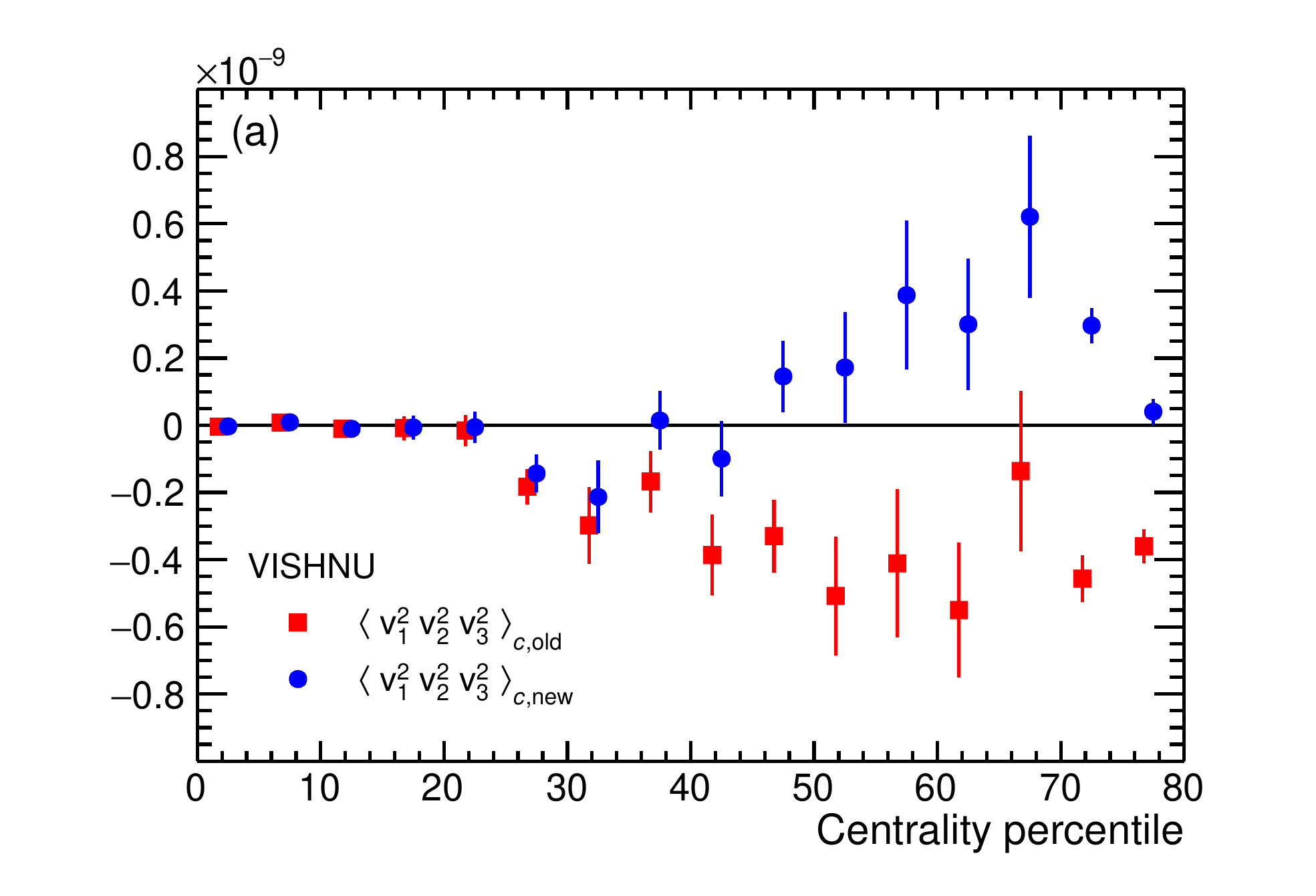} &
		\includegraphics[scale=0.43]{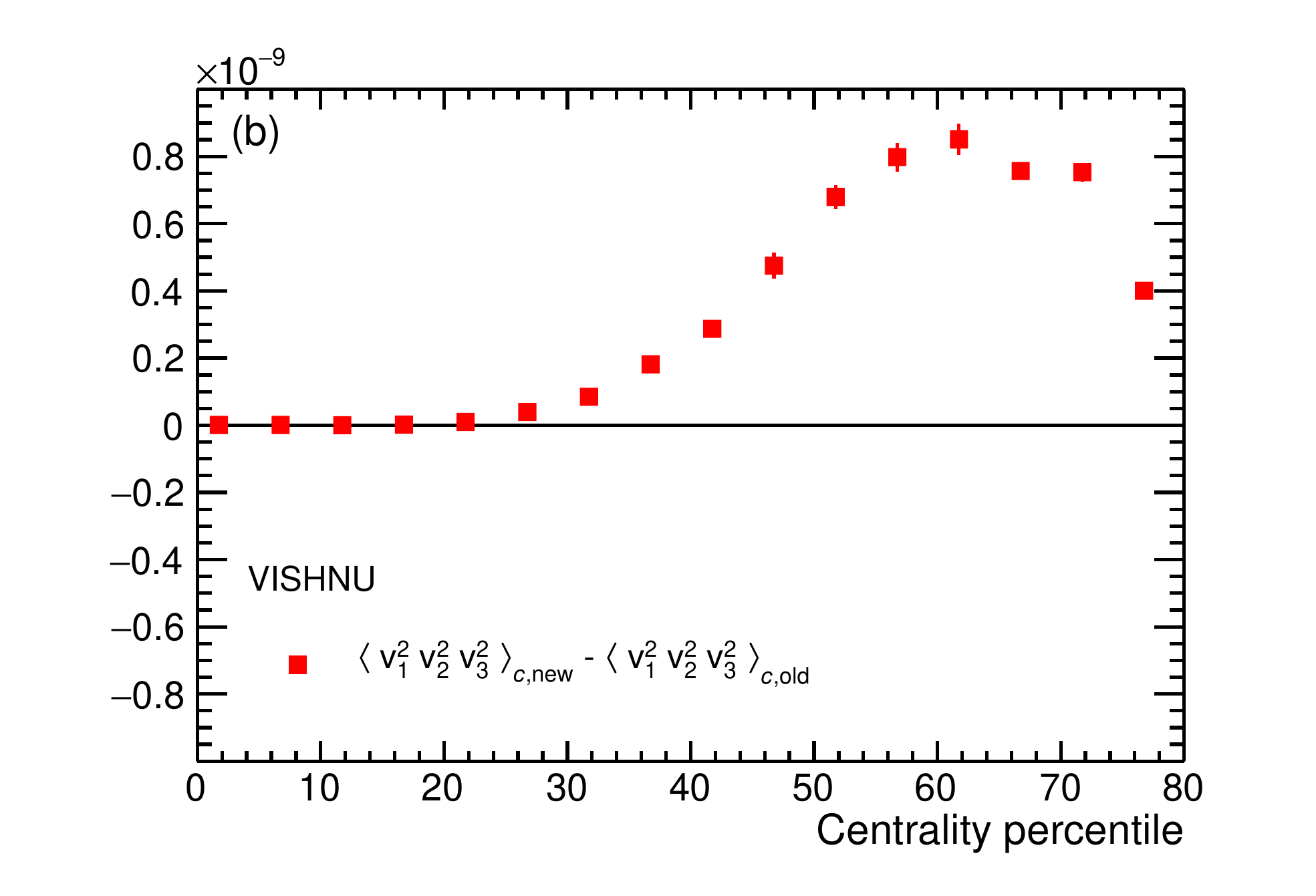}
	\end{tabular}
	\caption{(a) Comparison of centrality dependence of `new' (i.e. SC($k,l,m$) as defined in Eq.~(\ref{SC(k,l,m)_flowHarmonics})) and `old' (Eq.~(\ref{eq:six3n2n1n3n2n1n_inTermsOfFlow})) approaches to calculate cumulants of three flow amplitudes $v_1$, $v_2$ and $v_3$; (b) Centrality dependence of the difference between `new' and `old' approach.}
	\label{fig:VishnuOLDvsNEW}
\end{figure}

The comparison of the centrality dependence of these two expressions for the realistic VISHNU model (the details of VISHNU setup are presented later in Sec.~\ref{s:Predictions-from-realistic-Monte-Carlo-studies}) is presented in Fig.~\ref{fig:VishnuOLDvsNEW}(a), while in Fig.~\ref{fig:VishnuOLDvsNEW}(b) we have presented only the centrality dependence of the 2nd line in Eq.~(\ref{eq:six3n2n1n3n2n1n_inTermsOfFlow}), i.e. the difference between `new' and `old' approach. Clearly, in this concrete and realistic example, this difference is not negligible in mid-central and in peripheral collisions.   
\begin{figure}
	\begin{tabular}{c c}
		\includegraphics[scale=0.43]{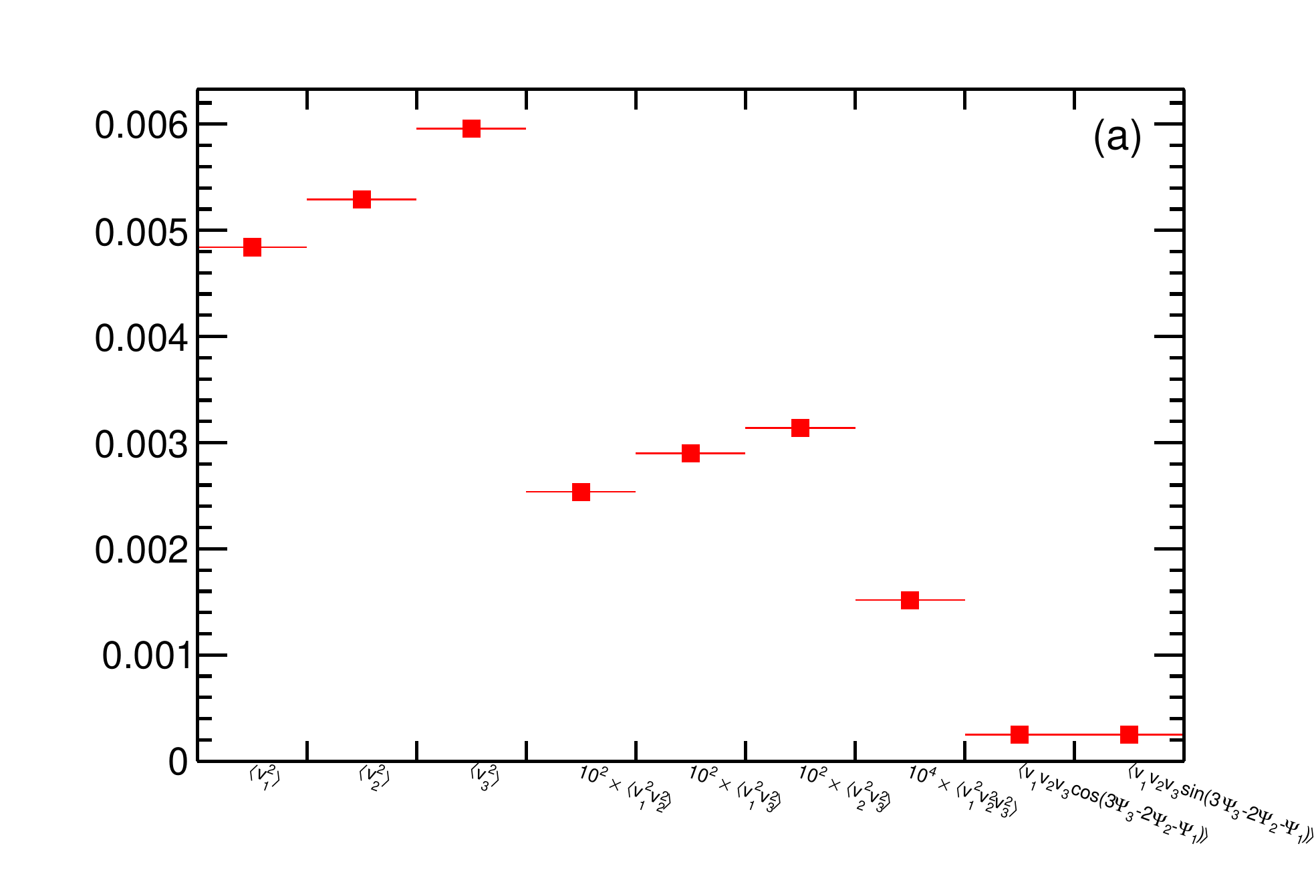} &
		\includegraphics[scale=0.43]{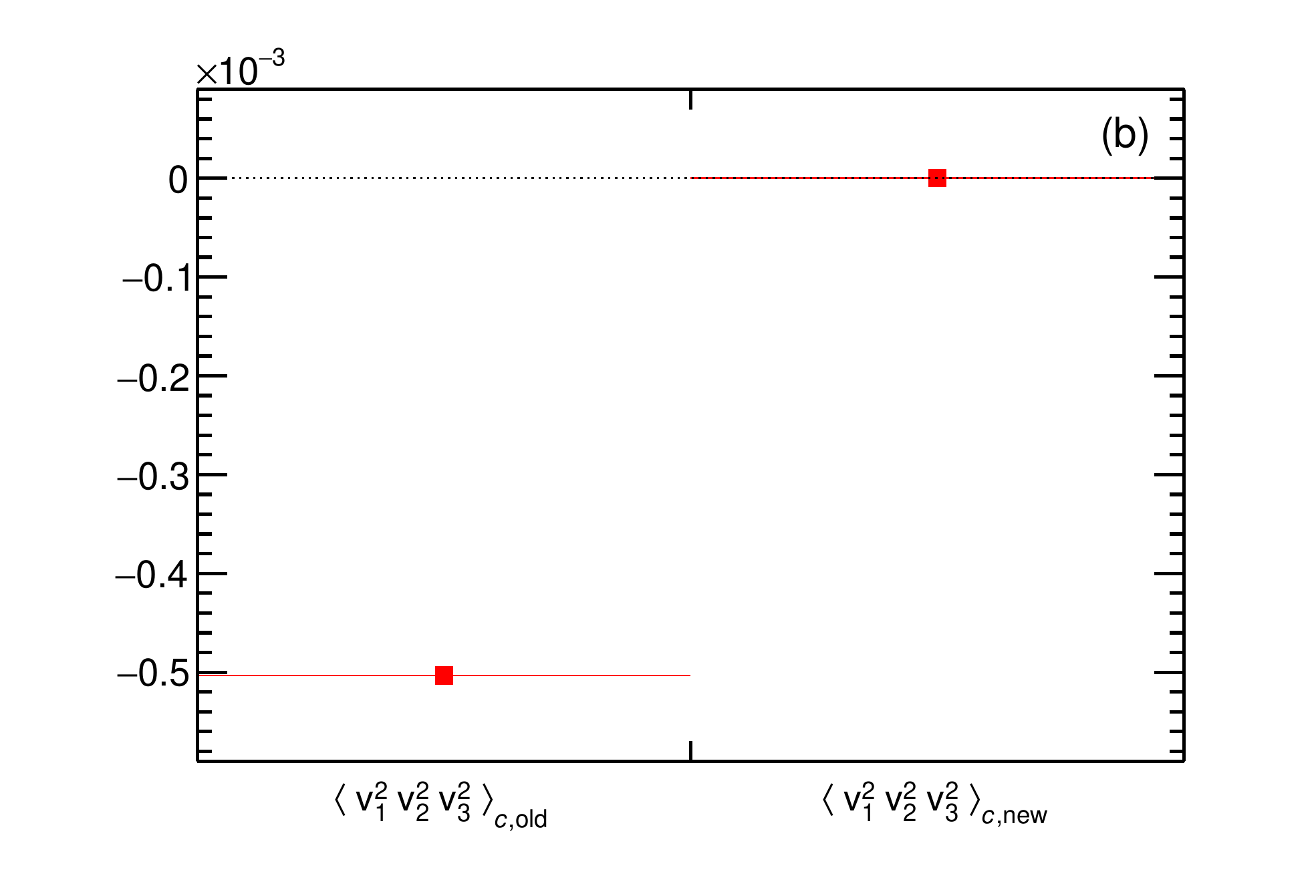}
	\end{tabular}
	\caption{(a) Average values of all individual correlators which appear in the definitions in Eqs.~(\ref{SC(k,l,m)_flowHarmonics})and (\ref{eq:six3n2n1n3n2n1n_inTermsOfFlow}) in the simple Toy Monte Carlo study described in the main text. In order to bring all the values onto the same scale, only in this plot to improve the visibility the values of two-harmonic correlators were multiplied with 100, and the values of three-harmonic correlators with 10000 (as indicated in the bin labels). (b) Cumulants calculated from the averages presented in the LHS figure, by using the `old' approach in Eq.~(\ref{eq:six3n2n1n3n2n1n_inTermsOfFlow}) (with $n=1$) and `new' approach in Eq.~(\ref{SC(k,l,m)_flowHarmonics}) (with $k=1, l=2, m=3$). Only the cumulant calculated in the `new' approach is consistent with 0, as it should be for three independently sampled flow magnitudes $v_1$, $v_2$ and $v_3$.}
	\label{fig:averagesAndcumulants}
\end{figure}

The remaining question is to demonstrate unambiguously which of the two expressions above is a valid cumulant of flow amplitudes. To answer this question we have set up the following clear-cut Toy Monte Carlo study, which demonstrates that only the Eq.~(\ref{SC(k,l,m)_flowHarmonics}) is the valid cumulant of three flow amplitudes $v_1$, $v_2$ and $v_3$: a) The amplitudes $v_1$, $v_2$ and $v_3$ are sampled in each event uniformly and independently in the intervals (0.03,0.1), (0.04,0.1), and (0.05,0.1), respectively; b) The symmetry planes $\Psi_1$ and $\Psi_2$ are sampled in each event uniformly and independently in the interval (0,2$\pi$);
c) The symmetry plane $\Psi_3$ is determined in each event as: $\Psi_3 \equiv \frac{1}{3}(\frac{\pi}{4} + 2\Psi_2 + \Psi_1)$. With such a setup, in each event $\cos(3\Psi_3 - 2\Psi_2 - \Psi_1) = \cos\frac{\pi}{4} = \frac{\sqrt{2}}{2}$ (and similarly for the sine term), and therefore the 2nd line in Eq.~(\ref{eq:six3n2n1n3n2n1n_inTermsOfFlow}) is non-vanishing, which means that in this study Eqs.~(\ref{eq:six3n2n1n3n2n1n_inTermsOfFlow}) and (\ref{SC(k,l,m)_flowHarmonics}) will yield systematically different results. However, since the three flow amplitudes $v_1$, $v_2$ and $v_3$ are sampled independently, the corresponding cumulant must be 0 (Theorem~1 in \cite{Kubo}). After we have carried out this study for 10000 events, we have obtained the results presented in Fig.~\ref{fig:averagesAndcumulants}.
Since only $\left<v_{1}^2v_{2}^2v_{3}^2\right>_{c,{\rm new}}$ is consistent with 0 (see Fig.~\ref{fig:averagesAndcumulants}(b)), we conclude that this observable is a valid cumulant of three flow amplitudes $v_1$, $v_2$ and $v_3$. This simple Toy Monte Carlo demonstrates also that another observable, $\left<v_{1}^2v_{2}^2v_{3}^2\right>_{c,{\rm old}}$, is not a valid cumulant of the three flow amplitudes $v_1$, $v_2$ and $v_3$ due to the presence of 2nd line in Eq.~(\ref{eq:six3n2n1n3n2n1n_inTermsOfFlow}) involving symmetry planes, which spoils the cumulant property. 

Before presenting our first predictions for higher order SC in heavy-ion collisions at LHC energies, we discuss the systematic bias from nonflow contribution.

\subsection{Nonflow estimation with Toy Monte Carlo studies}
\label{ss:Nonflow-estimation-with-Toy-MonteCarlo-studies}

In this section, we discuss the generic scaling of the nonflow contribution in higher order SC. We begin with the implementation of the Fourier distribution $f(\varphi)$ parameterized as in Eq.~(\ref{eq:FourierSeries_vn_psin}), which we use to sample the azimuthal angle of each simulated particle. For simplicity, we define $f(\varphi)$ with three independent parameters, the flow amplitudes $v_2$, $v_3$ and $v_4$:
\begin{equation}
f(\varphi) = \frac{1}{2\pi} \left[ 1 + 2 v_2 \cos \left( 2\varphi \right) + 2 v_3 \cos \left( 3\varphi \right) + 2 v_4 \cos \left( 4\varphi \right) \right].
\label{sect-ToyMC_eq-Fourier}
\end{equation}
The setup of our Toy Monte Carlo simulations goes as follows: For each one of the input number of events $N$, we set the values of the multiplicity $M$ and the flow harmonics $v_2$, $v_3$ and $v_4$ according to the requirements of the current analysis. Examples of such conditions can be that these input parameters are kept constant for all events or that they are uniformly sampled event-by-event in given ranges. We indicate this second case with the notation ($\cdot$,$\cdot$). After insertion of the harmonics, Eq.~\eqref{sect-ToyMC_eq-Fourier} is used to sample the azimuthal angles of the $M$ particles. We then compute all the needed azimuthal correlators with the generic framework introduced in~\cite{Bilandzic:2013kga} with the possibility left open to choose which event weight to use during the transition from single- to all-event averages. Finally our SC observable is obtained using Eq.~\eqref{eq:3pSC}.

\begin{figure}
	\begin{tabular}{c}
        \includegraphics[scale=0.43]{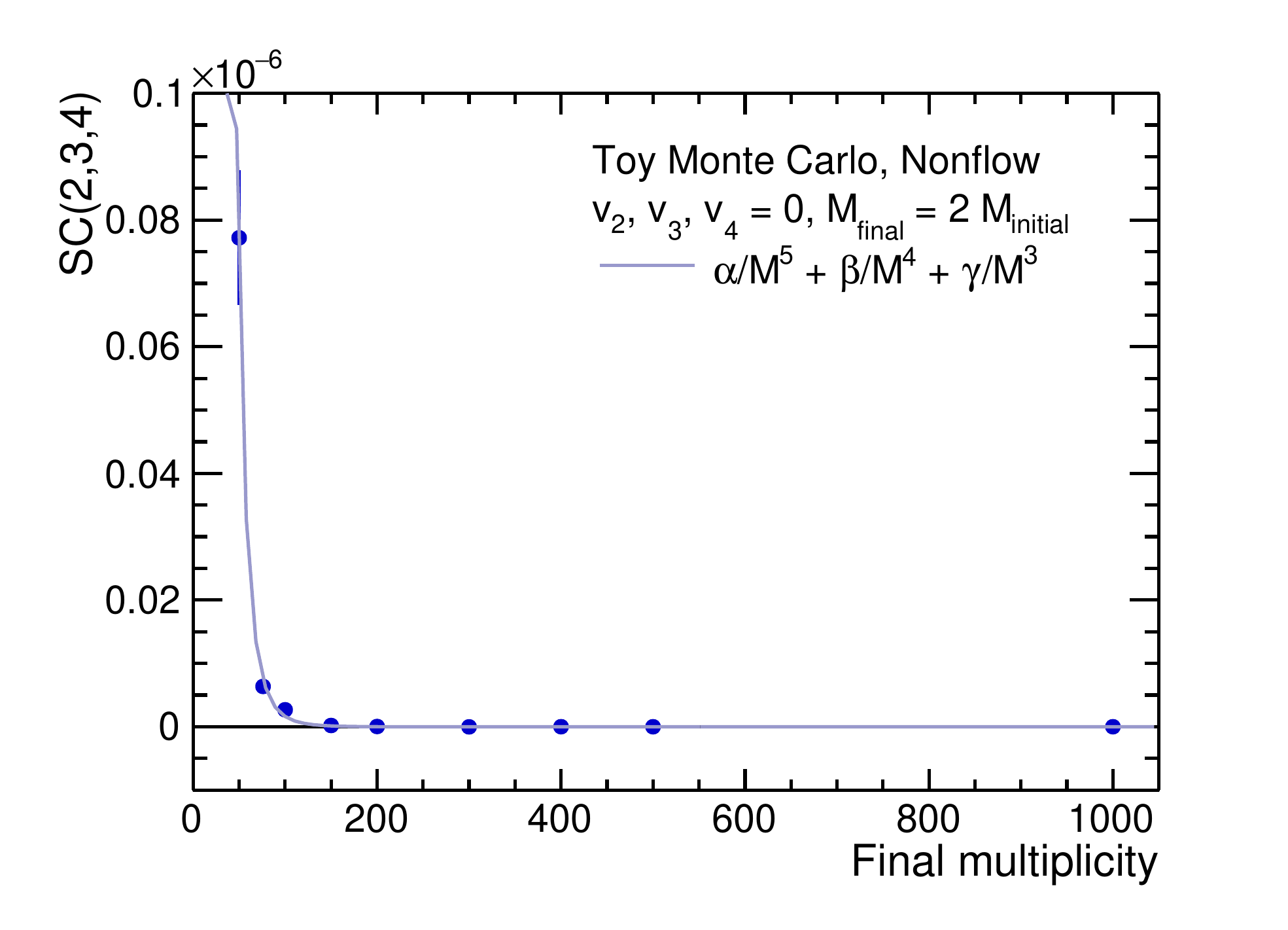}
	\end{tabular}
	\caption{Values of SC(2,3,4) obtained in the case of strong two-particle nonflow. The theoretical value of SC(2,3,4) is zero in this example.}
	\label{sect_ToyMC_fig_nonflow}
\end{figure}

Now that our Toy Monte Carlo setup is in place, we can use it to check if our SC observable has the needed properties. We simulate one example of simple nonflow correlations and look at the scaling of our expression as a function of the multiplicity. Our nonflow is described as follows: for each one of the $N = 10^8$ events we generate, we sample a fixed initial number of particles $M_{\textrm{initial}}$ amongst the possibilities: 25, 38, 50, 75, 100, 150, 200, 250 and 500. The flow harmonics are all set to zero. We then introduce strong two-particle correlations by taking each particle two times in the computation of the two- and multi-particle correlations. This means our final multiplicity is given by
\begin{equation}
M_{\textrm{final}} = 2 M_{\textrm{initial}}.
\label{sect-ToyMC_eq-MultiNonFlow}
\end{equation}
As detailed in the list of requirements in Appendix~\ref{a:List-of-requirments}, the nonflow scaling of our three-harmonic SC can be described with the following expression:
\begin{equation}
\delta^{\rm SC}_3 = \frac{\alpha}{M^5} + \frac{\beta}{M^4} + \frac{\gamma}{M^3},
\label{sect_ToyMC_eq_fitNonFlow}
\end{equation}
where $M$ corresponds to the final multiplicity introduced in Eq.~\eqref{sect-ToyMC_eq-MultiNonFlow}. The fit is done over all simulated multiplicities and can be found in Fig.~\ref{sect_ToyMC_fig_nonflow}. The obtained fit parameters are as follows:
\begin{gather}
\begin{split}
\alpha & = 27.01 \pm 9.79, \\
\beta & = - 0.0947 \pm 0.1743, \\
\gamma & = (- 1.383 \pm 5.876)\cdot 10 ^{-4},
\end{split}
\end{gather}
with a goodness of the fit of $\chi ^2/ndf = 0.7$. We see that the results are well described with the chosen fit. Parameters $\beta$ and $\gamma$ are both consistent with zero, meaning the dominant contribution to the nonflow comes from the six-particle correlator.  This is the same leading order behavior as in the corresponding traditional cumulant expansion for this particular SC. Therefore SC($k$,$l$,$m$) is not sensitive to lower order correlations.
%
%

This simulation has been done in a totally controlled environment which is not a true representation of what happens in a heavy-ion collision. We check now SC properties in realistic Monte Carlo simulations in the next section, and provide first predictions.

\section{Predictions from realistic Monte Carlo studies}
\label{s:Predictions-from-realistic-Monte-Carlo-studies}

In this section we discuss the physics of our new observables. We provide studies and obtain separate predictions both for the coordinate and momentum space, by using two realistic Monte Carlo models: iEBE-VISHNU and HIJING.

\subsection{Predictions from iEBE-VISHNU}
\label{ss:Realistic_Monte_Carlo_studies}


To inspect to what extent the generalized SC can capture the collective behavior of the heavy-ion collision evolution, we use iEBE-VISHNU~\cite{Shen:2014vra}, a heavy-ion collision event generator based on the hydrodynamic calculations. In this event generator, after preparing the initial state using Monte Carlo approach, the evolution of the energy density is calculated via 2+1 causal hydrodynamics together with equation of state calculated from a combination of lattice QCD and hadronic resonance gas model (s95p-v1~\cite{Huovinen:2009yb}). After the hydrodynamic evolution, the Cooper-Frye formula~\cite{Cooper:1974mv} is used to convert each fluid patch to a distribution of hadrons. Using iEBE-VISHNU, one can study the hadronic gas evolution after hadronization by using Ultra-relativistic Quantum Molecular Dynamics (UrQMD) transport model~\cite{Schenke:2012wb}. The evolution stops if no collision or decay happen in the system. In the present study, we have not used UrQMD to decrease the simulation time.

	


In this paper, we study Pb--Pb collision with center-of-mass energy per nucleon pair $\sqrt{s_{\rm NN}}=2.76\,$TeV. For the initial state, MC-Glauber with $0.118$ as wounded nucleon/binary collision mixing ratio is used. For the hydrodynamic evolution, the DNMR formalism~\cite{Denicol:2010xn,Denicol:2012cn} with fixed shear viscosity over entropy density ($\eta/s=0.08$) and zero bulk viscosity is exploited. The hydrodynamic initial time is set to $0.6\;$fm/$c$. The events are divided into 16 centrality classes between 0--80\% with equally sized bins. For each centrality, we have generated 14000 events. Let us point out that in the present study we have taken into account the particles $\pi^{\pm}$, $K^{\pm}$, $p$ and $\bar{p}$ in the final particle distribution, since they are the most abundant particles in the final distribution.
In iEBE-VISHNU, the reaction plane angle, i.e. the  angle  between the impact parameter vector and a reference frame, is fixed to be zero for all events. One notes that here we are dealing with a 2+1 dimensional hydrodynamic calculations in which the boost invariance is considered in the third (longitudinal) direction. For that reason, there is no pseudorapidity dependence in the present simulation.

It is worth mentioning that the aim of the simulation in this paper is not to present a precise justification of the experimental observation. In fact, we would try to demonstrate that in the presence of flow the generalized SC have non-trivial values and that their measurements are feasible for Pb--Pb collisions at LHC energies in terms of required statistics. Nevertheless, to show that our simulations can be considered as at least a qualitative prediction for future experimental observation, we present our Monte Carlo simulation results of few well-studied two-harmonic SC and compare them to the published data from ALICE in Appendix~\ref{ss:AppendixD}.

Regarding the $p_T$ range, the range $0.28 < p_T < 4$~GeV has been chosen in the present study. We should say that in ALICE $p_T$ is in the range $0.2 < p_T < 5$~GeV.  It turns out that SC is very sensitive to the lower limit of $p_T$ range where we expect to have a considerable amount of particles. For the range $0.28 < p_T < 4$~GeV, which is more close to ALICE $p_T$ range, we have a qualitative agreement between simulation and data for SC of two flow amplitudes (see Appendix~\ref{ss:AppendixD}). The reason we have used $0.28$ (not $0.3$ or $0.2$) for the lower $p_T$ range is that the $p_T$ dependent output of VISHNU is reported in a fixed discrete range between $0 < p_T <4$~GeV. We should point out that our computations  with lower $p_T=0.28$~GeV show a reasonable agreement with Ref.~\cite{Zhu:2016puf} in which VISHNU simulation with MC-Glauber model for the initial state and constant $\eta/s=0.08$ has been studied. 


According to the collective picture of the produced matter in heavy-ion collision, the anisotropic flow corresponds to the anisotropy in the initial state, quantified by eccentricities $\epsilon_n e^{in\Phi_n}$. In fact, eccentricities  are defined as the moments of initial energy density~\cite{Teaney:2010vd}, 
\begin{equation}
\epsilon_n e^{in\Phi_n} \equiv-\frac{\int r^n e^{in\varphi} \rho(r,\varphi)rdrd\varphi}{\int r^n  \rho(r,\varphi)rdrd\varphi},\qquad n=2,3,\ldots,
\end{equation}
where $(r,\varphi)$ are the polar coordinates in the two dimensional transverse space, and $\rho(r,\varphi)$ is the energy density in this space. Similar to the anisotropic flow, the SC of the eccentricity distribution 
shown by $\text{SC}_{\epsilon}(n,m,p)$ can be studied by replacing $v_n$ with $\epsilon_n$ in Eq.~(\ref{eq:cumulantsInTermsOfHarmonics}). 

In addition, the flow fluctuations are the manifestation of initial state fluctuations after the collective evolution. One notes that the difference between initial state models (Glauber, CGC, TrENTo, etc.) is not only encoded in different cumulants of eccentricities $\epsilon_2$ or $\epsilon_3$ but also in the other details of the initial state fluctuations, for instance in the correlations between eccentricities. In other words, to discriminate between different models, we need to study different aspects of the eccentricity fluctuations and their correlations. The higher order SC observables introduced in this paper should be considered as an example of such a study. Modeling collective evolution as a linear/non-linear response to the initial state helps our understanding of both the initial state and the collective evolution models. Using it, one can explain the experimentally observed flow fluctuations, in the sense that we can qualitatively inspect how much of the observed value is a manifestation of the initial state and how much of it is the consequence of the collective evolution. Even more than that, it is also possible to examine the strength (couplings) of the linear/non-linear terms with this method. A more detailed investigation in this context is postponed to future studies.

		\begin{figure}[t!]
	\begin{center}
		\begin{tabular}{c c}
			\includegraphics[scale=0.43]{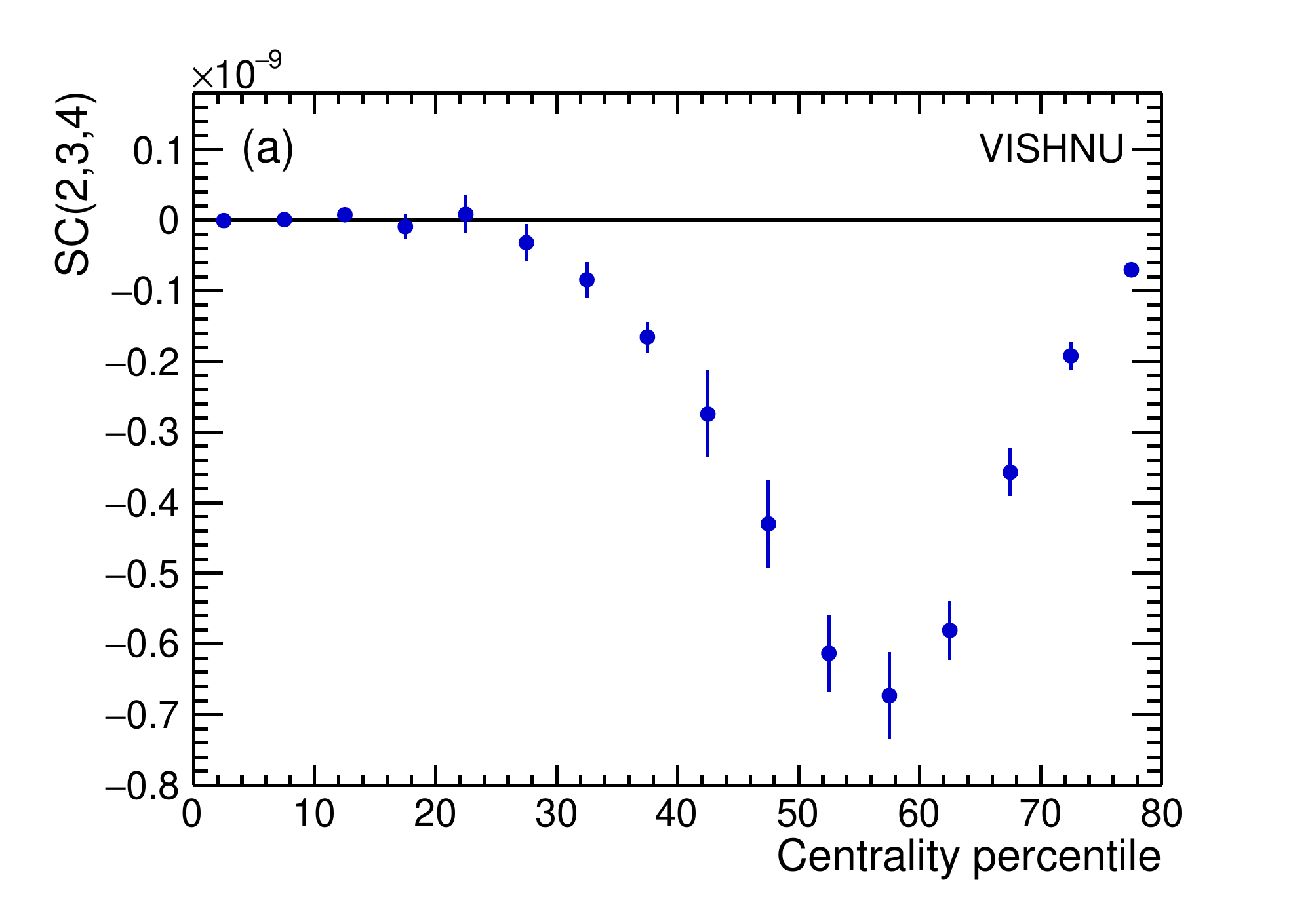} &
			\includegraphics[scale=0.43]{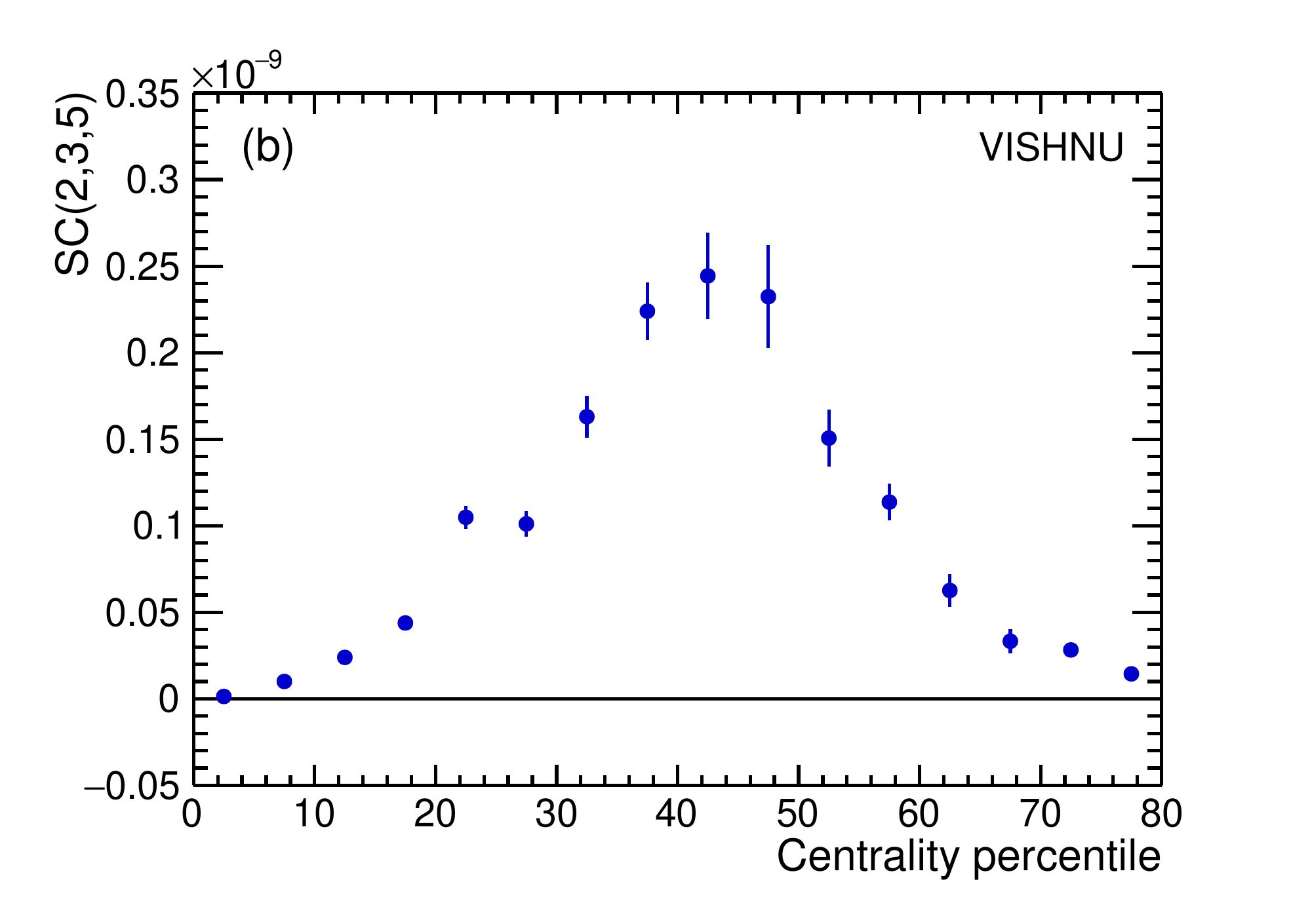} \\
			\includegraphics[scale=0.43]{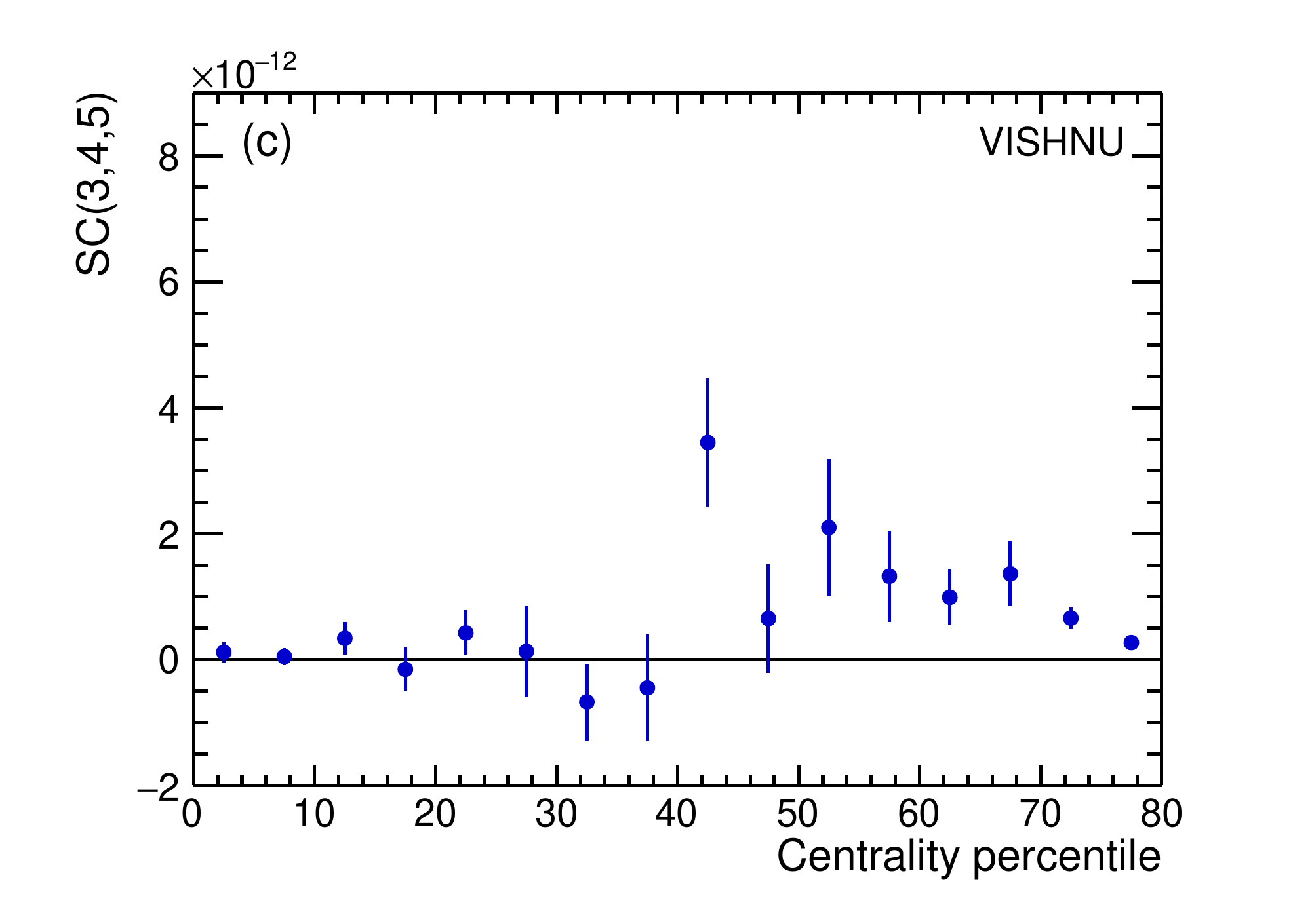} &
			\includegraphics[scale=0.43]{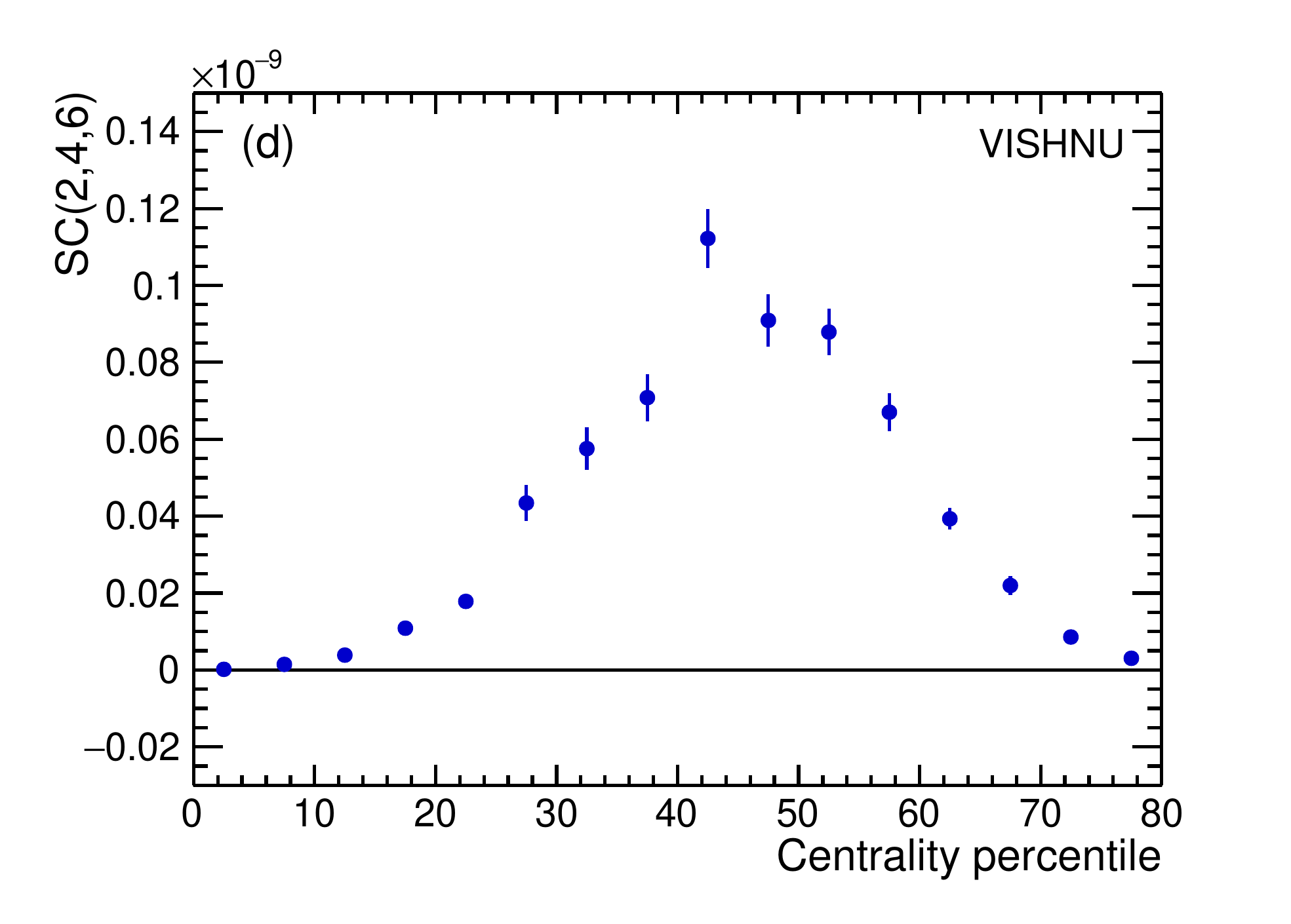} 

		\end{tabular}		
		\caption{ Four different generalized SC obtained from iEBE-VISHNU. } 
		\label{SCVmnp}
	\end{center}
\end{figure}	

\begin{figure}[t!]
	\begin{center}
		\begin{tabular}{c c}
			\includegraphics[scale=0.43]{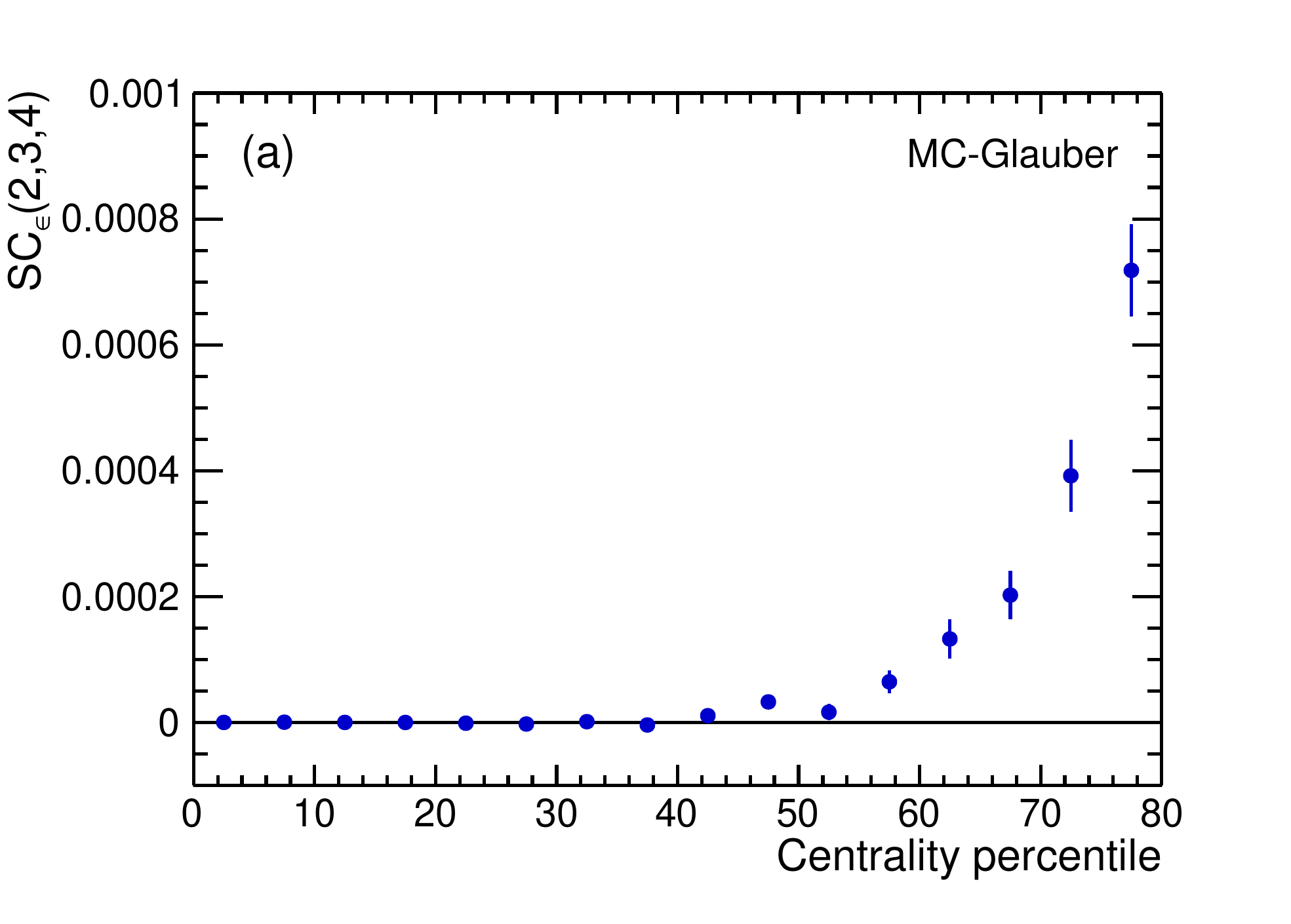} &
			\includegraphics[scale=0.43]{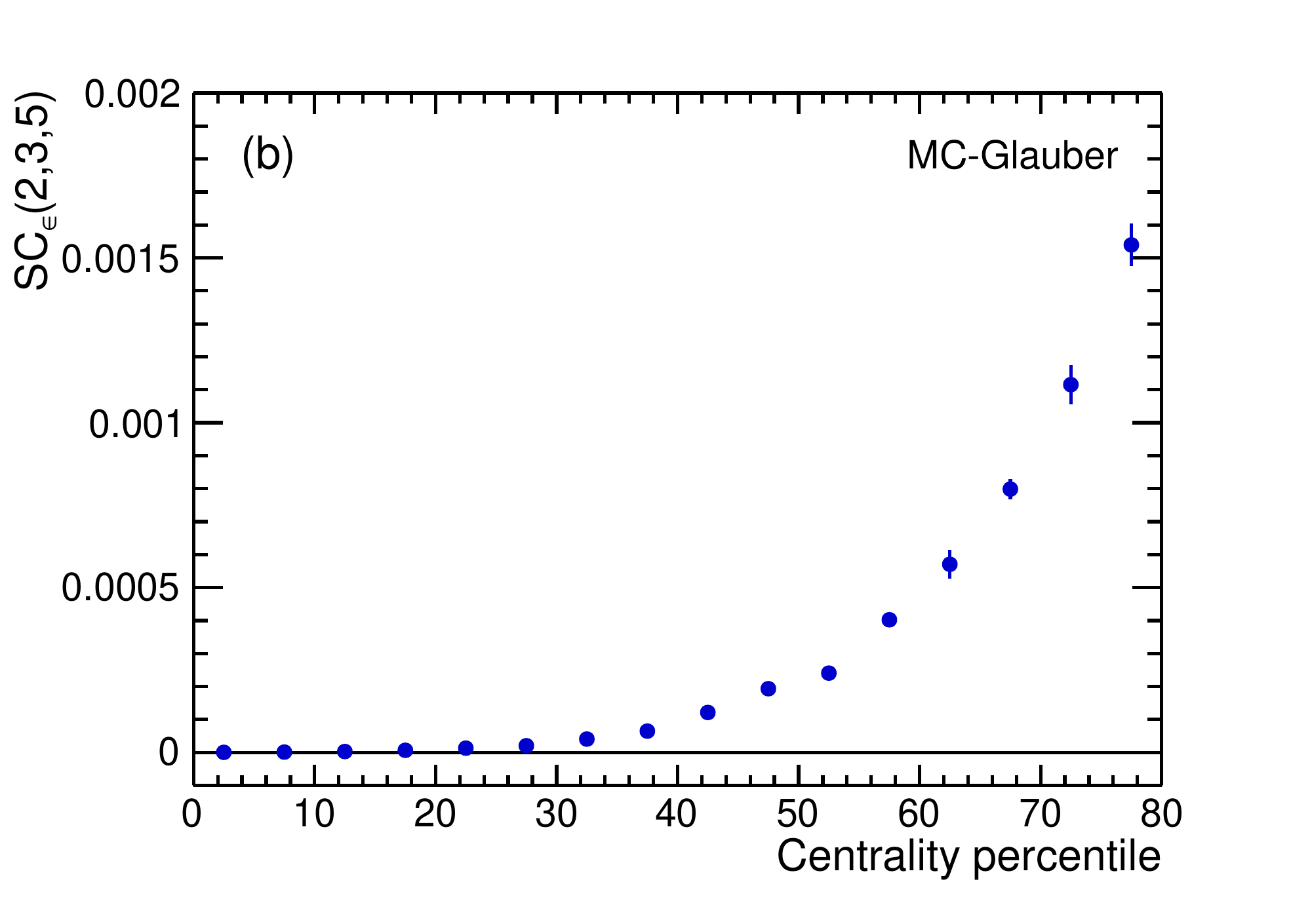} \\
			\includegraphics[scale=0.43]{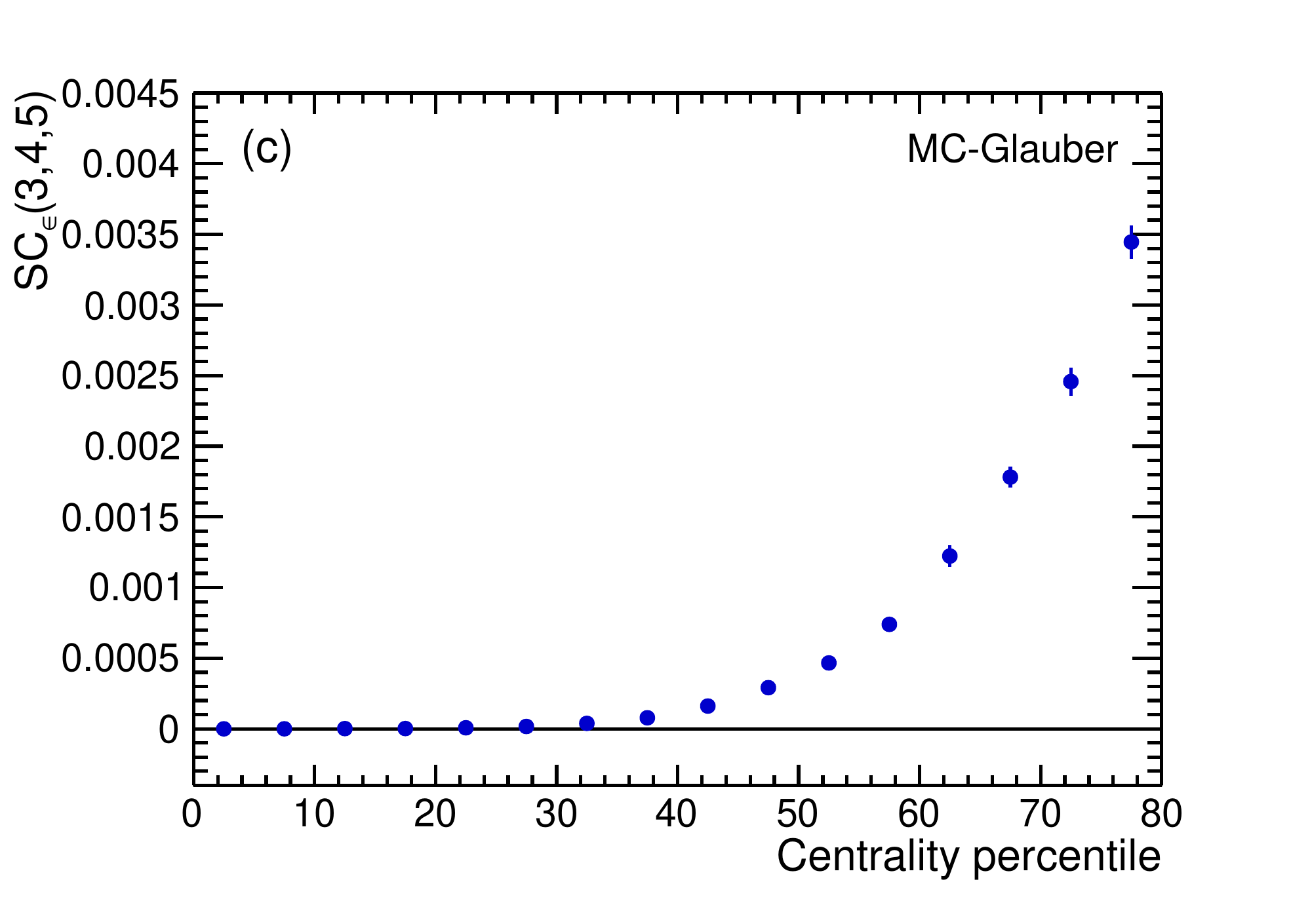} &
			\includegraphics[scale=0.43]{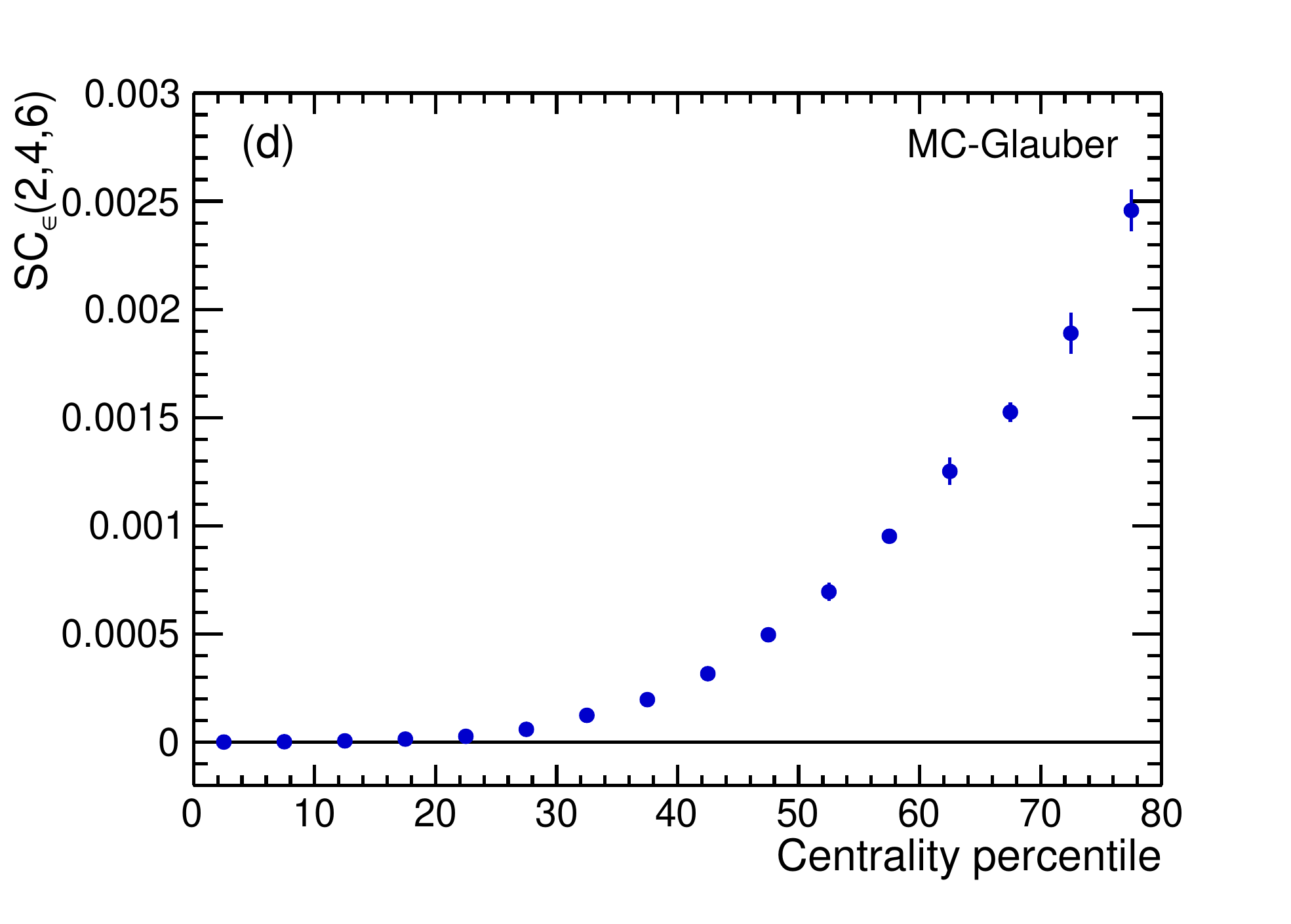} 

		\end{tabular}		
		\caption{ Four different generalized SC obtained from MC-Glauber. } 
		\label{SCepsmnp}
	\end{center}
\end{figure}


Using our Monte Carlo simulations, we have shown SC(2,3,4), SC(2,3,5), SC(3,4,5) and SC(2,4,6) in Fig.~\ref{SCVmnp}. Also the same quantities which are obtained from the initial state are presented in Fig.~\ref{SCepsmnp}.
One can see from the Figs.~\ref{SCVmnp} and \ref{SCepsmnp} that except for SC(3,4,5) in the final state (Figs.~\ref{SCVmnp}(c)) all the other presented cumulants show significant non-vanishing values. Moreover, one can observe that the cumulants obtained from the initial state (Fig.~\ref{SCepsmnp}) are monotonically increasing, while the same cumulants for the distribution after hydrodynamic evolution change the slope after centrality classes $45\%$ to $55\%$ and approach to zero. This can be due to the fact that, in the more peripheral collisions, the collective evolution duration is shorter, and small system size cannot transport the initially produced correlation into the final momentum distribution. Moreover, one can see a sign flip in $\text{SC}(2,3,4)$ and a suppression in $\text{SC}(3,4,5)$ in Fig.~\ref{SCVmnp}, compared to the initial state generalized SC in Fig.~\ref{SCepsmnp}. We will return to these points after defining the normalized generalized SC in the following.  

As it can be seen from Fig.~\ref{SCVmnp} and Fig.~\ref{SCepsmnp}, due to the different scale of the initial and final distributions, we are not able to compare $\text{SC}(n,m,p)$ and $\text{SC}_{\epsilon}(n,m,p)$. In order to clarify to what extent the observed values of SC are related to the correlations in the initial state, we will exploit the normalized generalized SC,
	\begin{eqnarray}\label{NGSC}
\text{NSC}(n,m,p)=\frac{\text{SC}(n,m,p)}{\langle v_n^2 \rangle \langle v_m^2 \rangle \langle v_p^2\rangle},\qquad \qquad \text{NSC}_{\epsilon}(n,m,p)=\frac{\text{SC}_{\epsilon}(n,m,p)}{\langle \epsilon_n^2 \rangle \langle \epsilon_m^2 \rangle \langle \epsilon_p^2\rangle}.
\label{NSCeq}
\end{eqnarray}
With such a study, for instance, one can answer what is the influence of the collective stage in the heavy-ion evolution. If NSC calculated in terms of $\epsilon_n$ and $v_n$ are the same, that means that the anisotropies in the initial state are the dominant source of anisotropies in the final state, and therefore NSC can be used to tune the details of initial conditions only. Using normalized (generalized) SC, we also avoid the sensitivity to the $p_T$ range in the final particle distribution. In fact, NSC clearly eliminates any dependence of multiplying amplitude of flow harmonics, which was obtained independently from correlators depending only on the same harmonics. Therefore, the independent information contained only in the correlations can be extracted best only from the normalized SC. While this is rather straightforward to achieve in models having only flow correlations, it is much more of a challenge in experimental analyses, due to difficulties in suppressing nonflow contribution in the denominator in Eq.~\eqref{NSCeq}. In \cite{ALICE:2016kpq}, where normalized SC were measured for the first time, the nonflow in denominator was suppressed by introducing pseudorapidity gaps in two-particle correlations.

\begin{figure}[t!]
				\begin{center}
					\begin{tabular}{c c}
						\includegraphics[scale=0.43]{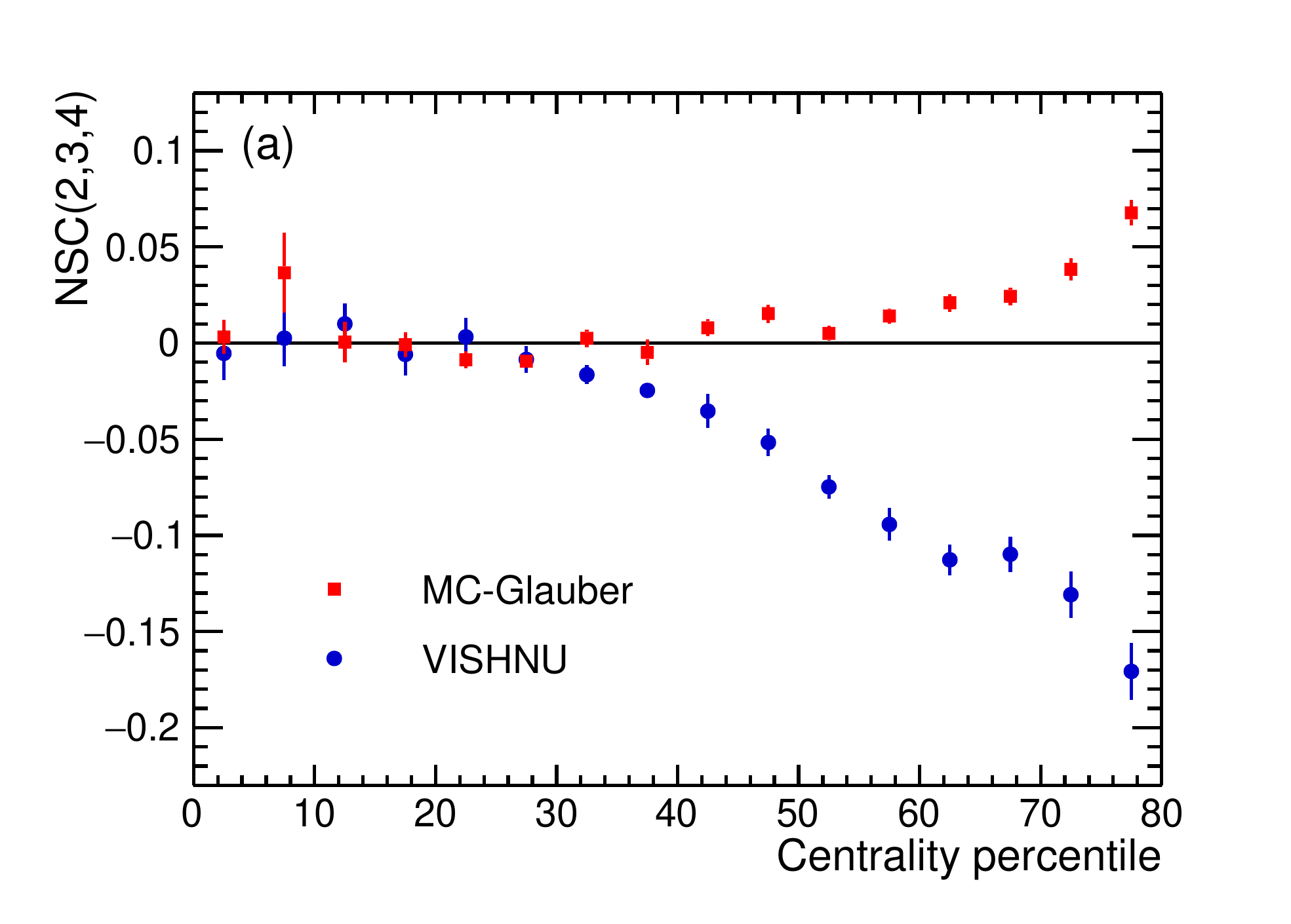} &
						\includegraphics[scale=0.43]{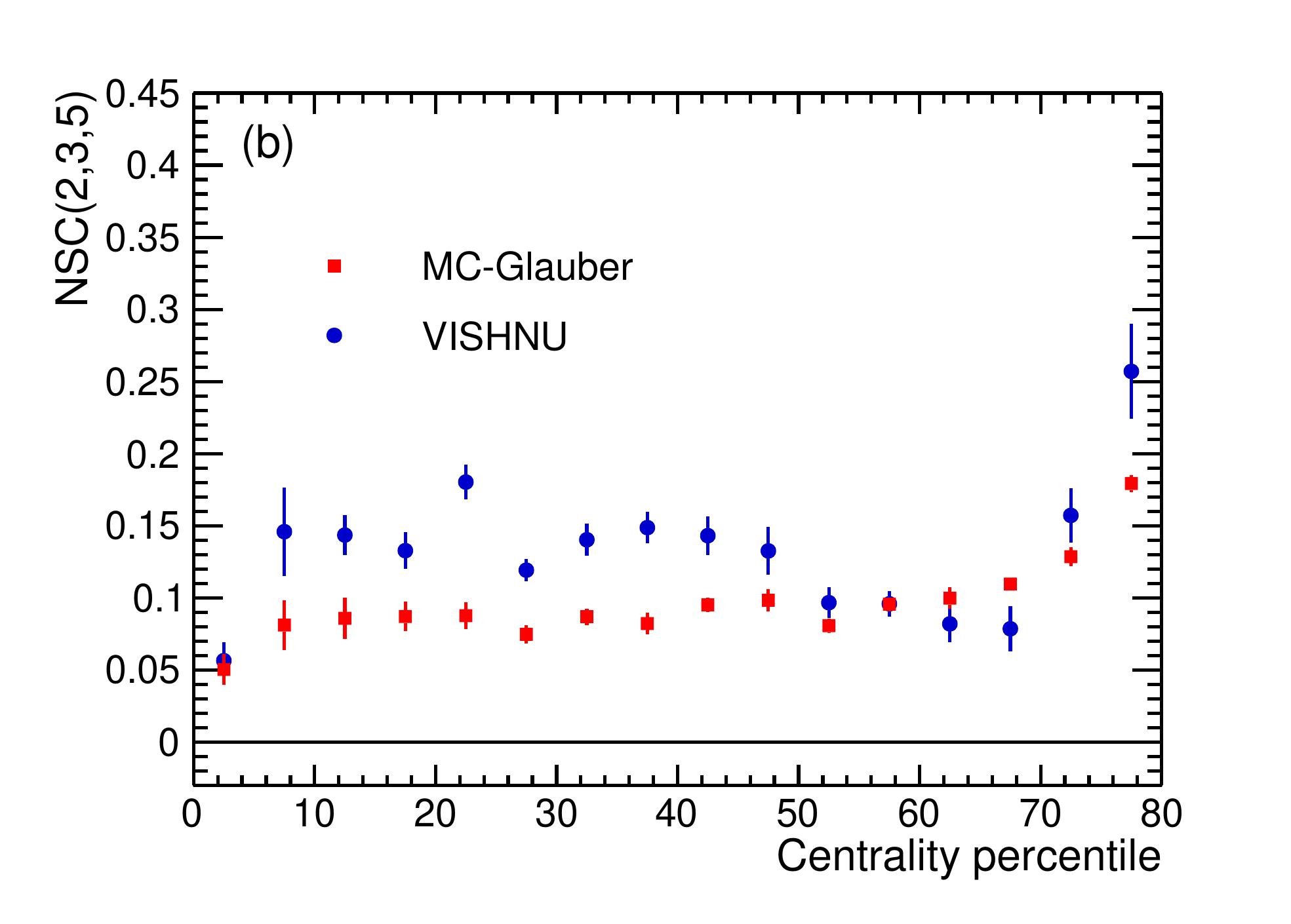} \\
						\includegraphics[scale=0.43]{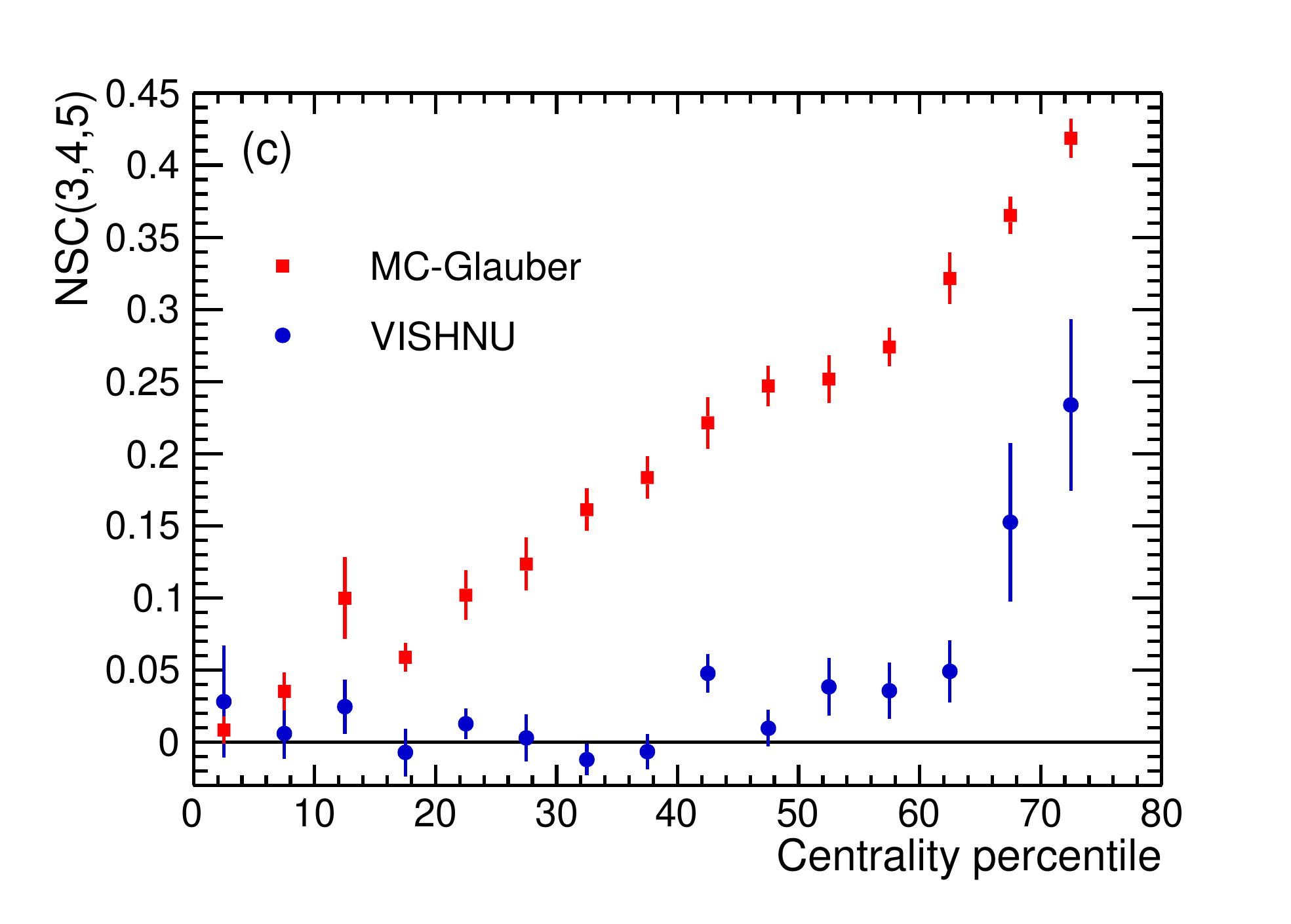} &
						\includegraphics[scale=0.43]{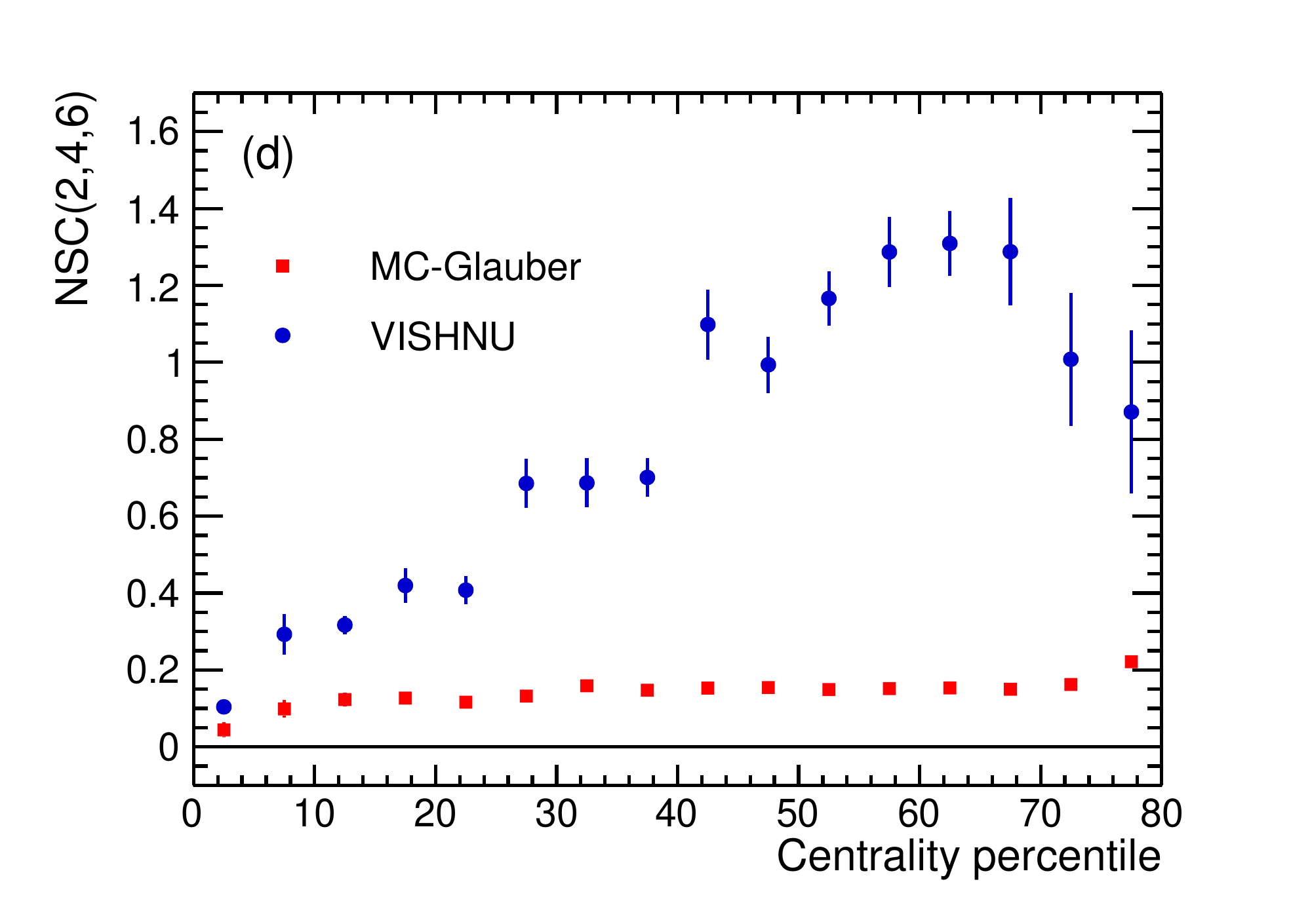} 

					\end{tabular}		
					\caption{ Comparison between four different normalized generalized SC obtained from MC-Glauber and iEBE-VISHNU. } 
					\label{NSCmnp}
				\end{center}
			\end{figure}

The generalized NSC are depicted in Fig.~\ref{NSCmnp}. The similarity and discrepancy between $\text{NSC}(n,m,p)$ and $\text{NSC}_{\epsilon}(n,m,p)$ can be explained qualitatively by considering the linear and non-linear hydrodynamic response. As a matter of fact, the linear response is approximately true for $\epsilon_2$ and $\epsilon_3$ in the central and mid-central collisions \cite{Gardim:2011xv}. It means the event with larger ellipticity $\epsilon_2$ has larger elliptic flow $v_2$, and the dependence is approximately linear. The same relation holds between triangularity $\epsilon_3$ and triangular flow $v_3$. However, this is not the case for higher harmonics. For instance, it has been shown that for $v_4 e^{4i\Psi_4}$ and $v_5e^{5i\Psi_5}$ we have the following relations \cite{Gardim:2011xv,Teaney:2012ke,Teaney:2013dta,Giacalone:2018syb},
\begin{equation}\label{nonlinear}
\begin{aligned}
v_4 e^{4i\Psi_4} &= k_4 \epsilon_4 e^{i4\Phi_4}+k'_4 \epsilon_2^2 e^{i4\Phi_2}\,, \\
v_5 e^{5i\Psi_5} &= k_5 \epsilon_5 e^{i5\Phi_5}+k'_5 \epsilon_2 \epsilon_3 e^{i(2\Phi_2+3\Phi_3)}\,,
\end{aligned}
\end{equation}
where $k_n$ and $k'_n$ are coefficients related to the hydrodynamic response.

Interestingly, compared to $\text{NSC}_{\epsilon}(2,3,4)$ a sign change can be seen in $\text{NSC(2,3,4)}$ (Fig.~\ref{NSCmnp}(a)) which is similar to what has been observed for $\text{NSC(3,4)}$ and $\text{NSC}_{\epsilon}(3,4)$ in Fig.~\ref{NSCmn}(c) in Appendix~\ref{ss:AppendixD}. In the present case, we are dealing with genuine three-harmonic observable. However, the main difference between generalized SC of the initial state and the final state comes from the contribution of the non-linear term $\epsilon_2^2$ in the $v_4$. In fact, the term $\epsilon_2^2$ and its anti-correlation with $\epsilon_3$ should be responsible for this sign change similar to the $\text{NSC(3,4)}$ and $\text{NSC}_{\epsilon}(3,4)$ case (see Appendix~\ref{ss:AppendixD}). The same logic can explain the suppression of the cumulant $\text{NSC}(3,4,5)$ in Fig.~\ref{NSCmnp}(c) too. We know that there is a non-linear contribution with the term $\epsilon_2 \epsilon_3$ in $v_5$ (see Eq.~\eqref{nonlinear}).
As a result, the terms $\epsilon_2^2$  and $\epsilon_2 \epsilon_3$ can explain the small value of $\text{SC(3,4,5)}$ (or suppression of $\text{NSC}(3,4,5)$ in comparison with $\text{NSC}_{\epsilon}(3,4,5)$).  However, in $\text{NSC}(2,3,5)$ only the term $\epsilon_2\epsilon_3$ plays the role. As can be seen in Fig.~\ref{NSCmnp}(b), the effect of the term $\epsilon_2\epsilon_3$ is small in $\text{NSC}(2,3,5)$.
It is worth to note that in even simpler cumulants $\text{NSC}(2,3)$ and $\text{NSC}_{\epsilon}(2,3)$ (see Fig.~\ref{NSCmn}(b) in Appendix~\ref{ss:AppendixD}) we do not have such an agreement in centrality classes above $40\%$. We think there must be a more complex reason such as the presence of extra non-linear terms behind the approximate agreement between $\text{NSC}(2,3,5)$ and $\text{NSC}_{\epsilon}(2,3,5)$ in a wide range of centrality classes. Finally, compared to $\text{NSC}_{\epsilon}(2,4,6)$, we observe a considerable enhancement in $\text{NSC}(2,4,6)$ in Fig.~\ref{NSCmnp}(d). This enhancement is even larger than what has been observed in Fig.~\ref{NSCmn}(b) for $\text{SC(2,4)}$. The reason would be due to the fact that in $\text{NSC}(2,4,6)$ the terms $\epsilon_2^2$ in $v_4$ and the $\epsilon_2^3$ in $v_6$ are responsible for this enhancement. It seems the term $\epsilon_3^2$ in $v_6$ and its anti-correlation with $\epsilon_2$ does not have enough power to compete with $\epsilon_2$, $\epsilon_2^2$ and $\epsilon_2^3$ trivial correlations. It is important to point out that we have explained the generalized SC from the initial state and hydrodynamic response only qualitatively. In this context, a further rigorous study is needed to be done in the future.

\subsection{Estimating nonflow contribution with HIJING}
\label{ss:Estimating-nonflow-contribution-with-HIJING}

As detailed in Appendix~\ref{a:List-of-requirments}, one of requirements for our observables to be called SC is that it should be robust against nonflow. We have already studied in a Toy Monte Carlo simulation the case of strong two-particle correlation for different multiplicities in Sec.~\ref{ss:Nonflow-estimation-with-Toy-MonteCarlo-studies}. We now investigate nonflow contribution further and introduce HIJING, which stands for Heavy-Ion Jet INteraction Generator~\cite{Gyulassy:1994ew}. It is a Monte Carlo model used to study particle and jets production in nuclear collisions. It contains models to describe mechanisms like jet production, jet fragmentation, nuclear shadowing, etc. The correlations these mechanisms introduce involve generally only few particles and should not be included in the analysis of collective effects like anisotropic flow. Since HIJING has all the phenomena produced in a heavy-ion collision except flow itself, we can use it to test the robustness of our SC observable against few-particle nonflow correlations.

In general, when one uses an azimuthal correlator in an expression like
\begin{equation}
\left<e^{in(\varphi_1\!-\!\varphi_2)}\right> = \left<\cos[n(\varphi_1\!-\!\varphi_2)]\right> = v_n^2\,,
\end{equation}
one assumes in the derivation that the underlying two-variate p.d.f. $f(\varphi_1,\varphi_2)$ fully factorizes into the product of two marginal p.d.f.'s, i.e.:
\begin{equation}
f(\varphi_1,\varphi_2) = f_{\varphi_1}(\varphi_1)f_{\varphi_2}(\varphi_2)\,.
\end{equation}
\begin{figure}
	\begin{tabular}{c c}
		\includegraphics[scale=0.43]{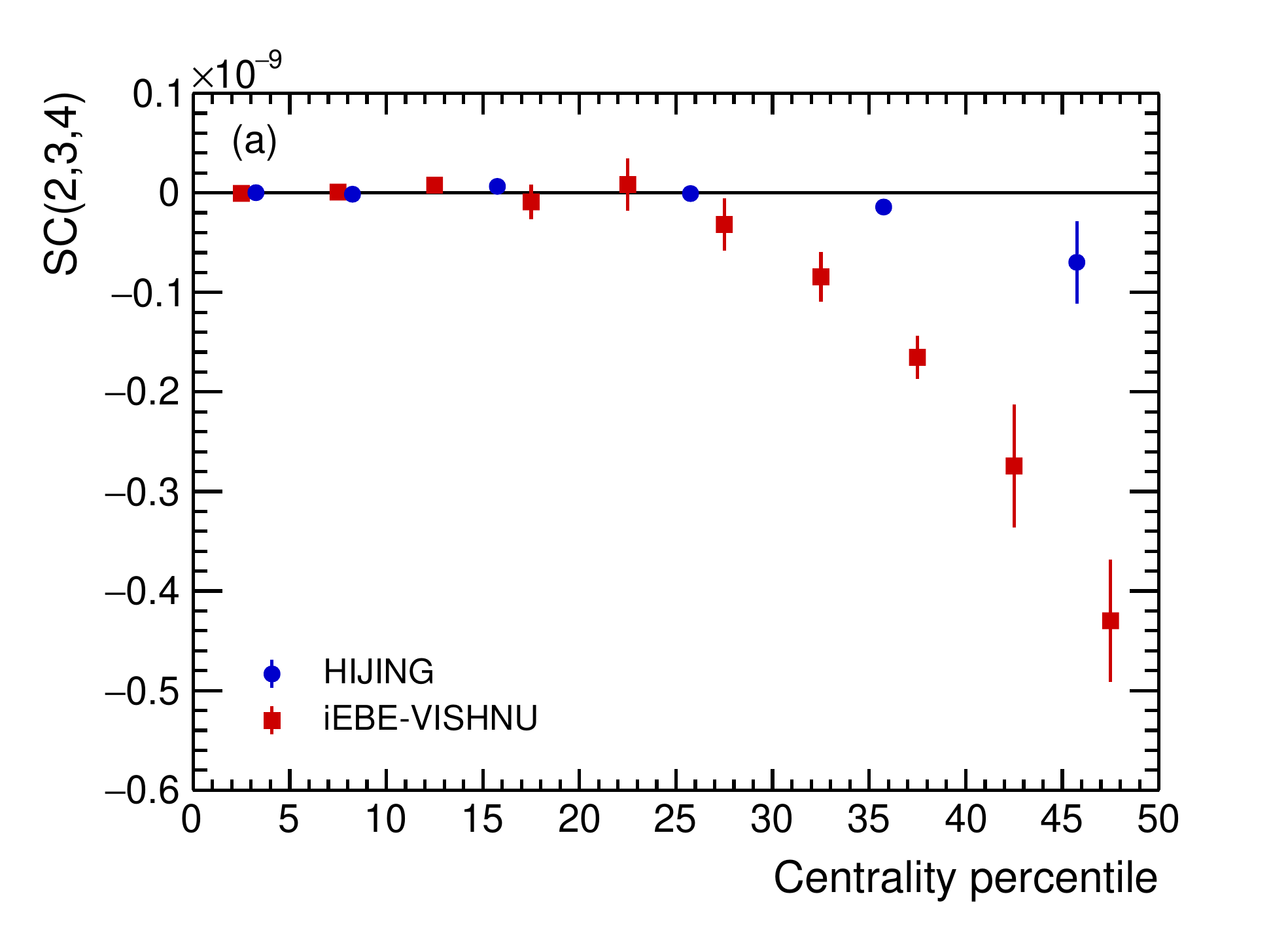} &
		\includegraphics[scale=0.43]{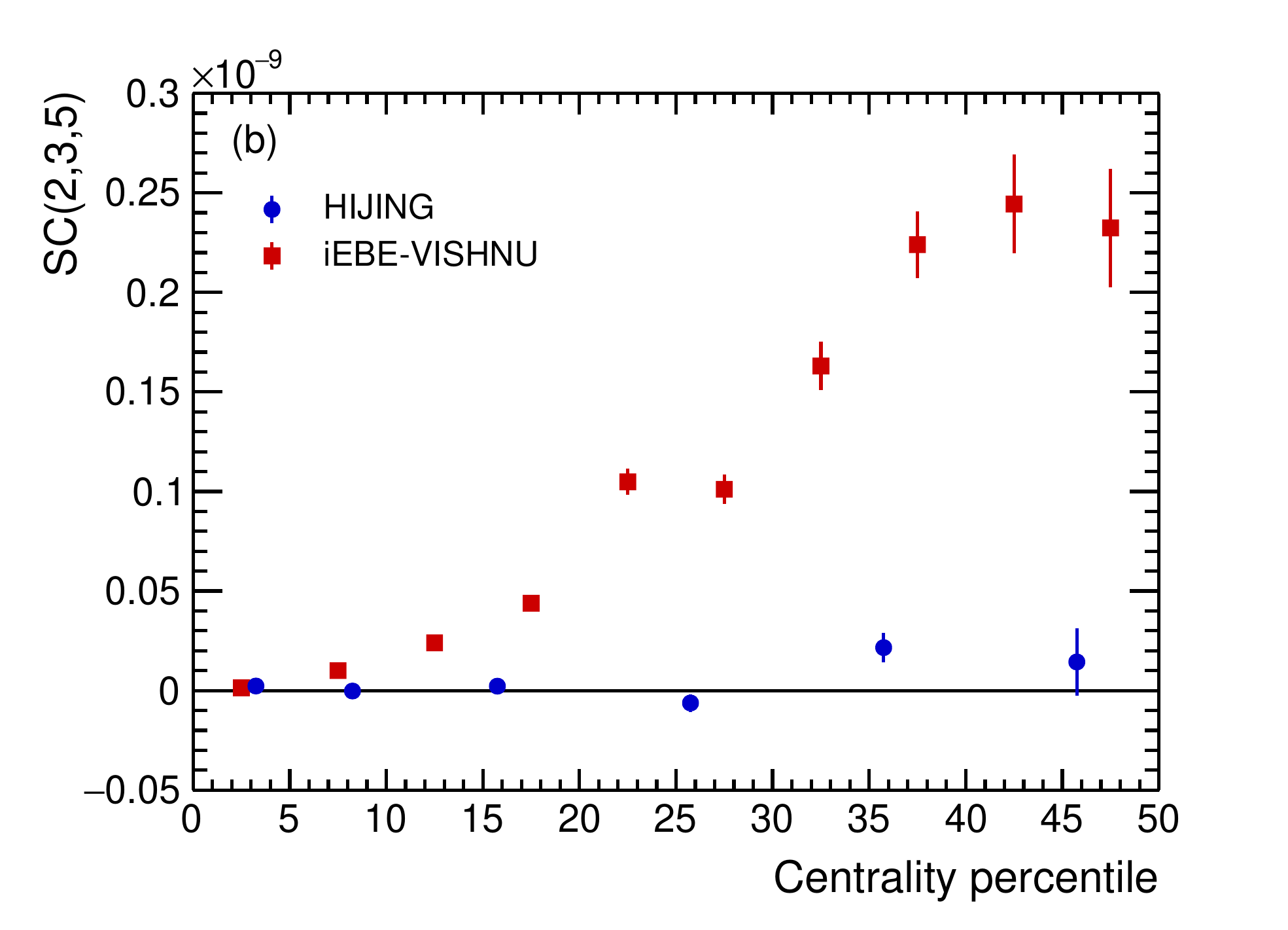}
	\end{tabular}
	\caption{Predictions for the centrality dependence of SC(2,3,4) (a) and SC(2,3,5) (b) from the HIJING and iEBE-VISHNU models. The $p_{T}$ range of the iEBE-VISHNU simulations (0.28 $< p_{T} <$ 4 GeV) have been chosen to be close to the HIJING one (0.2 $< p_{T} <$ 5 GeV).}
	\label{sect_Hijing_fig_HijingVishnu}
\end{figure}
If all azimuthal correlations are induced solely by anisotropic flow, such factorization is exactly satisfied. In the case of few-particle nonflow correlations the above factorization is broken, and flow amplitudes $v_n$ are now systematically biased when estimated with azimuthal correlators, i.e. at leading order we expect to have:
\begin{equation}
\left<e^{in(\varphi_1\!-\!\varphi_2)}\right> = v_n^2 + \delta_2\,,
\end{equation}
where $\delta_2$ denotes the systematic bias in 2-particle azimuthal correlations due to nonflow. We have written the above argument for the two-particle azimuthal correlations, but the argument can be trivially generalized for any larger number of particles (see the discussion on nonflow in Appendix~\ref{a:List-of-requirments}). It is impossible in practice to calculate and to quantify the systematic bias $\delta_2$ stemming from vastly different sources of nonflow correlations. Since HIJING has all relevant sources of correlations, except anisotropic flow, it is a great model to estimate solely the value of this systematic bias $\delta_2$, i.e. in HIJING we expect the following relation to hold:
\begin{equation}
\left<e^{in(\varphi_1\!-\!\varphi_2)}\right> = \delta_2\,,
\end{equation}
and analogously for higher order azimuthal correlators. If the given multi-particle azimuthal correlator, or some compound observable like SC which is estimated from azimuthal correlators, when measured in the HIJING dataset is consistent with 0, that means that nonflow contribution in that observable is negligible, and therefore it can be used as a reliable estimator of anisotropic flow properties. 

In this section we show the predictions from HIJING for the centrality dependence of two different combinations of harmonics: SC(2,3,4) and SC(2,3,5). The data used here correspond to Pb--Pb collisions taken at the center of mass energy of $\sqrt{s_{\text{NN}}}$ = 2.76 TeV. Two kinetic cuts have been applied: 0.2 $< p_{T} <$ 5 GeV and -0.8 $< \eta <$ 0.8. The results obtained with HIJING are shown on Fig.~\ref{sect_Hijing_fig_HijingVishnu} (a) for SC(2,3,4) and on Fig.~\ref{sect_Hijing_fig_HijingVishnu} (b) for SC(2,3,5), alongside with the VISHNU predictions for the same combinations of harmonics. We can see that in both cases, our new SC observable in HIJING is compatible with 0 for head-on and mid-central collisions, meaning our observable is robust against nonflow. The comparison also shows that flow has generally a more important contribution than nonflow. This means that any observation of nonzero values of SC in real data could be attributed to collective effects.

\section{Summary}
\label{s:Summary}

In summary, we have presented the generalization of recently introduced flow observables, dubbed Symmetric Cumulants. We have presented the unique way how the genuine multi-harmonic correlations can be estimated with multi-particle azimuthal correlators. All desired properties of higher order Symmetric Cumulants were tested with the carefully designed Toy Monte Carlo studies. By using the realistic iEBE-VISHNU model, we have demonstrated that their measurements are feasible and we have provided the first predictions for their centrality dependence in Pb--Pb collisions at LHC energies. A separate study has been presented for their values in the coordinate space. Study based on HIJING demonstrated that these new observables are robust against systematic biases due to nonflow correlations. These generalized, higher order observables, can be therefore used to extract reliably new and independent information about the initial conditions and the properties of QGP in high-energy nuclear collisions.

\acknowledgments{

This project has received funding from the European Research Council (ERC) under the European Union’s Horizon 2020 research and innovation programme (grant agreement No 759257).

This research was supported by the Munich Institute for Astro- and Particle Physics (MIAPP) of the DFG cluster of excellence ``Origin and Structure of the Universe.''
}

\appendix
\section{List of requirements for the generalization of Symmetric Cumulants}
\label{a:List-of-requirments}

In this self-contained appendix we establish how SC need to be generalized for the case of correlated fluctuations involving more than two flow amplitudes. Out of plenty of ways to perform such a generalization, we narrow down the one which uniquely satisfies the following list of requirements, which must hold for any number of flow amplitudes: 
\begin{enumerate}
	\item \textit{No built-in trivial contribution.} For constant flow amplitudes $v_k$, $v_l$, $v_m$, ..., the observable SC($k,l,m,\ldots$) is identically zero. 
	\item \textit{Genuine multi-harmonic correlations (cumulants).} If fluctuations only of a subset of amplitudes $v_k$, $v_l$, $v_m$, ..., are correlated, SC($k,l,m,\ldots$) is identically zero. This requirement ensures that the SC($k,l,m,\ldots$) observable is a well defined cumulant, i.e. it isolates only the genuine correlation among all flow amplitudes in question. This cumulant property secures that the SC extracts unique and independent information at each order: SC constructed from $n$ distinct flow amplitudes provides information which cannot be accessed from SC constructed from smaller number of $n-1$, $n-2$, ..., flow amplitudes. This requirement also ensures in particular that if fluctuations of amplitudes $v_k$, $v_l$, $v_m$, ..., are all mutually uncorrelated, SC($k,l,m,\ldots$) is identically zero. This requirement also ensures that all amplitudes in the SC definition must be different (e.g. if one starts with the definition of two-harmonic SC$(k,l) \equiv \left<v_k^2v_l^2\right>-\left<v_k^2\right>\left<v_l^2\right>$ and identifies in the resulting expression $l=k$, then SC($k,l$) is trivially non-zero as it reduces to the variance $\left<v_k^4\right> - \left<v_k^2\right>^2$ of one flow amplitude). For other formal mathematical properties which any cumulants must satisfy we refer to~\cite{Kubo}. 
	\item \textit{Symmetry}. Observable SC$(k,l,m,\ldots)$ must be the same by definition for any permutation of amplitudes $v_k,v_l,v_m,\ldots$. Example: $\mathrm{SC}(k,l) = \mathrm{SC}(l,k)$, $\mathrm{SC}(k,l,m) = \mathrm{SC}(k,m,l) = \mathrm{SC}(l,k,m) = \mathrm{SC}(l,m,k) = \mathrm{SC}(m,l,k) = \mathrm{SC}(m,k,l)$, etc. This requirement distinguishes SC from other possibilities of building cumulants from flow amplitudes. As a consequence of this requirement, all amplitudes in the definition of SC must be raised to the same power.
	\item \textit{Cleanliness}. Any dependence on the symmetry planes must be canceled out by definition in SC. This is an important requirement, since in general in Fourier series parameterizations of anisotropic distributions (see Eq.~(\ref{eq:FourierSeries_vn_psin})), both flow amplitudes $v_n$ and symmetry planes $\Psi_n$ appear at equal footing, since they are two independent degrees of freedom to quantify anisotropic flow phenomenon. This requirement restricts tremendously the number of azimuthal correlators which are suitable to build the SC. There are two cases of interest in this context which need to be clarified:
	\begin{enumerate}
		\item In all azimuthal correlators from which SC are built, one always have {\it equal} number of positive and negative harmonics (see e.g. Eqs.~(\ref{eq:3pSC}), (\ref{eq:SC(k,l,m,n)_exp}) or (\ref{eq:A}-\ref{eq:E})). Therefore, if there are $k$ azimuthal angles in the multi-particle correlator, we can always label the corresponding harmonics as $n_1,n_2,\ldots, n_{k/2}$ and $-n_1,-n_2,\ldots,-n_{k/2}$. It follows directly from the general result in Eq.~(\ref{eq:generalResult}):
		\begin{eqnarray}
		\left<e^{i(n_1\varphi_1+\ldots+n_{k/2}\varphi_{k/2}-n_1\varphi_{k/2 + 1}-\ldots-n_{k/2}\varphi_{k})}\right> &=& v_{n_1}^2\cdots v_{n_{k/2}}^2
		e^{i[n_1(\Psi_{n_1}-\Psi_{n_1})+\ldots+n_{k/2}(\Psi_{n_{k/2}}-\Psi_{n_{k/2}})]}\nonumber\\
		&=& v_{n_1}^2\cdots v_{n_{k/2}}^2\,.
		\end{eqnarray}
		In the 1st line above we have used the fact that for any Fourier harmonic $n$ we have $v_n = v_{-n}$ and $\Psi_n = \Psi_{-n}$, which can be derived straight from the formal properties of Fourier series. Therefore, for the case when symmetry planes are all different and do not fluctuate within the same event, by definition SC observables do not have any contribution from symmetry planes;
		\item Experimentally, if the symmetry planes $\Psi_n$ change with $\eta$ or $p_T$, which is known as decorrelation or factorization breaking~\cite{Jia:2014vja,Pang:2014pxa,Khachatryan:2015oea,Bozek:2017qir}, the differential analysis of SC observables might still exhibit dependence on symmetry planes. In order to avoid this problem, we advocate the differential SC analyses in the narrow phase-space windows of interest, e.g. narrow $\Delta\eta$ bins, when all particles from the multiplet must be taken from the same $\Delta\eta$ bin.
	\end{enumerate}
	\item \textit{Isotropy.} All azimuthal correlators used in estimating SC experimentally must be isotropic since these are the only correlators which for a detector with uniform acceptance give non-trivial contributions, after the single-event averages of azimuthal correlators have been extended to all-event averages. Isotropic azimuthal correlators are the correlators for which the harmonics satisfy the relation: $\sum_{i} n_i$ = 0~\cite{Bhalerao:2011yg}. Such correlators are invariant under the arbitrary rotation of the coordinate system in which azimuthal angles are being measured.   
	\item \textit{Uniqueness.} No other combination of multi-particle azimuthal correlators shall yield to the same final expression for flow amplitudes. For instance, out of two possibilities to estimate SC$(m,n)$: 
	\begin{eqnarray}
	{\rm SC}(m,n)_{\rm correct} &\equiv&\left<\left<\cos[m\varphi_1\!+\!n\varphi_2\!-\!m\varphi_3\!-\!n\varphi_4\right>\right>\!-\!\left<\left<\cos[m(\varphi_1\!-\!\varphi_2)]\right>\right>\left<\left<\cos[n(\varphi_1\!-\!\varphi_2)]\right>\right>\,,\label{eq:SC(m,n)_standard}\\
	{\rm SC}(m,n)_{\rm wrong} &\equiv& \left<\left<\cos[m(\varphi_1\!-\!\varphi_2)]\right>\left<\cos[n(\varphi_1\!-\!\varphi_2)]\right>\right>\!-\!\left<\left<\cos[m(\varphi_1\!-\!\varphi_2)]\right>\right>\left<\left<\cos[n(\varphi_1\!-\!\varphi_2)]\right>\right>,\label{eq:SC(m,n)_fake}
	\end{eqnarray}
	only the definition in Eq.~(\ref{eq:SC(m,n)_standard}) leads to the desired properties of SC. While it is clear, by using Eq.~(\ref{eq:generalResult}), that definition in Eq.~(\ref{eq:SC(m,n)_standard}) evaluates to the desired expression $\left<v_m^2v_n^2\right>-\left<v_m^2\right>\left<v_n^2\right>$, it is not so obvious that the alternative definition in Eq.~(\ref{eq:SC(m,n)_fake}) evaluates to something else. We clarify this technical point further in Appendix~\ref{a:Checking-the-other-requirements-for-the-Symmetric-Cumulants}. This property implies that SC must be estimated experimentally in terms of all-event averages of each individual azimuthal correlator.
	\item \textit{Robustness against nonflow.} SC must be consistent with zero for a system containing only a few-particle nonflow correlations. Mathematically, nonflow correlations are all correlations for which the factorization of joint multi-variate p.d.f. of $k$-azimuthal angles cannot be written as a product of single-particle p.d.f.'s., i.e. for nonflow correlations:
	\begin{equation}
	f(\varphi_1,\ldots,\varphi_k)\neq f(\varphi_1)\cdots f(\varphi_k)\,.
	\end{equation}
	If the particle emission is dominated by anisotropic flow, in the above expression we have equality. For $k$-particle azimuthal correlator, the nonflow contribution $\delta_k$ exhibits the universal generic scaling as a function of multiplicity $M$:
	\begin{equation}
	\delta_k\sim\frac{1}{M^{k-1}}\,.
	\end{equation}
	By using the above result, it follows that in general one obtains the generic scaling also for the nonflow contribution $\delta^{\rm SC}_{k}$ in $k$-harmonic SC observable:
	\begin{equation}
	\delta^{\rm SC}_{k} \sim \delta_{2k} + \cdots + \prod_{i=1}^k \delta_2 \sim \sum_{i=k}^{2k-1} \frac{1}{M^i}\,.
	\label{eq:nonflowScalingSC}
	\end{equation}
    The leading order nonflow contribution of $k$-harmonic SC observable is the same as for $2k$-azimuthal correlator. For instance, the scaling of nonflow contribution to SC($k,l,m$) can be described well for any choice of harmonics $k$, $l$ and $m$ with the following relation:
	\begin{equation}
	\delta^{\rm SC}_3 = \frac{\alpha}{M^5} + \frac{\beta}{M^4} + \frac{\gamma}{M^3}\,,
	\label{eq:nonflowScalingSC_3-harmonic}
	\end{equation}
	where constants $\alpha$, $\beta$ and $\gamma$ quantify various sources of nonflow contribution and can be obtained from the fit of multiplicity dependence of corresponding SC observable. Any violation from the few-particle nonflow scaling in~Eq.~(\ref{eq:nonflowScalingSC}) indicates that SC is dominated by contribution from collective correlations, and robust against nonflow. In practice, the nonflow contribution can be estimated by evaluating the SC observable over the data simulated with HIJING event generator (see Sec.~\ref{ss:Estimating-nonflow-contribution-with-HIJING}), which contains only nonflow correlations. 
	\item \textit{Event weights.} When estimating SC, it is essential that all azimuthal correlators are expressed as all-event averages $\left<\left<\cdots\right>\right>$. This raises a non-trivial question of what is the correct weight to use to weight the event-by-event averages $\left<\cdots\right>$ of azimuthal correlators. Namely, in the following generic expression:
	\begin{equation}
	\left<\left<\cdots\right>\right> = \frac{\sum_i w_i \left<\cdots\right>_i}{\sum_i w_i}\,,
	\label{eq:weightsGeneric}
	\end{equation}
	where the two sums go over all events, there are various possibilities to choose event weights $w_i$ from. Different choices will in general yield both to the different statistical and systematical properties of the final SC results. In the present work, we have only tested that the standard choice, namely `number of combinations' event weight~\cite{Bilandzic:2010jr}, yields to the smallest statistical spread of SC (see Appendix~\ref{a:Checking-the-other-requirements-for-the-Symmetric-Cumulants}). There is no performance penalty due to double counting (e.g. by treating $\cos[n(\varphi_1\!-\!\varphi_2)]$ and $\cos[n(\varphi_2\!-\!\varphi_1)]$ separately, despite the fact that cosine is an even function) when using this weight, since all expressions for multi-particle azimuthal correlations (which by definition include double counting) are analytically expressed in terms of $Q$-vectors~\cite{Bilandzic:2013kga}. For subevent cumulants, the optimal weight is `number of unique combinations'. For the specific case when the event weights depends only on multiplicity, one can completely circumvent this problem in the following two independent ways:
	\begin{enumerate}
		\item One bins all available data in bins of multiplicity, where the bin width is 1. The final SC results in each multiplicity bin are independent and therefore can be trivially combined with the standard $1/\sigma^2$ weighting, where $\sigma$ is the spread of SC results in a given unit multiplicity bin;
		\item For a given subsample of the whole dataset (e.g. events corresponding to the specific centrality), one randomly selects for the analysis in each event the same number of particles (typically corresponding to the smallest multiplicity in that subsample).
	\end{enumerate}
	In both cases a) and b), the weight $w_i$ in Eq.~(\ref{eq:weightsGeneric}) is a constant function of multiplicity by construction, and any constant weight is equivalent to the unit weight. We leave the further study of event weights for our future work. This non-trivial problem was addressed recently also in~\cite{Gardim:2016nrr}.
\end{enumerate}

As an example, we now demonstrate how the Requirements 1--3 are satisfied for the SC$(k,l,m)$ defined in Eq.~(\ref{SC(k,l,m)_flowHarmonics}). It is trivial to verify that Eq.~(\ref{SC(k,l,m)_flowHarmonics}) is identically zero if all three amplitudes are constant (Requirement~1). When all three amplitudes fluctuate independently, all averages in Eq.~(\ref{SC(k,l,m)_flowHarmonics}) can be factorized completely and it follows: 
\begin{eqnarray}
\mathrm{SC}(k,l,m)\!=\!\left<v_k^2\right>\left<v_l^2\right>\left<v_m^2\right> 
\!-\!\left<v_k^2\right>\left<v_l^2\right>\left<v_m^2\right>
\!-\!\left<v_k^2\right>\left<v_m^2\right>\left<v_l^2\right>
\!-\!\left<v_l^2\right>\left<v_m^2\right>\left<v_k^2\right>
\!+\!2 \left<v_k^2\right>\left<v_l^2\right>\left<v_m^2\right> 
\!=\!0.
\label{SC(k,l,m)_threeIndependnentHarmonics}
\end{eqnarray}
When all three amplitudes fluctuate, but only two fluctuations are correlated (e.g. among harmonics $k$ and $l$), one can only partially factorize some of the averages in Eq.~(\ref{SC(k,l,m)_flowHarmonics}), but the1 resulting observable is nevertheless still identically zero: 
\begin{equation}
\mathrm{SC}(k,l,m) = \left<v_k^2v_l^2\right>\left<v_m^2\right> 
- \left<v_k^2v_l^2\right>\left<v_m^2\right>
- \left<v_k^2\right>\left<v_m^2\right>\left<v_l^2\right>
- \left<v_l^2\right>\left<v_m^2\right>\left<v_k^2\right>
+ 2 \left<v_k^2\right>\left<v_l^2\right>\left<v_m^2\right> 
= 0.
\label{SC(k,l,m)_twoIndependnentHarmonics}
\end{equation}
The same conclusion follows if only correlations among the other two amplitudes, namely $v_k$ and $v_m$, or $v_m$ and $v_l$, are correlated. The results in Eqs.~(\ref{SC(k,l,m)_threeIndependnentHarmonics}) and (\ref{SC(k,l,m)_twoIndependnentHarmonics}) demonstrate that $\mathrm{SC}(k,l,m)$ observable is well defined cumulant of flow amplitudes (Requirement~2). Trivially, $\mathrm{SC}(k,l,m)$ is symmetric in all three harmonics (Requirement~3). The remaining requirements are discussed in detail in Appendix~\ref{a:Checking-the-other-requirements-for-the-Symmetric-Cumulants} with Monte Carlo studies.

\section{Definitions for higher order Symmetric Cumulants}
\label{a:Definitions-for-higher-order-Symmetric-Cumulants}

In this appendix, we write down explicitly the equations of few higher order SC, which satisfy all requirements set upon their generalization in Appendix~\ref{a:List-of-requirments}. We provide first the theoretical definitions in terms of flow amplitudes, and then the corresponding relations in terms of azimuthal correlators, which can be used to estimate SC experimentally. 


The four-harmonic SC$(k,l,m,n)$ we define directly as:
\begin{eqnarray}
\mathrm{SC}(k,l,m,n) & = & \left<v_{k}^{2}v_{l}^{2}v_{m}^{2}v_{n}^{2}\right>
-\left<v_{k}^{2}v_{l}^{2}v_{m}^{2}\right>\left<v_{n}^{2}\right>
- \left<v_{k}^{2}v_{l}^{2}v_{n}^{2}\right>\left<v_{m}^{2}\right> 
- \left<v_{k}^{2}v_{m}^{2}v_{n}^{2}\right>\left<v_{l}^{2}\right> 
- \left<v_{l}^{2}v_{m}^{2}v_{n}^{2}\right>\left<v_{k}^{2}\right>\nonumber\\
&-&\left<v_{k}^{2}v_{l}^{2}\right>\left<v_{m}^{2}v_{n}^{2}\right> 
-\left<v_{k}^{2}v_{m}^{2}\right>\left<v_{l}^{2}v_{n}^{2}\right>
-\left<v_{k}^{2}v_{n}^{2}\right>\left<v_{l}^{2}v_{m}^{2}\right>\nonumber\\
&+& 2\big(\left<v_{k}^{2}v_{l}^{2}\right>\left<v_{m}^{2}\right>\left<v_{n}^{2}\right>
+\left<v_{k}^{2}v_{m}^{2}\right>\left<v_{l}^{2}\right>\left<v_{n}^{2}\right>
+\left<v_{k}^{2}v_{n}^{2}\right>\left<v_{l}^{2}\right>\left<v_{m}^{2}\right>\nonumber\\
&&+\left<v_{l}^{2}v_{m}^{2}\right>\left<v_{k}^{2}\right>\left<v_{n}^{2}\right>
+\left<v_{l}^{2}v_{n}^{2}\right>\left<v_{k}^{2}\right>\left<v_{m}^{2}\right>
+\left<v_{m}^{2}v_{n}^{2}\right>\left<v_{k}^{2}\right>\left<v_{l}^{2}\right>\big)\nonumber\\
&-&6\left<v_{k}^{2}\right>\left<v_{l}^{2}\right>\left<v_{m}^{2}\right>\left<v_{n}^{2}\right>\,.
\label{eq:SC(k,l,m,n)}
\end{eqnarray}
Experimentally, SC$(k,l,m,n)$ can be estimated with the following combination of azimuthal correlators:
\begin{eqnarray}
\mathrm{SC}(k,l,m,n) &=&\left<\left<\cos[k\varphi_1\!+\!l\varphi_2\!+\!m\varphi_3\!+\!n\varphi_4\!-\!k\varphi_5\!-\!l\varphi_6\!-\!m\varphi_7\!-\!n\varphi_8]\right>\right>\nonumber\\
&-&\left<\left<\cos[k\varphi_1\!+\!l\varphi_2\!+\!m\varphi_3\!-\!k\varphi_4\!-\!l\varphi_5\!-\!m\varphi_6]\right>\right>\left<\left<\cos[n(\varphi_1\!-\!\varphi_2)]\right>\right>\nonumber\\
&-&\left<\left<\cos[k\varphi_1\!+\!l\varphi_2\!+\!n\varphi_3\!-\!k\varphi_4\!-\!l\varphi_5\!-\!n\varphi_6]\right>\right>\left<\left<\cos[m(\varphi_1\!-\!\varphi_2)]\right>\right>\nonumber\\
&-&\left<\left<\cos[k\varphi_1\!+\!m\varphi_2\!+\!n\varphi_3\!-\!k\varphi_4\!-\!m\varphi_5\!-\!n\varphi_6]\right>\right>\left<\left<\cos[l(\varphi_1\!-\!\varphi_2)]\right>\right>\nonumber\\
&-&\left<\left<\cos[l\varphi_1\!+\!m\varphi_2\!+\!n\varphi_3\!-\!l\varphi_4\!-\!m\varphi_5\!-\!n\varphi_6]\right>\right>\left<\left<\cos[k(\varphi_1\!-\!\varphi_2)]\right>\right>\nonumber\\
&-&\left<\left<\cos[k\varphi_1\!+\!l\varphi_2\!-\!k\varphi_3\!-\!l\varphi_4]\right>\right>\left<\left<\cos[m\varphi_1\!+\!n\varphi_2\!-\!m\varphi_3\!-\!n\varphi_4]\right>\right>\nonumber\\
&-&\left<\left<\cos[k\varphi_1\!+\!m\varphi_2\!-\!k\varphi_3\!-\!m\varphi_4]\right>\right>\left<\left<\cos[l\varphi_1\!+\!n\varphi_2\!-\!l\varphi_3\!-\!n\varphi_4]\right>\right>\nonumber\\
&-&\left<\left<\cos[k\varphi_1\!+\!n\varphi_2\!-\!k\varphi_3\!-\!n\varphi_4]\right>\right>\left<\left<\cos[l\varphi_1\!+\!m\varphi_2\!-\!l\varphi_3\!-\!m\varphi_4]\right>\right>\nonumber\\
&+&2\left<\left<\cos[k\varphi_1\!+\!l\varphi_2\!-\!k\varphi_3\!-\!l\varphi_4]\right>\right>\left<\left<\cos[m(\varphi_1\!-\!\varphi_2)]\right>\right>\left<\left<\cos[n(\varphi_1\!-\!\varphi_2)]\right>\right>\nonumber\\
&+&2\left<\left<\cos[k\varphi_1\!+\!m\varphi_2\!-\!k\varphi_3\!-\!m\varphi_4]\right>\right>\left<\left<\cos[l(\varphi_1\!-\!\varphi_2)]\right>\right>\left<\left<\cos[n(\varphi_1\!-\!\varphi_2)]\right>\right>\nonumber\\
&+&2\left<\left<\cos[k\varphi_1\!+\!n\varphi_2\!-\!k\varphi_3\!-\!n\varphi_4]\right>\right>\left<\left<\cos[l(\varphi_1\!-\!\varphi_2)]\right>\right>\left<\left<\cos[m(\varphi_1\!-\!\varphi_2)]\right>\right>\nonumber\\
&+&2\left<\left<\cos[l\varphi_1\!+\!m\varphi_2\!-\!l\varphi_3\!-\!m\varphi_4]\right>\right>\left<\left<\cos[k(\varphi_1\!-\!\varphi_2)]\right>\right>\left<\left<\cos[n(\varphi_1\!-\!\varphi_2)]\right>\right>\nonumber\\
&+&2\left<\left<\cos[l\varphi_1\!+\!n\varphi_2\!-\!l\varphi_3\!-\!n\varphi_4]\right>\right>\left<\left<\cos[k(\varphi_1\!-\!\varphi_2)]\right>\right>\left<\left<\cos[m(\varphi_1\!-\!\varphi_2)]\right>\right>\nonumber\\
&+&2\left<\left<\cos[m\varphi_1\!+\!n\varphi_2\!-\!m\varphi_3\!-\!n\varphi_4]\right>\right>\left<\left<\cos[k(\varphi_1\!-\!\varphi_2)]\right>\right>\left<\left<\cos[l(\varphi_1\!-\!\varphi_2)]\right>\right>\nonumber\\
&-&6\left<\left<\cos[k(\varphi_1\!-\!\varphi_2)]\right>\right>
\left<\left<\cos[l(\varphi_1\!-\!\varphi_2)]\right>\right>
\left<\left<\cos[m(\varphi_1\!-\!\varphi_2)]\right>\right>
\left<\left<\cos[n(\varphi_1\!-\!\varphi_2)]\right>\right>\,.
\label{eq:SC(k,l,m,n)_exp}
\end{eqnarray}
%


The five-harmonic SC$(k,l,m,n,o)$ we define directly as:
\begin{eqnarray}
\mathrm{SC}(k,l,m,n,o) & = & \left<v_{k}^{2}v_{l}^{2}v_{m}^{2}v_{n}^{2}v_{o}^{2}\right>
-\left<v_{k}^{2}v_{l}^{2}v_{m}^{2}v_{n}^{2}\right>\left<v_{o}^{2}\right>
-\left<v_{k}^{2}v_{l}^{2}v_{m}^{2}v_{o}^{2}\right>\left<v_{n}^{2}\right>\nonumber\\
&-&\left<v_{k}^{2}v_{l}^{2}v_{n}^{2}v_{o}^{2}\right>\left<v_{m}^{2}\right> -
\left<v_{k}^{2}v_{m}^{2}v_{n}^{2}v_{o}^{2}\right>\left<v_{l}^{2}\right> - \left<v_{l}^{2}v_{m}^{2}v_{n}^{2}v_{o}^{2}\right>\left<v_{k}^{2}\right>\nonumber\\
&-&\left<v_{k}^{2}v_{l}^{2}v_{m}^{2}\right>\left<v_{n}^{2}v_{o}^{2}\right>
-\left<v_{k}^{2}v_{l}^{2}v_{n}^{2}\right>\left<v_{m}^{2}v_{o}^{2}\right>
-\left<v_{k}^{2}v_{l}^{2}v_{o}^{2}\right>\left<v_{m}^{2}v_{n}^{2}\right>
-\left<v_{k}^{2}v_{m}^{2}v_{n}^{2}\right>\left<v_{l}^{2}v_{o}^{2}\right>\nonumber\\
&-&\left<v_{k}^{2}v_{m}^{2}v_{o}^{2}\right>\left<v_{l}^{2}v_{n}^{2}\right>
-\left<v_{k}^{2}v_{n}^{2}v_{o}^{2}\right>\left<v_{l}^{2}v_{m}^{2}\right>
-\left<v_{l}^{2}v_{m}^{2}v_{n}^{2}\right>\left<v_{k}^{2}v_{o}^{2}\right>
-\left<v_{l}^{2}v_{m}^{2}v_{o}^{2}\right>\left<v_{k}^{2}v_{n}^{2}\right>\nonumber\\
&-&\left<v_{l}^{2}v_{n}^{2}v_{o}^{2}\right>\left<v_{k}^{2}v_{m}^{2}\right>
-\left<v_{m}^{2}v_{n}^{2}v_{o}^{2}\right>\left<v_{k}^{2}v_{l}^{2}\right>\nonumber\\
&+&2\big(\left<v_{k}^{2}v_{l}^{2}v_{m}^{2}\right>\left<v_{n}^{2}\right>\left<v_{o}^{2}\right>
+\left<v_{k}^{2}v_{l}^{2}v_{n}^{2}\right>\left<v_{m}^{2}\right>\left<v_{o}^{2}\right>
+\left<v_{k}^{2}v_{l}^{2}v_{o}^{2}\right>\left<v_{m}^{2}\right>\left<v_{n}^{2}\right>\nonumber\\
&&+\left<v_{k}^{2}v_{m}^{2}v_{n}^{2}\right>\left<v_{l}^{2}\right>\left<v_{o}^{2}\right>  
+\left<v_{k}^{2}v_{m}^{2}v_{o}^{2}\right>\left<v_{l}^{2}\right>\left<v_{n}^{2}\right>
+\left<v_{k}^{2}v_{n}^{2}v_{o}^{2}\right>\left<v_{l}^{2}\right>\left<v_{m}^{2}\right>\nonumber\\
&&+\left<v_{l}^{2}v_{m}^{2}v_{n}^{2}\right>\left<v_{k}^{2}\right>\left<v_{o}^{2}\right>  
+\left<v_{l}^{2}v_{m}^{2}v_{o}^{2}\right>\left<v_{k}^{2}\right>\left<v_{n}^{2}\right>
+\left<v_{l}^{2}v_{n}^{2}v_{o}^{2}\right>\left<v_{k}^{2}\right>\left<v_{m}^{2}\right>\nonumber\\
&&+\left<v_{m}^{2}v_{n}^{2}v_{o}^{2}\right>\left<v_{k}^{2}\right>\left<v_{l}^{2}\right>\big)\nonumber\\
&+&2\big(\left<v_{k}^{2}v_{n}^{2}\right>\left<v_{l}^{2}v_{m}^{2}\right>\left<v_{o}^{2}\right>
+\left<v_{k}^{2}v_{o}^{2}\right>\left<v_{l}^{2}v_{m}^{2}\right>\left<v_{n}^{2}\right>
+\left<v_{k}^{2}v_{m}^{2}\right>\left<v_{l}^{2}v_{n}^{2}\right>\left<v_{o}^{2}\right>\nonumber\\
&&+\left<v_{k}^{2}v_{o}^{2}\right>\left<v_{l}^{2}v_{n}^{2}\right>\left<v_{m}^{2}\right>
+\left<v_{k}^{2}v_{m}^{2}\right>\left<v_{l}^{2}v_{o}^{2}\right>\left<v_{n}^{2}\right>
+\left<v_{k}^{2}v_{n}^{2}\right>\left<v_{l}^{2}v_{o}^{2}\right>\left<v_{m}^{2}\right>\nonumber\\
&&+\left<v_{k}^{2}v_{l}^{2}\right>\left<v_{m}^{2}v_{n}^{2}\right>\left<v_{o}^{2}\right>
+\left<v_{k}^{2}v_{o}^{2}\right>\left<v_{m}^{2}v_{n}^{2}\right>\left<v_{l}^{2}\right>
+\left<v_{l}^{2}v_{o}^{2}\right>\left<v_{m}^{2}v_{n}^{2}\right>\left<v_{k}^{2}\right>\nonumber\\
&&+\left<v_{k}^{2}v_{l}^{2}\right>\left<v_{m}^{2}v_{o}^{2}\right>\left<v_{n}^{2}\right>
+\left<v_{k}^{2}v_{n}^{2}\right>\left<v_{m}^{2}v_{o}^{2}\right>\left<v_{l}^{2}\right>
+\left<v_{l}^{2}v_{n}^{2}\right>\left<v_{m}^{2}v_{o}^{2}\right>\left<v_{k}^{2}\right>\nonumber\\
&&+\left<v_{k}^{2}v_{l}^{2}\right>\left<v_{n}^{2}v_{o}^{2}\right>\left<v_{m}^{2}\right>
+\left<v_{k}^{2}v_{m}^{2}\right>\left<v_{n}^{2}v_{o}^{2}\right>\left<v_{l}^{2}\right>
+\left<v_{l}^{2}v_{m}^{2}\right>\left<v_{n}^{2}v_{o}^{2}\right>\left<v_{k}^{2}\right>\big)\nonumber\\
&-&6\big(\left<v_{k}^{2}v_{l}^{2}\right>\left<v_{m}^{2}\right>\left<v_{n}^{2}\right>\left<v_{o}^{2}\right> 
+\left<v_{k}^{2}v_{m}^{2}\right>\left<v_{l}^{2}\right>\left<v_{n}^{2}\right>\left<v_{o}^{2}\right>
+\left<v_{k}^{2}v_{n}^{2}\right>\left<v_{l}^{2}\right>\left<v_{m}^{2}\right>\left<v_{o}^{2}\right>\nonumber\\
&&+\left<v_{k}^{2}v_{o}^{2}\right>\left<v_{l}^{2}\right>\left<v_{m}^{2}\right>\left<v_{n}^{2}\right> 
+\left<v_{l}^{2}v_{m}^{2}\right>\left<v_{k}^{2}\right>\left<v_{n}^{2}\right>\left<v_{o}^{2}\right>
+\left<v_{l}^{2}v_{n}^{2}\right>\left<v_{k}^{2}\right>\left<v_{m}^{2}\right>\left<v_{o}^{2}\right>\nonumber\\
&&+\left<v_{l}^{2}v_{o}^{2}\right>\left<v_{k}^{2}\right>\left<v_{m}^{2}\right>\left<v_{n}^{2}\right> 
+\left<v_{m}^{2}v_{n}^{2}\right>\left<v_{k}^{2}\right>\left<v_{l}^{2}\right>\left<v_{o}^{2}\right>
+\left<v_{m}^{2}v_{o}^{2}\right>\left<v_{k}^{2}\right>\left<v_{l}^{2}\right>\left<v_{n}^{2}\right>\nonumber\\
&&+\left<v_{n}^{2}v_{o}^{2}\right>\left<v_{k}^{2}\right>\left<v_{l}^{2}\right>\left<v_{m}^{2}\right>\big)\nonumber\\
&+& 24\left<v_{k}^{2}\right>\left<v_{l}^{2}\right>\left<v_{m}^{2}\right>\left<v_{n}^{2}\right>\left<v_{o}^{2}\right>\,.
\label{eq:SC(k,l,m,n,o)}
\end{eqnarray}
Experimentally, SC$(k,l,m,n,o)$ can be estimated with the following expression:
\begin{equation}
\mathrm{SC}(k,l,m,n,o) = A + 2B + 2C - 6D + 24E\,,  
\label{eq:SC(k,l,m,n,o)_exp_inter}
\end{equation}
where
\begin{eqnarray}
A &\equiv&\left<\left<\cos[k\varphi_1\!+\!l\varphi_2\!+\!m\varphi_3\!+\!n\varphi_4\!+\!o\varphi_5\!-\!k\varphi_6\!-\!l\varphi_7\!-\!m\varphi_8\!-\!n\varphi_9\!-\!o\varphi_{10}]\right>\right>\nonumber\\
&-&\left<\left<\cos[k\varphi_1\!+\!l\varphi_2\!+\!m\varphi_3\!+\!n\varphi_4\!-\!k\varphi_5\!-\!l\varphi_6\!-\!m\varphi_7\!-\!n\varphi_8]\right>\right>\left<\left<\cos[o(\varphi_1\!-\!\varphi_2)]\right>\right>\nonumber\\
&-&\left<\left<\cos[k\varphi_1\!+\!l\varphi_2\!+\!m\varphi_3\!+\!o\varphi_4\!-\!k\varphi_5\!-\!l\varphi_6\!-\!m\varphi_7\!-\!o\varphi_8]\right>\right>\left<\left<\cos[n(\varphi_1\!-\!\varphi_2)]\right>\right>\nonumber\\
&-&\left<\left<\cos[k\varphi_1\!+\!l\varphi_2\!+\!n\varphi_3\!+\!o\varphi_4\!-\!k\varphi_5\!-\!l\varphi_6\!-\!n\varphi_7\!-\!o\varphi_8]\right>\right>\left<\left<\cos[m(\varphi_1\!-\!\varphi_2)]\right>\right>\nonumber\\
&-&\left<\left<\cos[k\varphi_1\!+\!m\varphi_2\!+\!n\varphi_3\!+\!o\varphi_4\!-\!k\varphi_5\!-\!m\varphi_6\!-\!n\varphi_7\!-\!o\varphi_8]\right>\right>\left<\left<\cos[l(\varphi_1\!-\!\varphi_2)]\right>\right>\nonumber\\
&-&\left<\left<\cos[l\varphi_1\!+\!m\varphi_2\!+\!n\varphi_3\!+\!o\varphi_4\!-\!l\varphi_5\!-\!m\varphi_6\!-\!n\varphi_7\!-\!o\varphi_8]\right>\right>\left<\left<\cos[k(\varphi_1\!-\!\varphi_2)]\right>\right>\nonumber\\
&-&\left<\left<\cos[k\varphi_1\!+\!l\varphi_2\!+\!m\varphi_3\!-\!k\varphi_4\!-\!l\varphi_5\!-\!m\varphi_6]\right>\right>\left<\left<\cos[n\varphi_1\!+\!o\varphi_2\!-\!n\varphi_3\!-\!o\varphi_4]\right>\right>\nonumber\\
&-&\left<\left<\cos[k\varphi_1\!+\!l\varphi_2\!+\!n\varphi_3\!-\!k\varphi_4\!-\!l\varphi_5\!-\!n\varphi_6]\right>\right>\left<\left<\cos[m\varphi_1\!+\!o\varphi_2\!-\!m\varphi_3\!-\!o\varphi_4]\right>\right>\nonumber\\
&-&\left<\left<\cos[k\varphi_1\!+\!l\varphi_2\!+\!o\varphi_3\!-\!k\varphi_4\!-\!l\varphi_5\!-\!o\varphi_6]\right>\right>\left<\left<\cos[m\varphi_1\!+\!n\varphi_2\!-\!m\varphi_3\!-\!n\varphi_4]\right>\right>\nonumber\\
&-&\left<\left<\cos[k\varphi_1\!+\!m\varphi_2\!+\!n\varphi_3\!-\!k\varphi_4\!-\!m\varphi_5\!-\!n\varphi_6]\right>\right>\left<\left<\cos[l\varphi_1\!+\!o\varphi_2\!-\!l\varphi_3\!-\!o\varphi_4]\right>\right>\nonumber\\
&-&\left<\left<\cos[k\varphi_1\!+\!m\varphi_2\!+\!o\varphi_3\!-\!k\varphi_4\!-\!m\varphi_5\!-\!o\varphi_6]\right>\right>\left<\left<\cos[l\varphi_1\!+\!n\varphi_2\!-\!l\varphi_3\!-\!n\varphi_4]\right>\right>\nonumber\\
&-&\left<\left<\cos[k\varphi_1\!+\!n\varphi_2\!+\!o\varphi_3\!-\!k\varphi_4\!-\!n\varphi_5\!-\!o\varphi_6]\right>\right>\left<\left<\cos[l\varphi_1\!+\!m\varphi_2\!-\!l\varphi_3\!-\!m\varphi_4]\right>\right>\nonumber\\
&-&\left<\left<\cos[l\varphi_1\!+\!m\varphi_2\!+\!n\varphi_3\!-\!l\varphi_4\!-\!m\varphi_5\!-\!n\varphi_6]\right>\right>\left<\left<\cos[k\varphi_1\!+\!o\varphi_2\!-\!k\varphi_3\!-\!o\varphi_4]\right>\right>\nonumber\\
&-&\left<\left<\cos[l\varphi_1\!+\!m\varphi_2\!+\!o\varphi_3\!-\!l\varphi_4\!-\!m\varphi_5\!-\!o\varphi_6]\right>\right>\left<\left<\cos[k\varphi_1\!+\!n\varphi_2\!-\!k\varphi_3\!-\!n\varphi_4]\right>\right>\nonumber\\
&-&\left<\left<\cos[l\varphi_1\!+\!n\varphi_2\!+\!o\varphi_3\!-\!l\varphi_4\!-\!n\varphi_5\!-\!o\varphi_6]\right>\right>\left<\left<\cos[k\varphi_1\!+\!m\varphi_2\!-\!k\varphi_3\!-\!m\varphi_4]\right>\right>\nonumber\\
&-&\left<\left<\cos[m\varphi_1\!+\!n\varphi_2\!+\!o\varphi_3\!-\!m\varphi_4\!-\!n\varphi_5\!-\!o\varphi_6]\right>\right>\left<\left<\cos[k\varphi_1\!+\!l\varphi_2\!-\!k\varphi_3\!-\!l\varphi_4]\right>\right>\,,
\label{eq:A}
\end{eqnarray}
\begin{eqnarray}
B &\equiv&\left<\left<\cos[k\varphi_1\!+\!l\varphi_2\!+\!m\varphi_3\!-\!k\varphi_4\!-\!l\varphi_5\!-\!m\varphi_6]\right>\right>\left<\left<\cos[n(\varphi_1\!-\!\varphi_2)]\right>\right>\left<\left<\cos[o(\varphi_1\!-\!\varphi_2)]\right>\right>\nonumber\\
&+&\left<\left<\cos[k\varphi_1\!+\!l\varphi_2\!+\!n\varphi_3\!-\!k\varphi_4\!-\!l\varphi_5\!-\!n\varphi_6]\right>\right>\left<\left<\cos[m(\varphi_1\!-\!\varphi_2)]\right>\right>\left<\left<\cos[o(\varphi_1\!-\!\varphi_2)]\right>\right>\nonumber\\
&+&\left<\left<\cos[k\varphi_1\!+\!l\varphi_2\!+\!o\varphi_3\!-\!k\varphi_4\!-\!l\varphi_5\!-\!o\varphi_6]\right>\right>\left<\left<\cos[m(\varphi_1\!-\!\varphi_2)]\right>\right>\left<\left<\cos[n(\varphi_1\!-\!\varphi_2)]\right>\right>\nonumber\\
&+&\left<\left<\cos[k\varphi_1\!+\!m\varphi_2\!+\!n\varphi_3\!-\!k\varphi_4\!-\!m\varphi_5\!-\!n\varphi_6]\right>\right>\left<\left<\cos[l(\varphi_1\!-\!\varphi_2)]\right>\right>\left<\left<\cos[o(\varphi_1\!-\!\varphi_2)]\right>\right>\nonumber\\
&+&\left<\left<\cos[k\varphi_1\!+\!m\varphi_2\!+\!o\varphi_3\!-\!k\varphi_4\!-\!m\varphi_5\!-\!o\varphi_6]\right>\right>\left<\left<\cos[l(\varphi_1\!-\!\varphi_2)]\right>\right>\left<\left<\cos[n(\varphi_1\!-\!\varphi_2)]\right>\right>\nonumber\\
&+&\left<\left<\cos[k\varphi_1\!+\!n\varphi_2\!+\!o\varphi_3\!-\!k\varphi_4\!-\!n\varphi_5\!-\!o\varphi_6]\right>\right>\left<\left<\cos[l(\varphi_1\!-\!\varphi_2)]\right>\right>\left<\left<\cos[m(\varphi_1\!-\!\varphi_2)]\right>\right>\nonumber\\
&+&\left<\left<\cos[l\varphi_1\!+\!m\varphi_2\!+\!n\varphi_3\!-\!l\varphi_4\!-\!m\varphi_5\!-\!n\varphi_6]\right>\right>\left<\left<\cos[k(\varphi_1\!-\!\varphi_2)]\right>\right>\left<\left<\cos[o(\varphi_1\!-\!\varphi_2)]\right>\right>\nonumber\\
&+&\left<\left<\cos[l\varphi_1\!+\!m\varphi_2\!+\!o\varphi_3\!-\!l\varphi_4\!-\!m\varphi_5\!-\!o\varphi_6]\right>\right>\left<\left<\cos[k(\varphi_1\!-\!\varphi_2)]\right>\right>\left<\left<\cos[n(\varphi_1\!-\!\varphi_2)]\right>\right>\nonumber\\
&+&\left<\left<\cos[l\varphi_1\!+\!n\varphi_2\!+\!o\varphi_3\!-\!l\varphi_4\!-\!n\varphi_5\!-\!o\varphi_6]\right>\right>\left<\left<\cos[k(\varphi_1\!-\!\varphi_2)]\right>\right>\left<\left<\cos[m(\varphi_1\!-\!\varphi_2)]\right>\right>\nonumber\\
&+&\left<\left<\cos[m\varphi_1\!+\!n\varphi_2\!+\!o\varphi_3\!-\!m\varphi_4\!-\!n\varphi_5\!-\!o\varphi_6]\right>\right>\left<\left<\cos[k(\varphi_1\!-\!\varphi_2)]\right>\right>\left<\left<\cos[l(\varphi_1\!-\!\varphi_2)]\right>\right>\,,
\label{eq:B}
\end{eqnarray}
\begin{eqnarray}
C &\equiv&\left<\left<\cos[k\varphi_1\!+\!n\varphi_2\!-\!k\varphi_3\!-\!n\varphi_4]\right>\right>\left<\left<\cos[l\varphi_1\!+\!m\varphi_2\!-\!l\varphi_3\!-\!m\varphi_4]\right>\right>\left<\left<\cos[o(\varphi_1\!-\!\varphi_2)]\right>\right>\nonumber\\
&+&\left<\left<\cos[k\varphi_1\!+\!o\varphi_2\!-\!k\varphi_3\!-\!o\varphi_4]\right>\right>\left<\left<\cos[l\varphi_1\!+\!m\varphi_2\!-\!l\varphi_3\!-\!m\varphi_4]\right>\right>\left<\left<\cos[n(\varphi_1\!-\!\varphi_2)]\right>\right>\nonumber\\
&+&\left<\left<\cos[k\varphi_1\!+\!m\varphi_2\!-\!k\varphi_3\!-\!m\varphi_4]\right>\right>\left<\left<\cos[l\varphi_1\!+\!n\varphi_2\!-\!l\varphi_3\!-\!n\varphi_4]\right>\right>\left<\left<\cos[o(\varphi_1\!-\!\varphi_2)]\right>\right>\nonumber\\
&+&\left<\left<\cos[k\varphi_1\!+\!o\varphi_2\!-\!k\varphi_3\!-\!o\varphi_4]\right>\right>\left<\left<\cos[l\varphi_1\!+\!n\varphi_2\!-\!l\varphi_3\!-\!n\varphi_4]\right>\right>\left<\left<\cos[m(\varphi_1\!-\!\varphi_2)]\right>\right>\nonumber\\
&+&\left<\left<\cos[k\varphi_1\!+\!m\varphi_2\!-\!k\varphi_3\!-\!m\varphi_4]\right>\right>\left<\left<\cos[l\varphi_1\!+\!o\varphi_2\!-\!l\varphi_3\!-\!o\varphi_4]\right>\right>\left<\left<\cos[n(\varphi_1\!-\!\varphi_2)]\right>\right>\nonumber\\
&+&\left<\left<\cos[k\varphi_1\!+\!n\varphi_2\!-\!k\varphi_3\!-\!n\varphi_4]\right>\right>\left<\left<\cos[l\varphi_1\!+\!o\varphi_2\!-\!l\varphi_3\!-\!o\varphi_4]\right>\right>\left<\left<\cos[m(\varphi_1\!-\!\varphi_2)]\right>\right>\nonumber\\
&+&\left<\left<\cos[k\varphi_1\!+\!l\varphi_2\!-\!k\varphi_3\!-\!l\varphi_4]\right>\right>\left<\left<\cos[m\varphi_1\!+\!n\varphi_2\!-\!m\varphi_3\!-\!n\varphi_4]\right>\right>\left<\left<\cos[o(\varphi_1\!-\!\varphi_2)]\right>\right>\nonumber\\
&+&\left<\left<\cos[k\varphi_1\!+\!o\varphi_2\!-\!k\varphi_3\!-\!o\varphi_4]\right>\right>\left<\left<\cos[m\varphi_1\!+\!n\varphi_2\!-\!m\varphi_3\!-\!n\varphi_4]\right>\right>\left<\left<\cos[l(\varphi_1\!-\!\varphi_2)]\right>\right>\nonumber\\
&+&\left<\left<\cos[l\varphi_1\!+\!o\varphi_2\!-\!l\varphi_3\!-\!o\varphi_4]\right>\right>\left<\left<\cos[m\varphi_1\!+\!n\varphi_2\!-\!m\varphi_3\!-\!n\varphi_4]\right>\right>\left<\left<\cos[k(\varphi_1\!-\!\varphi_2)]\right>\right>\nonumber\\
&+&\left<\left<\cos[k\varphi_1\!+\!l\varphi_2\!-\!k\varphi_3\!-\!l\varphi_4]\right>\right>\left<\left<\cos[m\varphi_1\!+\!o\varphi_2\!-\!m\varphi_3\!-\!o\varphi_4]\right>\right>\left<\left<\cos[n(\varphi_1\!-\!\varphi_2)]\right>\right>\nonumber\\
&+&\left<\left<\cos[k\varphi_1\!+\!n\varphi_2\!-\!k\varphi_3\!-\!n\varphi_4]\right>\right>\left<\left<\cos[m\varphi_1\!+\!o\varphi_2\!-\!m\varphi_3\!-\!o\varphi_4]\right>\right>\left<\left<\cos[l(\varphi_1\!-\!\varphi_2)]\right>\right>\nonumber\\
&+&\left<\left<\cos[l\varphi_1\!+\!n\varphi_2\!-\!l\varphi_3\!-\!n\varphi_4]\right>\right>\left<\left<\cos[m\varphi_1\!+\!o\varphi_2\!-\!m\varphi_3\!-\!o\varphi_4]\right>\right>\left<\left<\cos[k(\varphi_1\!-\!\varphi_2)]\right>\right>\nonumber\\
&+&\left<\left<\cos[k\varphi_1\!+\!l\varphi_2\!-\!k\varphi_3\!-\!l\varphi_4]\right>\right>\left<\left<\cos[n\varphi_1\!+\!o\varphi_2\!-\!n\varphi_3\!-\!o\varphi_4]\right>\right>\left<\left<\cos[m(\varphi_1\!-\!\varphi_2)]\right>\right>\nonumber\\
&+&\left<\left<\cos[k\varphi_1\!+\!m\varphi_2\!-\!k\varphi_3\!-\!m\varphi_4]\right>\right>\left<\left<\cos[n\varphi_1\!+\!o\varphi_2\!-\!n\varphi_3\!-\!o\varphi_4]\right>\right>\left<\left<\cos[l(\varphi_1\!-\!\varphi_2)]\right>\right>\nonumber\\
&+&\left<\left<\cos[l\varphi_1\!+\!m\varphi_2\!-\!l\varphi_3\!-\!m\varphi_4]\right>\right>\left<\left<\cos[n\varphi_1\!+\!o\varphi_2\!-\!n\varphi_3\!-\!o\varphi_4]\right>\right>\left<\left<\cos[k(\varphi_1\!-\!\varphi_2)]\right>\right>\,,
\label{eq:C}
\end{eqnarray}
\begin{eqnarray}
D &\equiv&\left<\left<\cos[k\varphi_1\!+\!l\varphi_2\!-\!k\varphi_3\!-\!l\varphi_4]\right>\right>\left<\left<\cos[m(\varphi_1\!-\!\varphi_2)]\right>\right>\left<\left<\cos[n(\varphi_1\!-\!\varphi_2)]\right>\right>\left<\left<\cos[o(\varphi_1\!-\!\varphi_2)]\right>\right>\nonumber\\
&+&\left<\left<\cos[k\varphi_1\!+\!m\varphi_2\!-\!k\varphi_3\!-\!m\varphi_4]\right>\right>\left<\left<\cos[l(\varphi_1\!-\!\varphi_2)]\right>\right>\left<\left<\cos[n(\varphi_1\!-\!\varphi_2)]\right>\right>\left<\left<\cos[o(\varphi_1\!-\!\varphi_2)]\right>\right>\nonumber\\
&+&\left<\left<\cos[k\varphi_1\!+\!n\varphi_2\!-\!k\varphi_3\!-\!n\varphi_4]\right>\right>\left<\left<\cos[l(\varphi_1\!-\!\varphi_2)]\right>\right>\left<\left<\cos[m(\varphi_1\!-\!\varphi_2)]\right>\right>\left<\left<\cos[o(\varphi_1\!-\!\varphi_2)]\right>\right>\nonumber\\
&+&\left<\left<\cos[k\varphi_1\!+\!o\varphi_2\!-\!k\varphi_3\!-\!o\varphi_4]\right>\right>\left<\left<\cos[l(\varphi_1\!-\!\varphi_2)]\right>\right>\left<\left<\cos[m(\varphi_1\!-\!\varphi_2)]\right>\right>\left<\left<\cos[n(\varphi_1\!-\!\varphi_2)]\right>\right>\nonumber\\
&+&\left<\left<\cos[l\varphi_1\!+\!m\varphi_2\!-\!l\varphi_3\!-\!m\varphi_4]\right>\right>\left<\left<\cos[k(\varphi_1\!-\!\varphi_2)]\right>\right>\left<\left<\cos[n(\varphi_1\!-\!\varphi_2)]\right>\right>\left<\left<\cos[o(\varphi_1\!-\!\varphi_2)]\right>\right>\nonumber\\
&+&\left<\left<\cos[l\varphi_1\!+\!n\varphi_2\!-\!l\varphi_3\!-\!n\varphi_4]\right>\right>\left<\left<\cos[k(\varphi_1\!-\!\varphi_2)]\right>\right>\left<\left<\cos[m(\varphi_1\!-\!\varphi_2)]\right>\right>\left<\left<\cos[o(\varphi_1\!-\!\varphi_2)]\right>\right>\nonumber\\
&+&\left<\left<\cos[l\varphi_1\!+\!o\varphi_2\!-\!l\varphi_3\!-\!o\varphi_4]\right>\right>\left<\left<\cos[k(\varphi_1\!-\!\varphi_2)]\right>\right>\left<\left<\cos[m(\varphi_1\!-\!\varphi_2)]\right>\right>\left<\left<\cos[n(\varphi_1\!-\!\varphi_2)]\right>\right>\nonumber\\
&+&\left<\left<\cos[m\varphi_1\!+\!n\varphi_2\!-\!m\varphi_3\!-\!n\varphi_4]\right>\right>\left<\left<\cos[k(\varphi_1\!-\!\varphi_2)]\right>\right>\left<\left<\cos[l(\varphi_1\!-\!\varphi_2)]\right>\right>\left<\left<\cos[o(\varphi_1\!-\!\varphi_2)]\right>\right>\nonumber\\
&+&\left<\left<\cos[m\varphi_1\!+\!o\varphi_2\!-\!m\varphi_3\!-\!o\varphi_4]\right>\right>\left<\left<\cos[k(\varphi_1\!-\!\varphi_2)]\right>\right>\left<\left<\cos[l(\varphi_1\!-\!\varphi_2)]\right>\right>\left<\left<\cos[n(\varphi_1\!-\!\varphi_2)]\right>\right>\nonumber\\
&+&\left<\left<\cos[n\varphi_1\!+\!o\varphi_2\!-\!n\varphi_3\!-\!o\varphi_4]\right>\right>\left<\left<\cos[k(\varphi_1\!-\!\varphi_2)]\right>\right>\left<\left<\cos[l(\varphi_1\!-\!\varphi_2)]\right>\right>\left<\left<\cos[m(\varphi_1\!-\!\varphi_2)]\right>\right>\,,
\label{eq:D}
\end{eqnarray}
\begin{eqnarray}
E &\equiv&\left<\left<\cos[k(\varphi_1\!-\!\varphi_2)]\right>\right>
\left<\left<\cos[l(\varphi_1\!-\!\varphi_2)]\right>\right>
\left<\left<\cos[m(\varphi_1\!-\!\varphi_2)]\right>\right>
\left<\left<\cos[n(\varphi_1\!-\!\varphi_2)]\right>\right>
\left<\left<\cos[o(\varphi_1\!-\!\varphi_2)]\right>\right>\,.
\label{eq:E}
\end{eqnarray}
Analogously, one can proceed with all higher order SC.

\section{Checking the other requirements for the Symmetric Cumulants}
\label{a:Checking-the-other-requirements-for-the-Symmetric-Cumulants}

In this appendix, we develop the argumentation about the remaining requirements outlined in Appendix~\ref{a:List-of-requirments}. 
In what follows, all simulations are done using the Toy Monte Carlo setup described in Sec.~\ref{ss:Nonflow-estimation-with-Toy-MonteCarlo-studies}.

\begin{figure}
\begin{tabular}{c c}
	\includegraphics[scale=0.43]{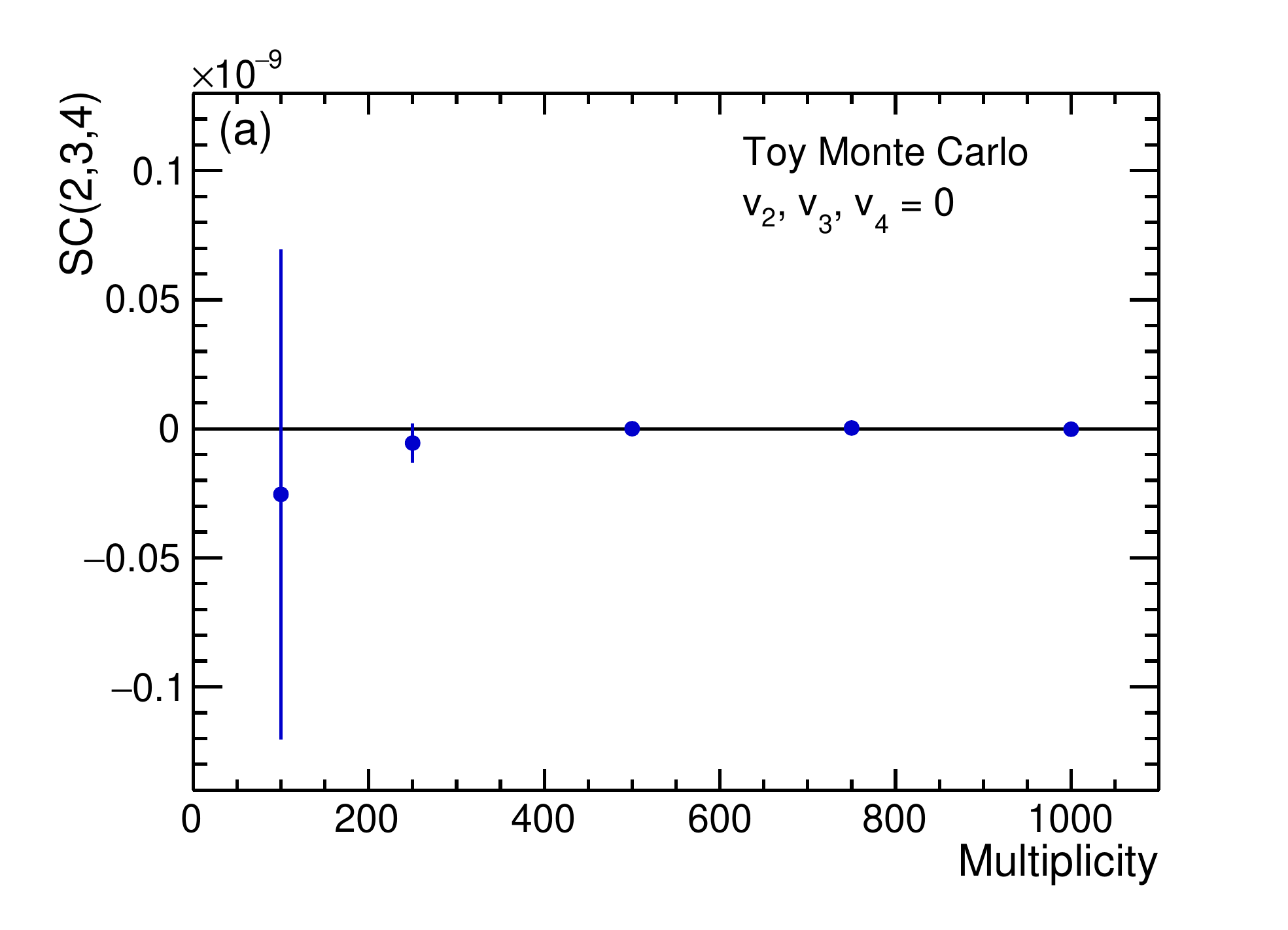} &
	\includegraphics[scale=0.43]{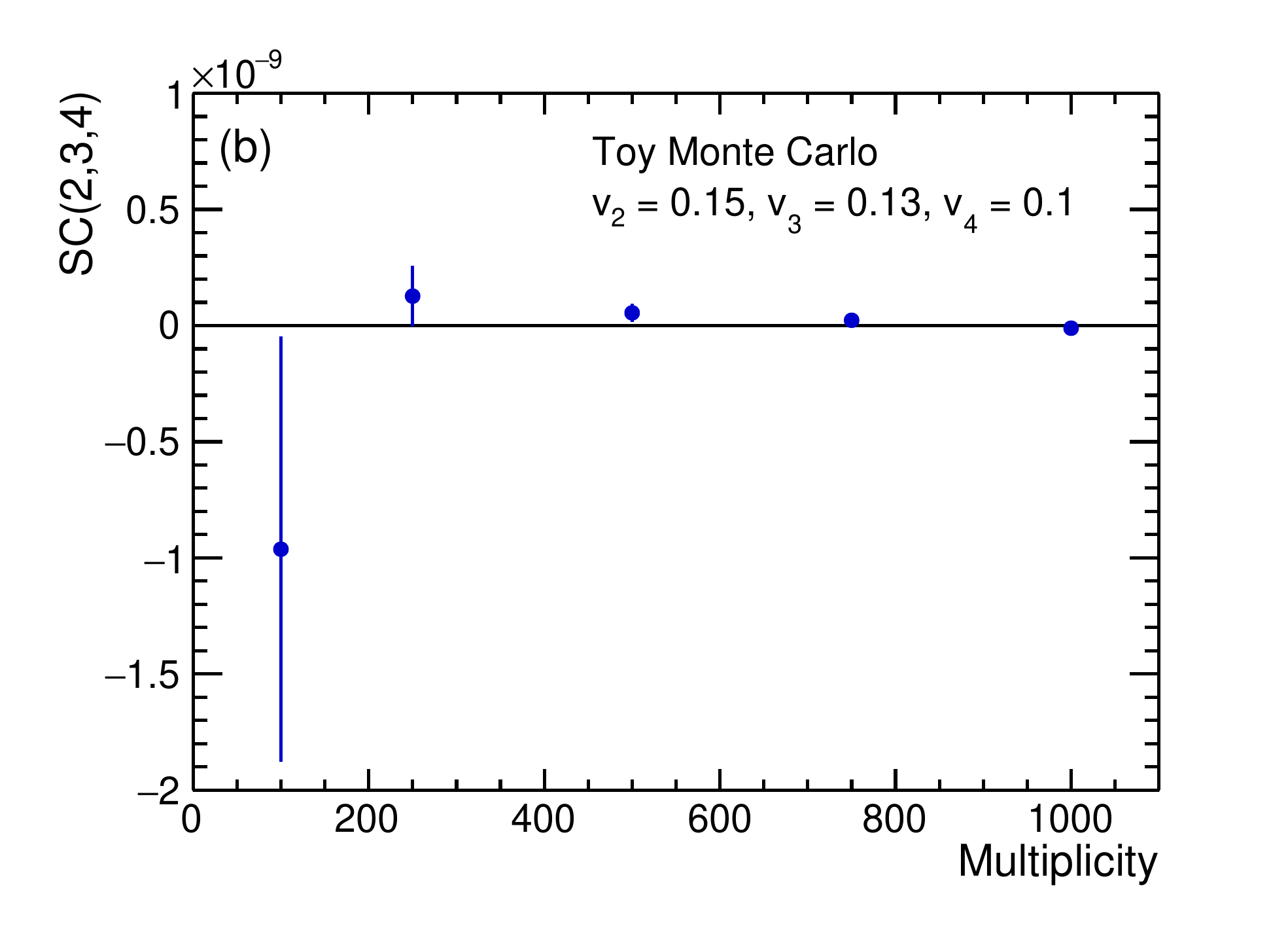} \\
	\includegraphics[scale=0.43]{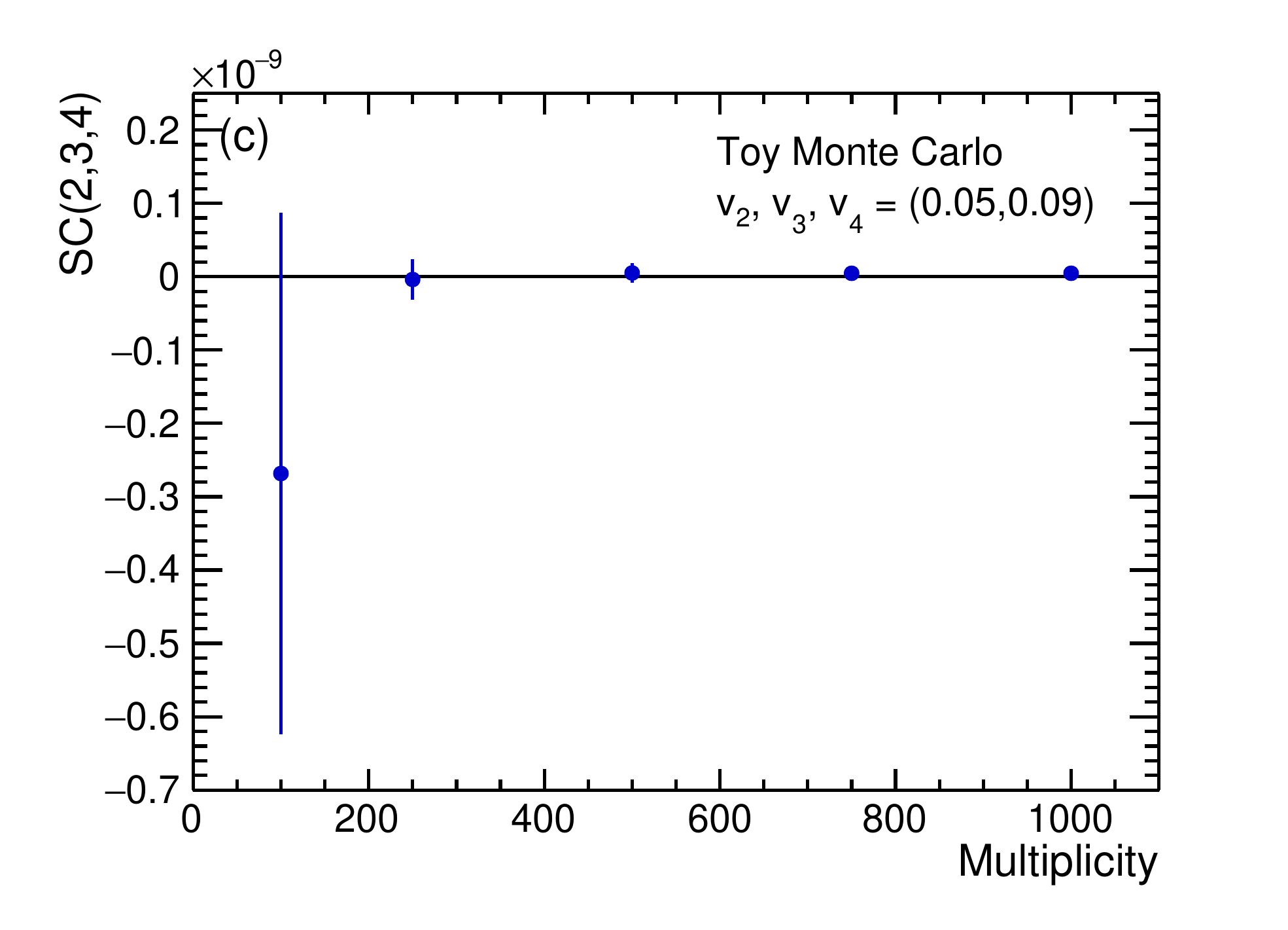} &
	\includegraphics[scale=0.43]{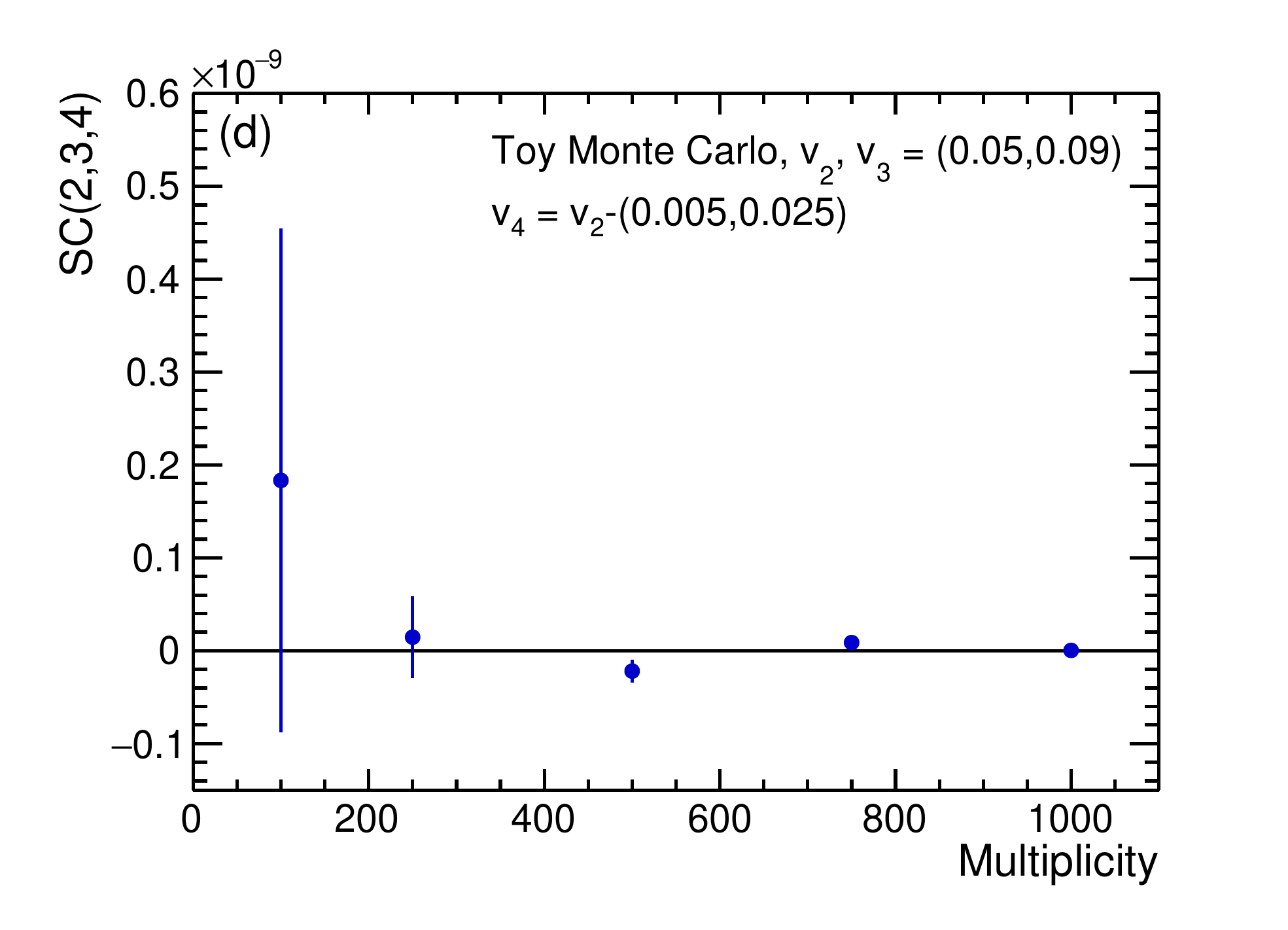} \\
\end{tabular}
	\caption{Values of the observable SC(2,3,4) obtained for five different multiplicities and four different inputs for the flow amplitudes in the Toy Monte Carlo setup: all amplitudes are set to zero (a), constant amplitudes (b), uncorrelated nonzero amplitudes (c) and only two correlated amplitudes (d). The notation (.,.) indicates the range for the uniform sampling event-by-event.}
\label{sect_ToyMC_fig_basictests}
\end{figure}

We first look at the effects of the correlations between less than three flow amplitudes. We consider $N = 10^8$ events in total and set the values of $v_2$, $v_3$ and $v_4$ according to the current test for five different multiplicities: 100, 250, 500, 750 and 1000. As we want to study the multiplicity fluctuations and the event weights separately, we keep constant the multiplicity. This implies that any event weight is equal to unit weight. The configurations of flow amplitudes used in our basic tests are the following: $v_2$, $v_3$ and $v_4$ are set to zero (Fig.~\ref{sect_ToyMC_fig_basictests} (a)), constant nonzero amplitudes fixed to $v_2 = 0.15$, $v_3 = 0.13$, $v_4 = 0.1$, respectively (Fig.~\ref{sect_ToyMC_fig_basictests} (b)), uncorrelated amplitudes uniformly sampled event-by-event in the interval (0.05, 0.09) (Fig.~\ref{sect_ToyMC_fig_basictests} (c)) and finally $v_2$ and $v_3$ uniformly sampled in (0.05, 0.09) and $v_4$ correlated to $v_2$ with the relation $v_4 = v_2 - (0.005,0.025)$ (Fig.~\ref{sect_ToyMC_fig_basictests} (d)).
All four studies lead to the same conclusion that SC(2,3,4) is compatible with zero for all multiplicities. This means our definition in Eq.~\eqref{eq:3pSC} does not contain any built-in bias in absence of fluctuations in the amplitudes, but also that it is insensitive to any subset of amplitudes where the fluctuations of at least one of the $v_n$ are not correlated with the others.
%

\begin{figure}
\begin{tabular}{c c}
	\includegraphics[scale=0.43]{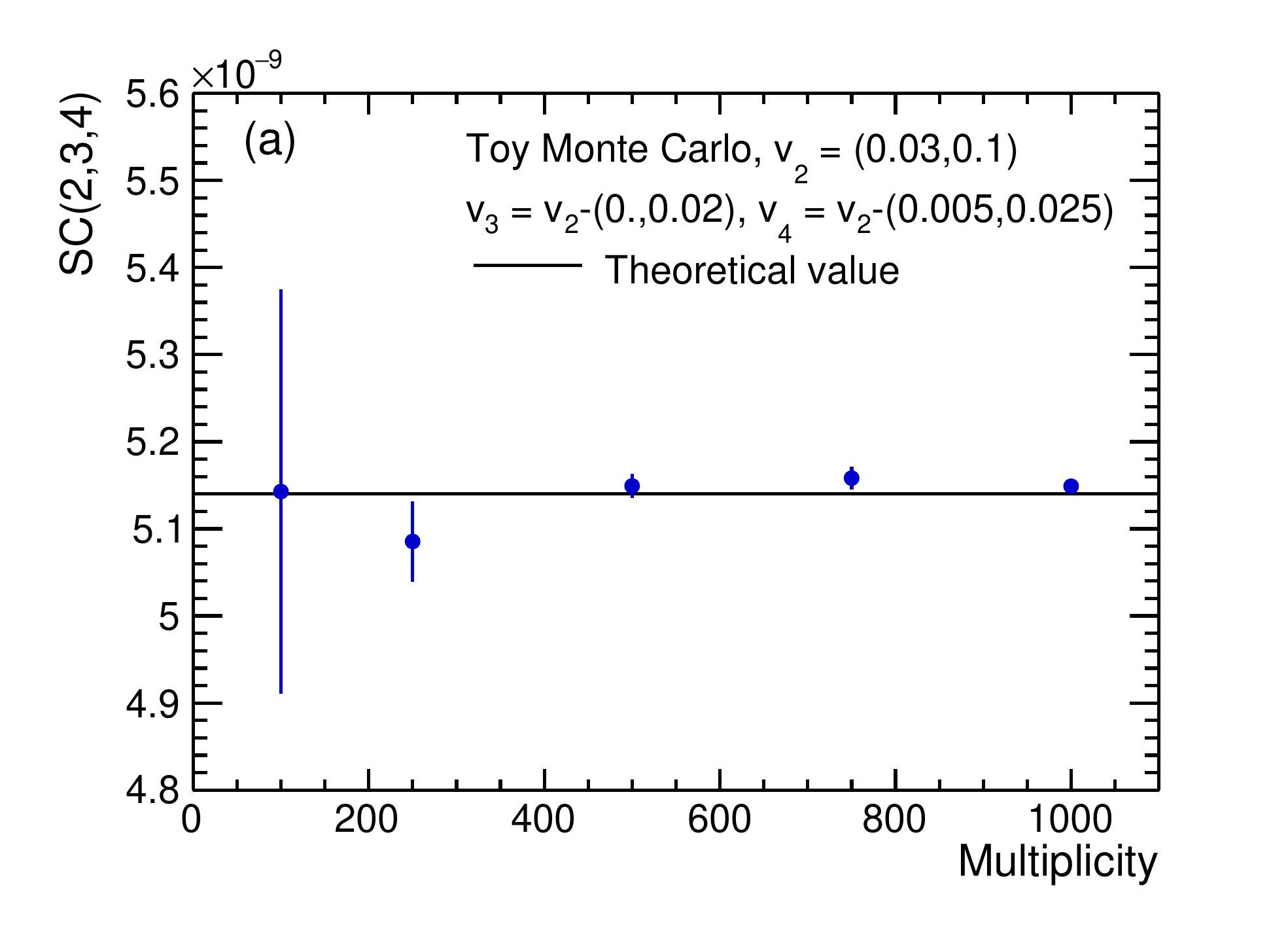} &
		\includegraphics[scale=0.43]{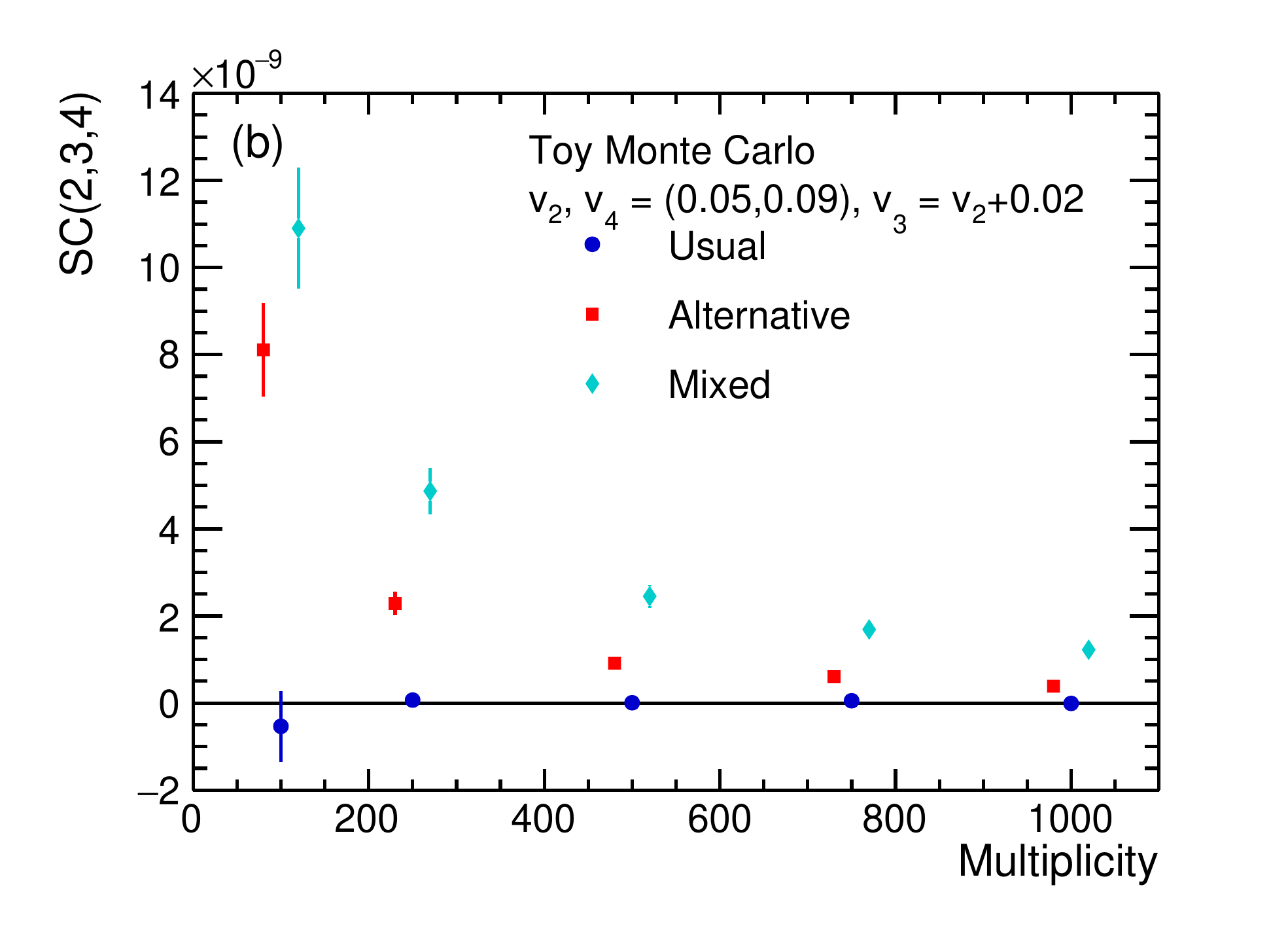}
\end{tabular}
	\caption{Values obtained for SC(2,3,4) when the three amplitudes are genuinely correlated. The notation (.,.) in the expressions for $v_3$ and $v_4$ indicates again the range for the uniform fluctuations (a). Values of SC(2,3,4) in the case of two anti-correlated amplitudes: $v_2, v_4$ are uniformly sampled event-by-event in (0.05, 0.09) and $v_3$ is correlated to $v_2$ by $v_3 = v_2 + 0.02$ for the usual expression (blue circles), the alternative expression (red squares) and the mixed expression (cyan diamonds) (b).}
	\label{sect_ToyMC_fig_allCorrel}
\end{figure}

We now look at an example of genuine correlations among the three flow amplitudes. We study the case where $v_2$ is uniformly sampled in (0.03, 0.1), $v_3 = v_2 - (0, 0.02)$ and $v_4 = v_2 - (0.005, 0.025)$. These values are input in our Toy Monte Carlo setup for $M$ = 100, 250, 500, 750 and 1000, and for $N = 10^8$. The theoretical value of SC(2,3,4) is $5.13992 \cdot 10^{-9}$. We can see on Fig.~\ref{sect_ToyMC_fig_allCorrel}(a) that for all considered values of multiplicities, SC(2,3,4) obtained through simulations is compatible with the theoretical result within the statistical errors. This result, combined with the ones obtained in the previous tests (Fig.~\ref{sect_ToyMC_fig_basictests}), implies that our definition of SC given by Eq.~\eqref{eq:3pSC} is sensitive only to the genuine correlation among all three flow amplitudes in question.

Now that we have demonstrated that our SC is sensitive only to three genuinely correlated flow amplitudes, we provide an argumentation about the uniqueness of its expression (Requirement~6 in Appendix~\ref{a:List-of-requirments}). We first show which expression must be chosen to solve the ambiguity between the two definitions in the case of SC($m$,$n$) using a mathematical derivation. As the exact computations are tedious for three different amplitudes, we then look at results for SC($k$,$l$,$m$) obtained with our Toy Monte Carlo setup and with HIJING.


We start by recalling here the definitions of the two-, three- and four-particle azimuthal correlators for a number of particle per event $M$ and for unit particle weights~\cite{Bilandzic:2010jr}:
\begin{eqnarray}
	\langle 2 \rangle _{m,-m} & = & \frac{1}{M(M-1)} \sum _{\substack{h, j = 1 \\ h \neq j}}^M {\rm e}^{im(\varphi_h - \varphi_j)}, \\
	\langle 3 \rangle _{m,n,o} & = & \frac{1}{M(M-1)(M-2)} \sum _{\substack{h, j, k = 1 \\ h \neq j \neq k}}^M {\rm e}^{i(m\varphi_h + n\varphi_j + o\varphi_k)}, \\
	\langle 4 \rangle _{m,-m,n,-n} & = & \frac{1}{M(M-1)(M-2)(M-3)} \sum _{\substack{h, j, k, l = 1 \\ h \neq j \neq k \neq l}}^M {\rm e}^{i(m(\varphi_h - \varphi_j) + n(\varphi_k - \varphi_l))},
\end{eqnarray}
where we have the condition that $m + n + o = 0$ for $\langle 3 \rangle _{m,n,o}$. The computation of the single-event average of the first term of Eq.~\eqref{eq:SC(m,n)_fake}, $\langle 2 \rangle _{m,-m} \langle 2 \rangle _{n,-n}$, gives us:
\begin{eqnarray}
	\langle 2 \rangle _{m,-m} \langle 2 \rangle _{n,-n} & = & \frac{1}{M^2 (M-1)^2} \left[ M(M-1)(M-2)(M-3) \langle 4 \rangle _{m,-m,n,-n} \right. \nonumber \\
		& + & M(M-1) \langle 2 \rangle _{m+n, -m-n} + M(M-1) \langle 2 \rangle _{m-n, -m+n} \nonumber \\
		& + & M(M-1)(M-2)	\langle 3 \rangle _{m+n, -m, -n} + M(M-1)(M-2) \langle 3 \rangle _{m-n, -m, n} \nonumber \\
		& + & \left. M(M-1)(M-2) \langle 3 \rangle _{m, -m+n,-n} + M(M-1)(M-2) \langle 3 \rangle _{m, -m-n, n} \right].
\end{eqnarray}
For $M \gg 3$, we can write that $M - k \simeq M$ for $k = 1,2,3$. This is the case in heavy-ion collisions where a large amount of particles are created in each event. After averaging over all the events, we obtain finally that
\begin{eqnarray}
	\langle \langle 2 \rangle _{m,-m} \langle 2 \rangle _{n,-n} \rangle & \simeq & \langle \langle 4 \rangle \rangle _{m,-m,n,-n} + \frac{1}{M^2} \left( \langle \langle 2 \rangle \rangle _{m+n, -m-n} + \langle \langle 2 \rangle \rangle _{m-n, -m+n}\right) \nonumber \\
		& + & \frac{1}{M} \left( \langle \langle 3 \rangle \rangle _{m+n, -m, -n} + \langle \langle 3 \rangle \rangle _{m-n, -m, n} \right. \nonumber \\
		& + & \left. \langle \langle 3 \rangle \rangle _{m, -m+n,-n} + \langle \langle 3 \rangle \rangle _{m, -m-n, n} \right).
\end{eqnarray}
We can therefore conclude that in the definitions in Eqs.~\eqref{eq:SC(m,n)_standard} and \eqref{eq:SC(m,n)_fake} the first terms are not equal. The reason is that in the alternative expression in Eq.~\eqref{eq:SC(m,n)_fake} the first term contains self-correlations between the particles. Such self-correlations must be removed exactly in correlations techniques as they trivially and non-negligibly bias the final results.


In order to look at the ambiguity in the correct definition of SC($k$,$l$,$m$), we introduce two other expressions leading to the same final theoretical result given by Eq.~\eqref{eq:3pSC}. These are:
\begin{eqnarray}
	\text{SC}(k,l,m)_{\rm alternative} & = & \langle\langle \cos(k(\varphi_1 - \varphi_2)) \rangle\langle \cos(l(\varphi_1 - \varphi_2)) \rangle\langle \cos(m(\varphi_1 - \varphi_2)) \rangle\rangle \nonumber \\
	& + & 2\langle\langle \cos(k(\varphi_1 - \varphi_2)) \rangle\rangle \langle\langle \cos(l(\varphi_1 - \varphi_2)) \rangle\rangle \langle\langle \cos(m(\varphi_1 - \varphi_2)) \rangle\rangle \nonumber \\
	& - & \left\lbrace \langle\langle \cos(k(\varphi_1 - \varphi_2)) \rangle\langle \cos(l(\varphi_1 - \varphi_2)) \rangle\rangle \langle\langle \cos(m(\varphi_1 - \varphi_2)) \rangle\rangle \right. \nonumber \\
	& + & \langle\langle \cos(l(\varphi_1 - \varphi_2)) \rangle\langle \cos(m(\varphi_1 - \varphi_2)) \rangle\rangle \langle\langle \cos(k(\varphi_1 - \varphi_2)) \rangle\rangle \nonumber \\
	& + & \left. \langle\langle \cos(k(\varphi_1 - \varphi_2)) \rangle\langle \cos(m(\varphi_1 - \varphi_2)) \rangle\rangle \langle\langle \cos(l(\varphi_1 - \varphi_2)) \rangle\rangle \right\rbrace, \label{sect_ToyMC_eq-alternative}\\
	\text{SC}(k,l,m)_{\rm mixed} & = & \langle\langle \cos(k\varphi_1 + l\varphi_2 - k\varphi_3 - l\varphi_4) \rangle\langle \cos(m(\varphi_1 - \varphi_2)) \rangle\rangle \nonumber \\
	& + & 2\langle\langle \cos(k(\varphi_1 - \varphi_2)) \rangle\rangle \langle\langle \cos(l(\varphi_1 - \varphi_2)) \rangle\rangle \langle\langle \cos(m(\varphi_1 - \varphi_2)) \rangle\rangle \nonumber \\
	& - & \left\lbrace \langle\langle \cos(k(\varphi_1 - \varphi_2)) \rangle\langle \cos(l(\varphi_1 - \varphi_2)) \rangle\rangle\langle\langle \cos(m(\varphi_1 - \varphi_2)) \rangle\rangle \right. \nonumber \\
	& + & \langle\langle \cos(l(\varphi_1 - \varphi_2)) \rangle\langle \cos(m(\varphi_1 - \varphi_2)) \rangle\rangle \langle\langle \cos(k(\varphi_1 - \varphi_2)) \rangle\rangle \nonumber \\
	& + & \left. \langle\langle \cos(k(\varphi_1 - \varphi_2)) \rangle\langle \cos(m(\varphi_1 - \varphi_2)) \rangle\rangle \langle\langle \cos(l(\varphi_1 - \varphi_2)) \rangle\rangle \right\rbrace. \label{sect_ToyMC_eq-mixed}
\end{eqnarray}
In what follows, we will refer to Eq.~\eqref{eq:3pSC} as the usual expression of SC($k,l,m$), to Eq.~\eqref{sect_ToyMC_eq-alternative} as its alternative expression and to Eq.~\eqref{sect_ToyMC_eq-mixed} as its mixed expression. Like for the case of SC($m$,$n$) in Eq.~\eqref{eq:SC(m,n)_fake}, the alternative definition contains only two-particle correlators for every harmonic. The mixed formula is called this way as it is a combination of two- and four-particle correlators.
Since the exact derivation made for two amplitudes becomes too tedious in the case for three differents amplitudes, we will try to establish which one is the unique expression to use for SC($k$,$l$,$m$) both by using the Toy Monte Carlo setup described in Sec.~\ref{ss:Nonflow-estimation-with-Toy-MonteCarlo-studies} and by using HIJING.


We first look at the Toy Monte Carlo simulation. We sample uniformly event-by-event, for $N = 10^7$ events, $v_2$ and $v_4$ in the interval (0.05, 0.09) and set that $v_3 = v_2 + 0.02$, before computing the three expressions described above. Like in our basic tests, the multiplicity is kept constant. Since not all three amplitudes are correlated, we expect that SC(2,3,4) is zero. The results of the simulations are shown on Fig.~\ref{sect_ToyMC_fig_allCorrel}(b). The usual expression is zero for all considered multiplicities. However, this is not the case for the alternative and the mixed definitions which show a clear multiplicity dependence.
Similarly as in the expression presented above, one can expect the terms of the form $\langle\langle \cos(a\varphi_1 + b\varphi_2 - a\varphi_3 - b\varphi_4) \rangle\langle \cos(c(\varphi_1 - \varphi_2)) \rangle\rangle$ for any $a,b,c = k,l,m$ and $\langle\langle \cos(k(\varphi_1 - \varphi_2)) \rangle\langle \cos(l(\varphi_1 - \varphi_2)) \rangle\langle \cos(m(\varphi_1 - \varphi_2)) \rangle\rangle$ to introduce self-correlations which are not visible in the definition of SC($k$,$l$,$m$) with flow amplitudes.
%
%

\begin{figure}
\begin{tabular}{c c}
	\includegraphics[scale=0.43]{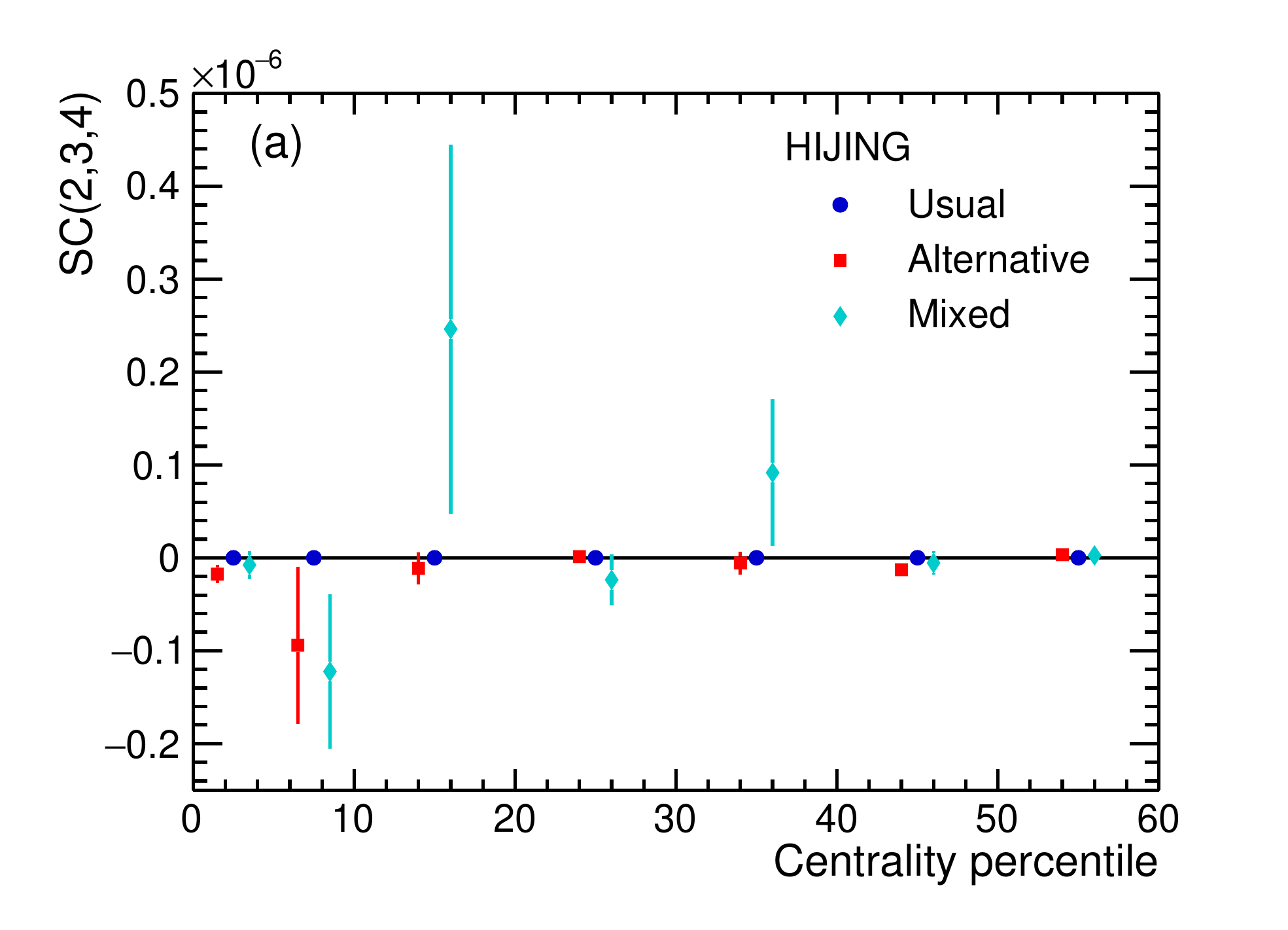} &
	\includegraphics[scale=0.43]{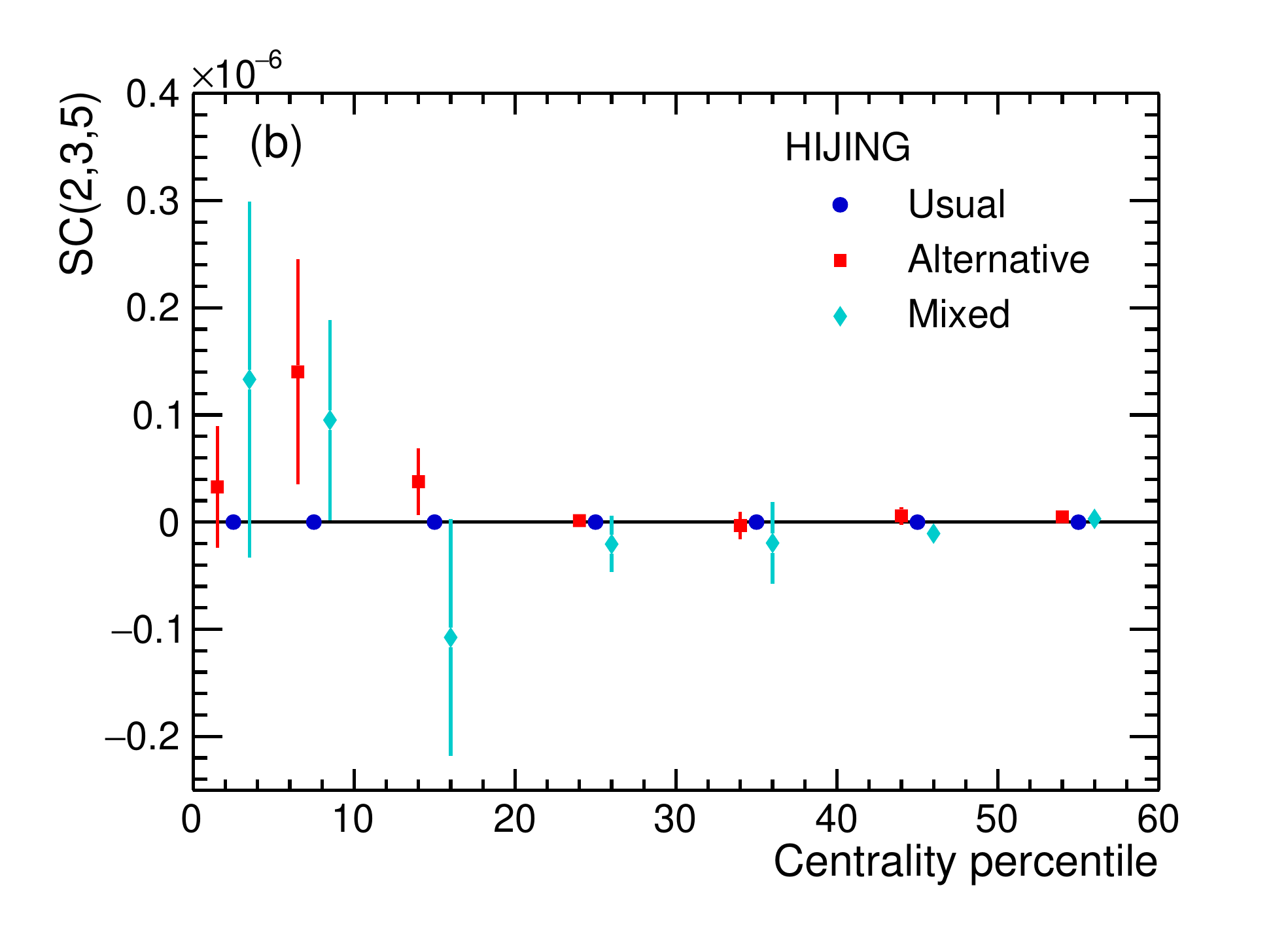}	
\end{tabular}
	\caption{Centrality dependence of the observables SC(2,3,4) (a) and SC(2,3,5) (b) for the usual expression of Eq.~\eqref{eq:3pSC} (blue circles), the alternative expression of Eq. \ref{sect_ToyMC_eq-alternative} (red squares) and the mixed expression of Eq. \ref{sect_ToyMC_eq-mixed} (cyan diamonds).}
\label{sect_Hijing_fig_GSC}
\end{figure}

We now look at the results obtained with HIJING (the description of the HIJING setup and resulting dataset can be found in Sec.~(\ref{ss:Estimating-nonflow-contribution-with-HIJING})). Figure \ref{sect_Hijing_fig_GSC} shows the centrality dependence of SC(2,3,4) and SC(2,3,5) in the case of the usual expression in Eq.~\eqref{eq:3pSC}, the alternative expression in Eq.~\eqref{sect_ToyMC_eq-alternative} and the mixed expression in Eq.~\eqref{sect_ToyMC_eq-mixed}.
Both alternative and mixed expressions show large statistical errors, which prevent us to rule out the compatibility of these results with zero. One possible reason of the size of these errors can be that the multiplicity dependence of the terms with the self-correlations is also present for their statistical errors.
However, the usual expression has smaller statistical errors and, therefore, seems to be more consistent to zero for all centrality bins and less sensitive to nonflow.

\begin{figure}
	\begin{tabular}{c}
		\includegraphics[scale=0.43]{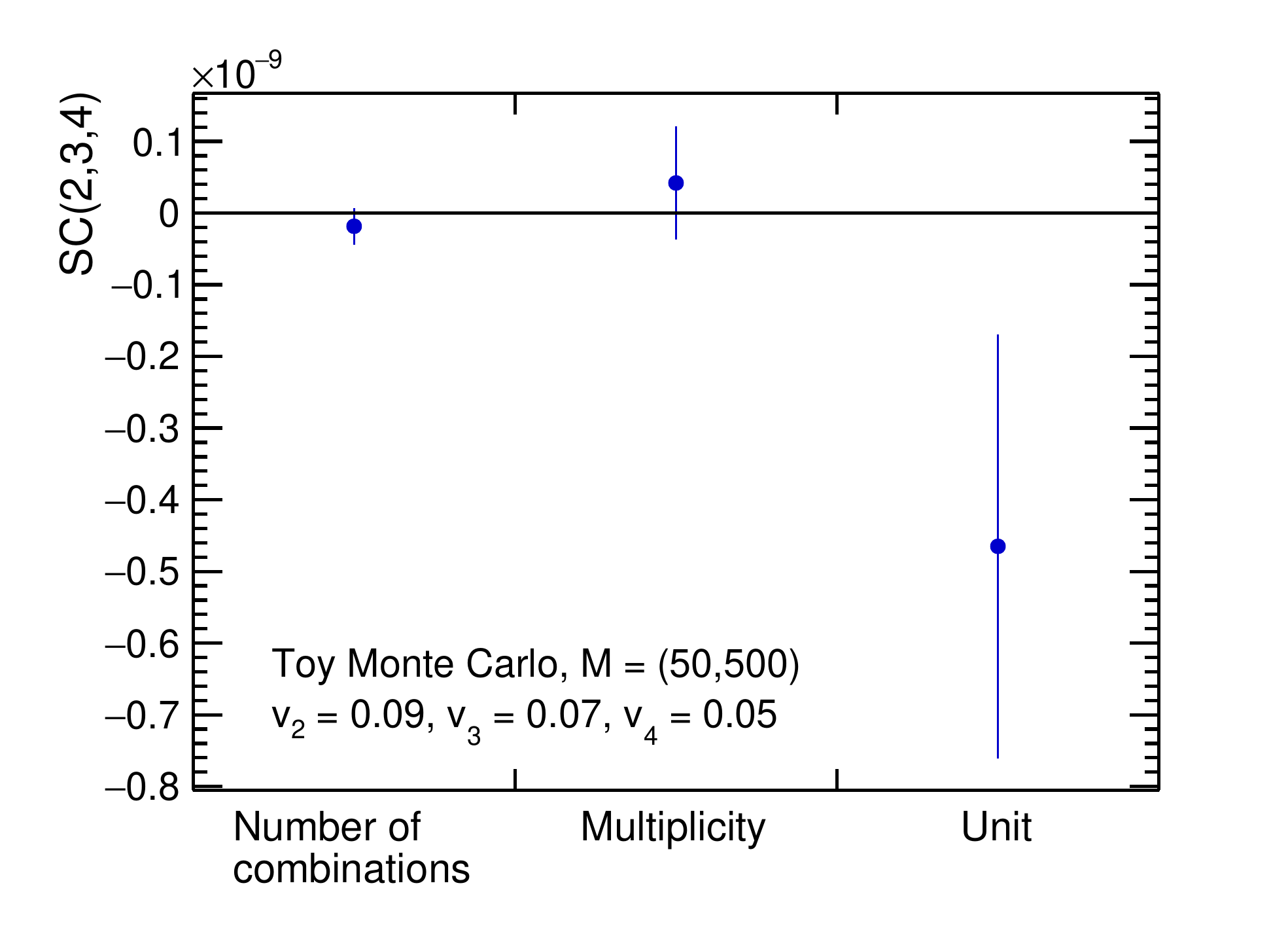}
	\end{tabular}
	\caption{Values of SC(2,3,4) obtained for three different event weights: the number of combinations, the multiplicity of the event and the unit weight. The multiplicity is sampled uniformly in the interval (50, 500). The theoretical value is zero as the flow amplitudes are uncorrelated.}
	\label{fig:weights}
\end{figure}

Finally, we conclude this appendix with the discussion on event weights (Requirement~8 in Appendix~\ref{a:List-of-requirments}). We therefore simulate $N = 10 ^8$ events with $v_2 = 0.09$, $v_3 = 0.07$ and $v_4 = 0.05$ with the setup described in Sec.~\ref{ss:Nonflow-estimation-with-Toy-MonteCarlo-studies}. Since all event weights are equal to unity if the multiplicity is kept constant, we sample $M$ uniformly in the interval (50, 500). We consider three different event weights. The first one is the weights used in~\cite{ALICE:2016kpq} and which was named the number of combinations. In our case, this weight is equal to $M(M-1)$ for the two-particle correlators, $M(M-1)(M-2)(M-3)$ for the four-particle correlators and finally $M(M-1)(M-2)(M-3)(M-4)(M-5)$ for the six-particle correlator. The other two event weights we introduce are the multiplicity of the event itself and the unit weight.
On Fig.~\ref{fig:weights} we see that the number of combinations weight has the smallest statistical spread. This result is consistent with what has been done in the analyses of individual flow harmonics with correlation techniques. This implies that the number of combinations is still the event weight to use in the case of the three-harmonic SC observable.

\section{Vishnu predictions for SC($m,n$)}
\label{ss:AppendixD}

	\begin{figure}[t!]
		\begin{center}
			\begin{tabular}{c}
				\includegraphics[scale=0.43]{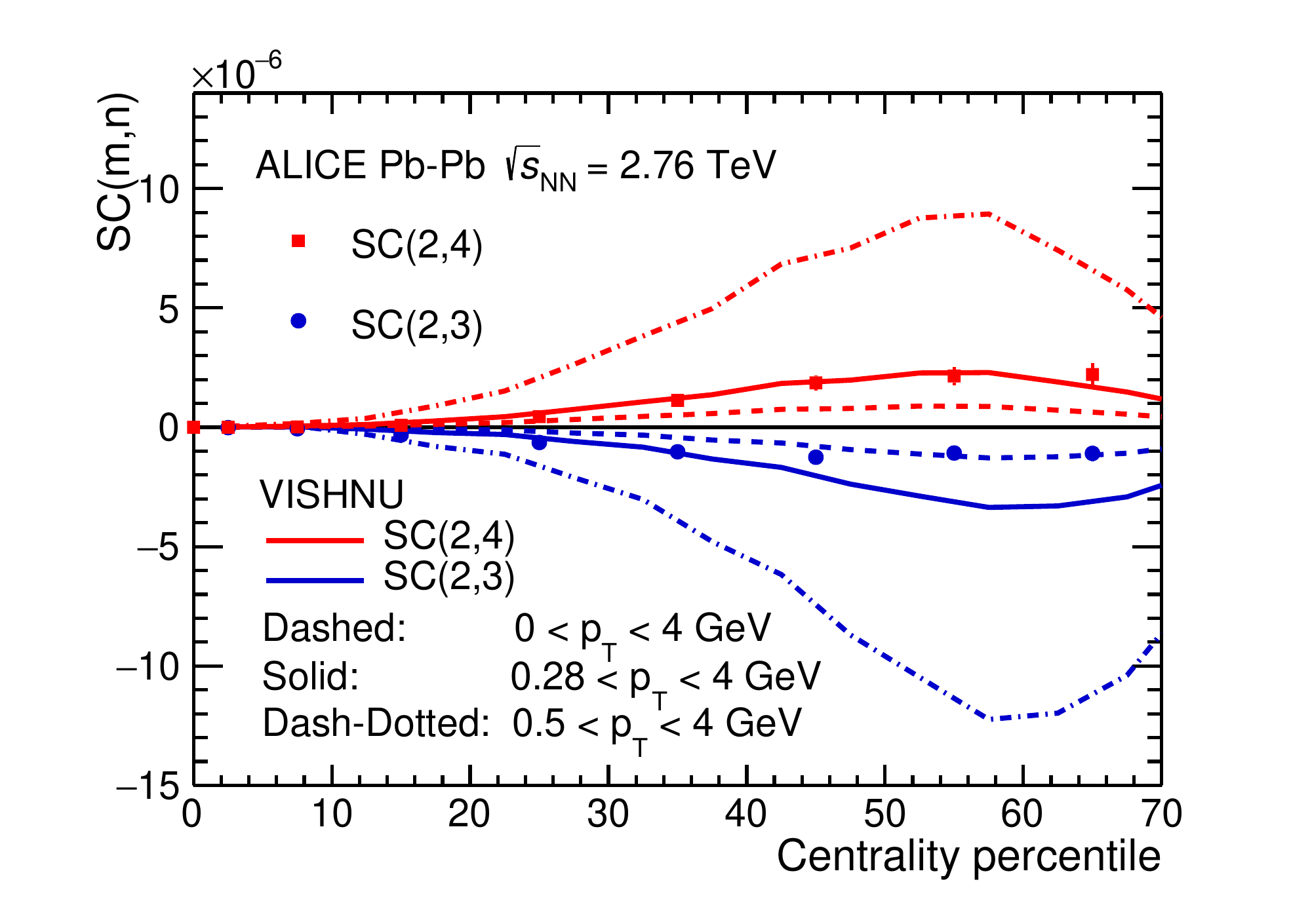} 
			\end{tabular}		
			\caption{$\text{SC}(2,3)$ and $\text{SC}(2,4)$ from the published results of ALICE (markers) and iEBE-VISHNU event generator (curves). The simulation is presented in three different $p_T$ ranges.} 
			\label{SC23SC24andALICE}
		\end{center}
	\end{figure}



\begin{figure}[t!]
	\begin{center}
		\begin{tabular}{c c}
			\includegraphics[scale=0.43]{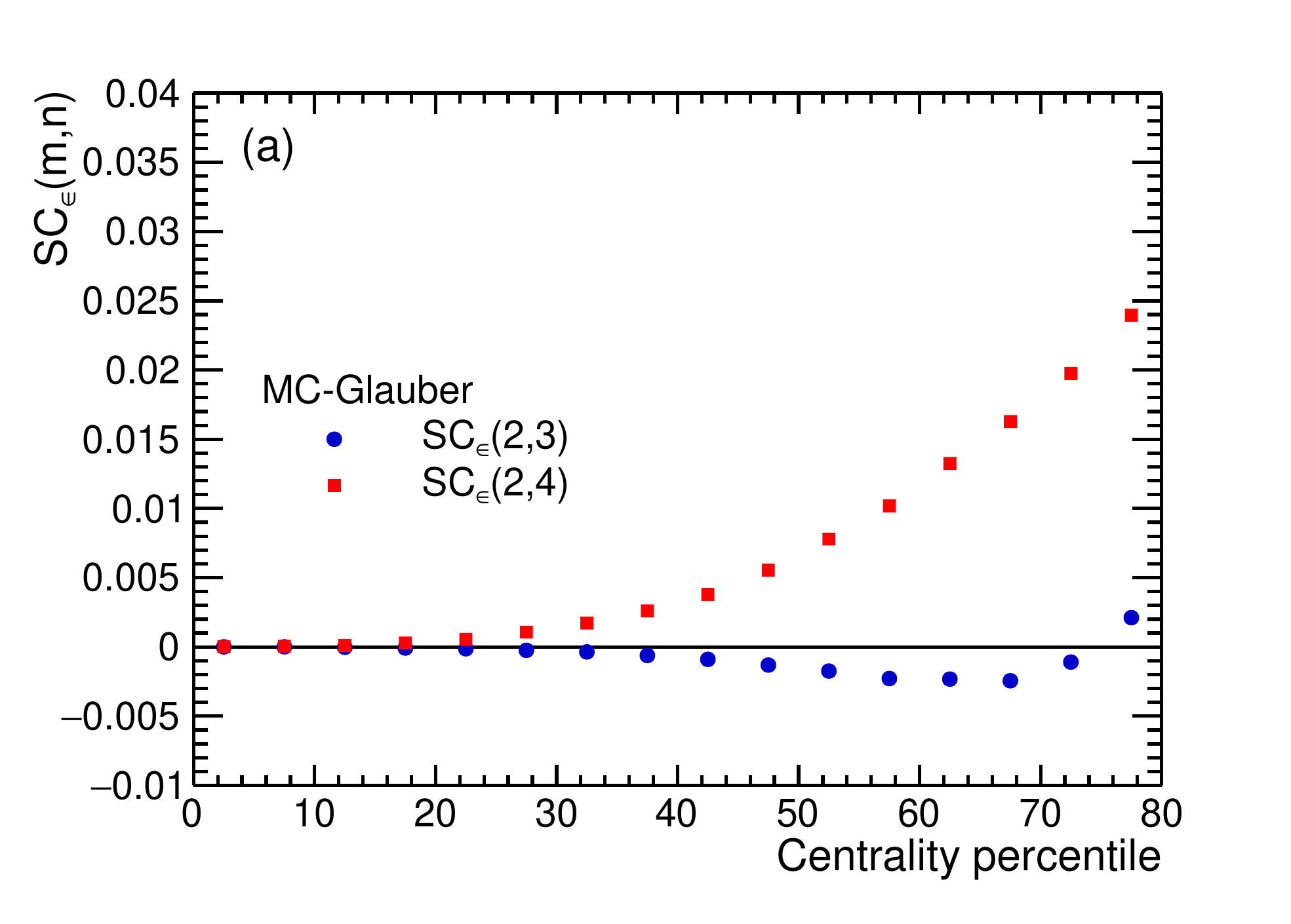} &
			\includegraphics[scale=0.43]{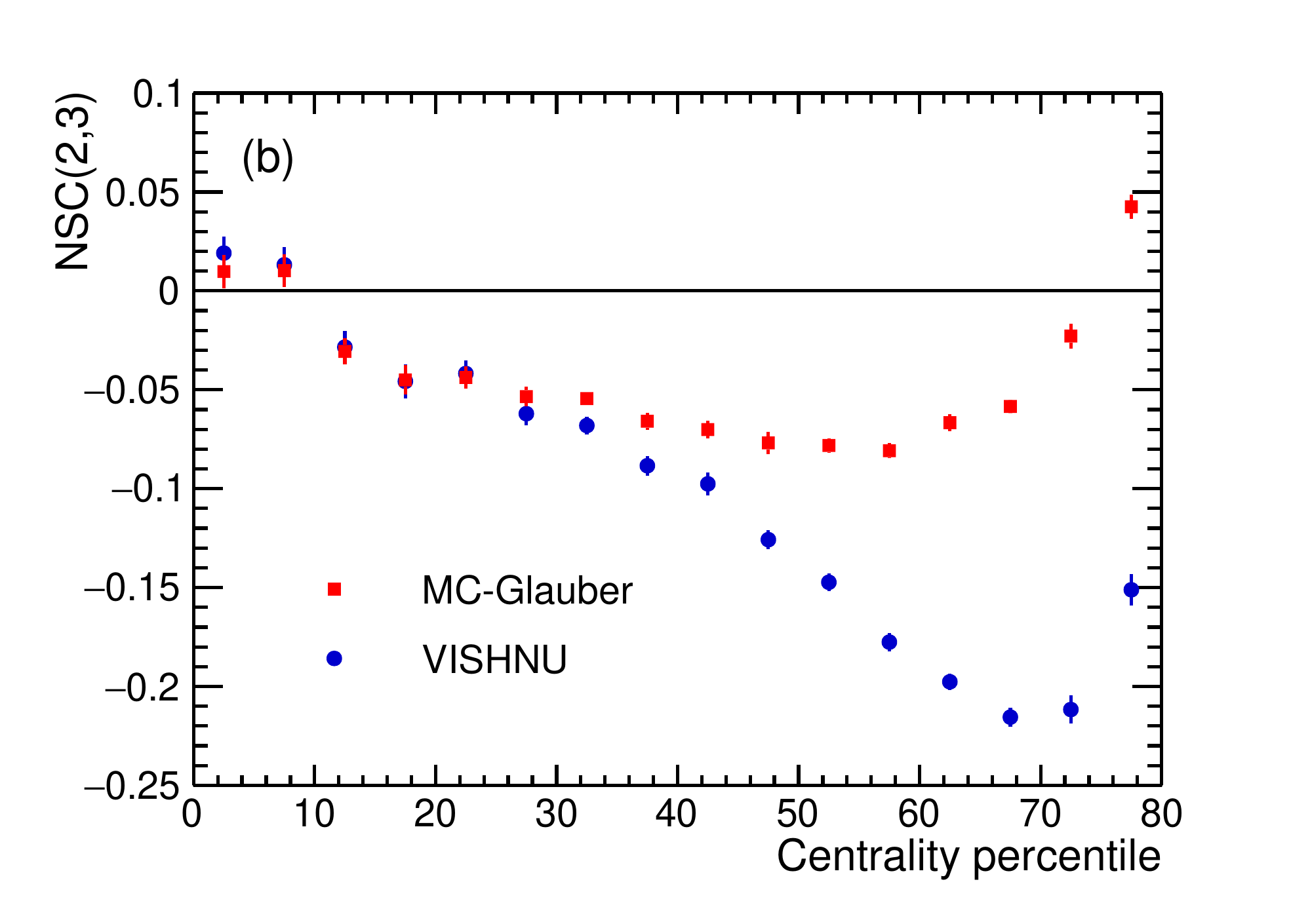} \\
			\includegraphics[scale=0.43]{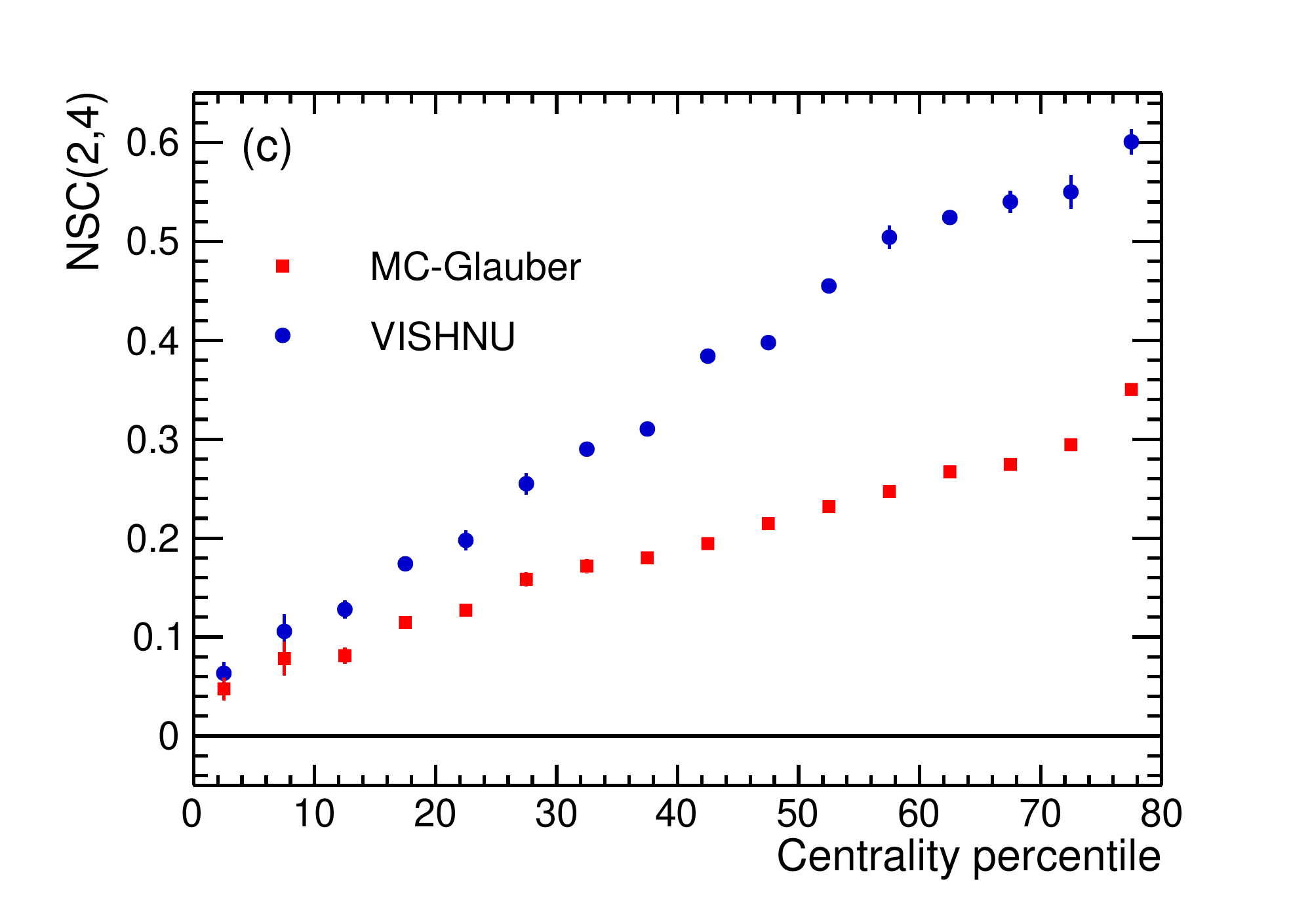} &
			\includegraphics[scale=0.43]{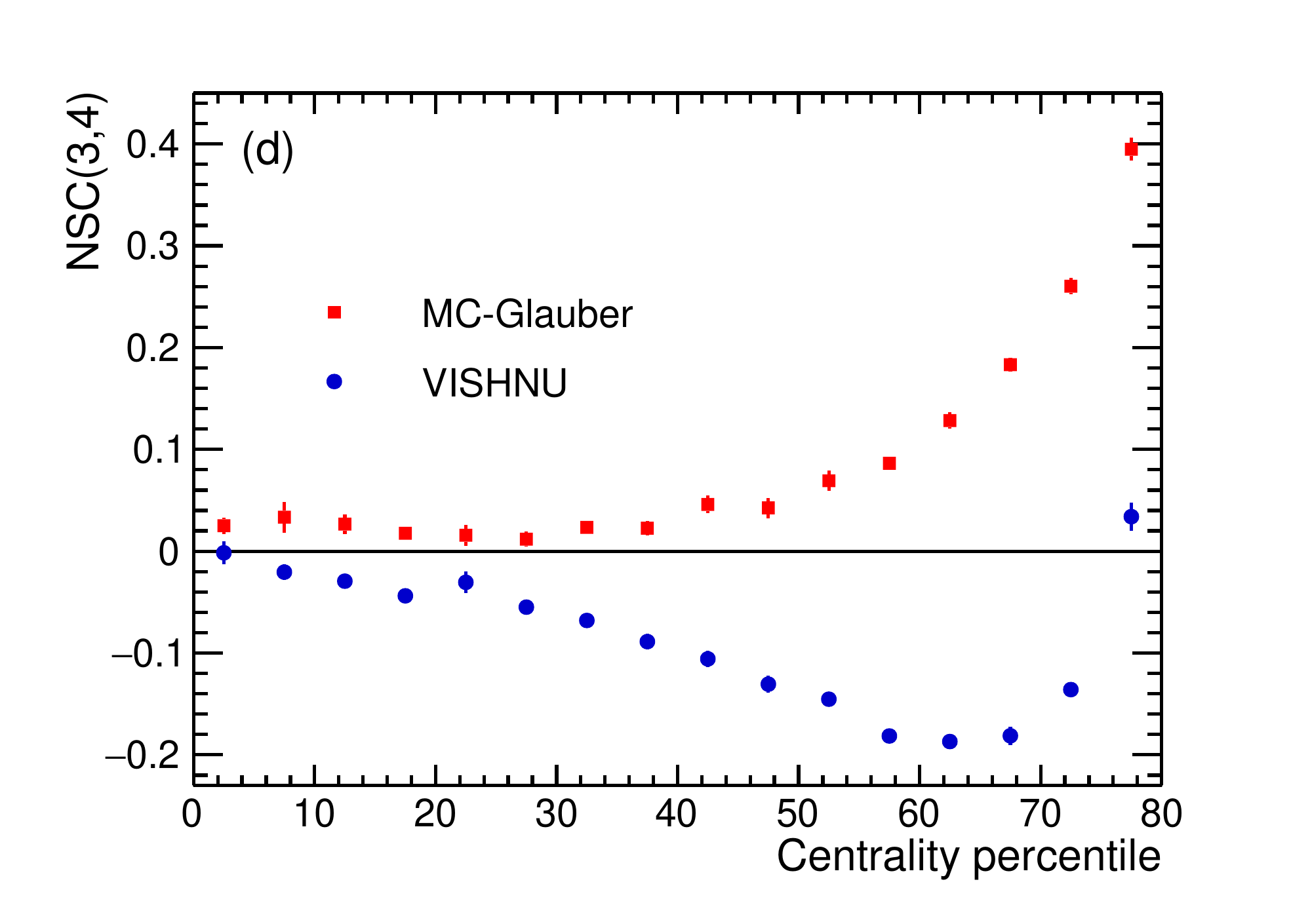} 

		\end{tabular}		
		\caption{ $\text{SC}_{\epsilon}(2,3)$ and $\text{SC}_{\epsilon}(2,4)$ obtained from initial distribution (a), and comparison between the initial and final normalized versions: $\text{NSC}(2,3)$ (b), $\text{NSC}(2,4)$ (c) and $\text{NSC}(3,4)$ (d).} 
		\label{NSCmn}
	\end{center}
\end{figure}

In this appendix, we reproduce few well-studied two-harmonic SC with our Monte Carlo simulations. In Fig.~\ref{SC23SC24andALICE}, we have shown centrality dependence of $\text{SC}(2,3)$ and $\text{SC}(2,4)$ obtained from iEBE-VISHNU in three different $p_T$ ranges, and we have compared it with the experimental results from ALICE at center-of-mass energy per nucleon pair $\sqrt{s_{\rm NN}}=2.76$~TeV~\cite{ALICE:2016kpq}. The ALICE $p_T$ is in the range $0.2 < p_T < 5$~GeV, and the range $0.28 < p_T < 4$~GeV in our simulation shows reasonable agreement with the ALICE data. Also we have shown few examples of normalized SC in Fig.~\ref{NSCmn}(b-d). One can see in Fig.~\ref{NSCmn}(b) that $\text{NSC}(2,3)$ and $\text{NSC}_{\epsilon}(2,3)$ are in good agreement in the centrality classes below $40\%$. It is due to the fact that the linear response is approximately true for $\epsilon_2$ and $\epsilon_3$ in this range \cite{Gardim:2011xv}. As matter of fact, the large discrepancy between $\text{NSC}(2,4)$ and $\text{NSC}_{\epsilon}(2,4)$ (Fig.~\ref{NSCmn}(c)) can be explained by non-linear term in $v_4$ equation. This effect has more non-trivial impact in the case of $\text{NSC}(3,4)$ and $\text{NSC}_{\epsilon}(3,4)$ (Fig.~\ref{NSCmn}(d)) where the normalized SC of the initial state shows a sign flip after the non-linear hydrodynamic response~\cite{Zhu:2016puf}. In this case, the $\epsilon_2$ contribution inside $v_4$ is anti-correlated with $\epsilon_3$. Therefore the non-linear term leads to the suppression of $\text{NSC}(3,4)$ compared to $\text{NSC}_{\epsilon}(3,4)$ (see Ref.~\cite{Zhu:2016puf}). This is fascinating because the same feature has been observed for the generalized SC in our study.

\section{Ellipse-like distributions}
\label{a:ellipse-like-distributions}
In this appendix, we demonstrate mathematically how the higher order SC observables can provide further and independent constraints. For clarity sake, we consider a simple Toy Monte Carlo model of ellipse-like distributions. Despite its simplicity, and since in mid-central heavy-ion collisions due to collision geometry the overlapping volume containing the strongly interacting nuclear matter is to leading order ellipsoidal, this toy model has some relevance also for the real-case scenarios. When the polar coordinates are measured from the center of ellipse, the ellipse is determined by the following relation:
\begin{equation}
f(\varphi) = \frac{b}{\sqrt{1-\varepsilon^2\cos^2\varphi}}\,,
\label{center}
\end{equation}
where $\varepsilon$ is the eccentricity defined as
\begin{equation}
\varepsilon^2\equiv 1-\frac{b^2}{a^2}\,,\label{eccentricity}
\end{equation}
and $a(b)$ is semimajor(semiminor) axis. Since we take $b \leq a$, it follows from (\ref{eccentricity}) that
eccentricity satisfies
\begin{equation}
0\leq\varepsilon<1\,.\label{limitse}
\end{equation}
For the distribution $f(\varphi)$ in Eq.~(\ref{center}) one can calculate the Fourier harmonics $v_n$ by using the following standard expression (due to symmetry of the problem all symmetry planes $\Psi_n$ are the same, and we can always span the coordinate system in such a way that $\Psi_n$ coincides with the orientation of semiminor axis $b$, which means that $\Psi_n = 0,\ \forall n$, in this simple model):
\begin{equation}
v_n =\left<\cos n\varphi\right> =\int_0^{2\pi}\frac{b\cos(n\varphi)}{\sqrt{1-\varepsilon^2\cos^2(\varphi)}}d\varphi\,.\label{harmonics:center}
\end{equation}
The integral in (\ref{harmonics:center}) cannot be expressed in terms of elementary functions. For instance, for the overall normalization $v_0$ and for the harmonic $v_2$ we have obtained the following results:
\begin{eqnarray}
v_0 &=& \frac{2\pi b}{\sqrt{1-\varepsilon^2}}\;{}_2F_1\left[\frac{1}{2},\frac{1}{2};1;\frac{\varepsilon^2}{\varepsilon^2-1}\right]\,,\\
v_2 &=& \frac{\sqrt{2}}{2}\,b\pi\frac{\varepsilon^2}{(2-\varepsilon^2)^{\frac{3}{2}}}\;{}_2F_1\left[\frac{3}{4},\frac{5}{4};2;\left(\frac{\varepsilon^2}{2-\varepsilon^2}\right)^2\right]\,,
\end{eqnarray}
where ${}_2F_1\left(a,b;c;z\right)$ is the hypergeometric function whose series expansion reads
\begin{equation}
{}_2F_1\left(a,b;c;z\right) = 1 + \frac{ab}{1!\,c}\, z + \frac{a(a+1)b(b+1)}{2!\,c(c+1)}\, z^2 + \cdots\,.
\end{equation}
We now inspect the leading order behavior of these expressions. Since the eccentricity satisfies relation~(\ref{limitse}), we can expand the results for harmonics obtained from (\ref{harmonics:center}) as a series in $\varepsilon$. After straightforward calculation, we have obtained the following results for the first non-vanishing harmonics,
\begin{eqnarray}
v_0 &=& 2b\pi +\frac{1}{2}\,b\pi\varepsilon^2 + \frac{9}{32}\,b\pi\varepsilon^4 +\frac{25}{128}\,b\pi\varepsilon^6+\frac{1225}{8192}\,b\pi\varepsilon^8+\mathcal{O}(\varepsilon^{10})\nonumber\,,\\
v_2 &=& \frac{1}{4}\,b\pi\varepsilon^2+\frac{3}{16}\,b\pi\varepsilon^4+\frac{75}{512}\,b\pi\varepsilon^6+
\frac{245}{2048}\,b\pi\varepsilon^8+\mathcal{O}(\varepsilon^{10})\,,\nonumber \\
v_4 &=& \frac{3}{64}\,b\pi\varepsilon^4+\frac{15}{256}\,b\pi\varepsilon^6+
\frac{245}{4096}\,b\pi\varepsilon^8+\mathcal{O}(\varepsilon^{10})\,,\nonumber\\
v_6 &=& \frac{5}{512}\,b\pi  \varepsilon^6+\frac{35}{2048}\,b\pi\varepsilon^8+\mathcal{O}(\varepsilon^{10})\,,\nonumber\\
v_8 &=& \frac{35}{16384}\,b\pi\varepsilon^8+\mathcal{O}(\varepsilon^{10})\,.\label{harmonics:series}
\end{eqnarray}
We stress out that all odd harmonics vanish for ellipse-like distributions when the polar coordinates are measured from the center due to the symmetry with respect to $y$-axis.

Using all above results, with the straightforward calculus we have obtained for the normalized ellipse-like distribution (normalization amounts simply to dividing with the overall normalization $v_0$ the starting expression in Eq.~(\ref{center}) and all final results for harmonics $v_n$ in Eq.~(\ref{harmonics:series})) the following relations among the even flow amplitudes $v_2, v_4, v_6, \ldots$
\begin{eqnarray}
\frac{v_4}{v_2^2}&=& \frac{3}{2} -\frac{1}{128}\,\varepsilon^4 +\mathcal{O}(\varepsilon^{6})\nonumber\,,\\
\frac{v_6}{v_2^3}&=& \frac{5}{2} -\frac{15}{512}\,\varepsilon^4 + \mathcal{O}(\varepsilon^{6})\nonumber\,,\\
\frac{v_8}{v_2^4}&=& \frac{35}{8} -\frac{21}{256}\,\varepsilon^4 + \mathcal{O}(\varepsilon^{6})\,.
\label{eq:evenCorrelations}
\end{eqnarray}
We conclude that when the distribution of particles in heavy-ion collisions is ellipsoidal, there is always a geometric correlation among all even harmonics, determined by relations in Eqs.~(\ref{eq:evenCorrelations}). If the fluctuations occur only in the magnitude of elliptical shape (e.g. only $\varepsilon$ and $b$ in Eq.~(\ref{center}) fluctuate event-by-event), the shape remains in each event elliptical, and therefore odd harmonics cannot develop~\cite{Kolb:2003zi}. This sort of fluctuations will contribute only to SC(2,4), SC(2,4,6), etc., but not to SC(2,3,4). The non-vanishing result for SC(2,3,4) implies that there are additional sources of fluctuations in the system  which couple all three amplitudes $v_2, v_3, v_4$. In particular, SC(2,3,4) can separate the fluctuations in the shape of ellipsoidal from the magnitude fluctuations of the persistent ellipsoidal shape. On the other hand, SC(2,4) cannot disentangle these two different sources of fluctuations.

\section{Comment on Parseval theorem}
\label{a:Comment-on-Parseval-theorem}

One may wonder to what extent correlated fluctuations of different flow harmonics might originate from some non-physical built-in mathematical property, solely from definitions. To clarify this point we start with the standard parametrization of Fourier series in Eq.~(\ref{eq:FourierSeries_vn_psin}). We can attach to the function $f(\varphi)$ a physical interpretation in the context of flow analyses only if we also promote $f(\varphi)$ to the level of p.d.f. This is achieved by imposing the following two constraints:
\begin{equation}
\int_{0}^{2\pi}f(\varphi)\,d\varphi = 1, \quad \mathrm{and} \quad f(\varphi) \geq 0, \quad \forall\varphi\,.
\label{eq:FourierSeries_PDF}
\end{equation}
From the purely mathematical point of view, flow amplitudes $v_n$ in Eq.~(\ref{eq:FourierSeries_vn_psin}) always satisfy the following relation, so-called Parseval theorem:
\begin{equation}
\frac{1}{2\pi}+\frac{1}{\pi}\sum_{n=1}^{\infty}v_n^2 = \int_{0}^{2\pi}|f(\varphi)|^2\,d\varphi\,.
\label{eq:ParsevalIdentity}
\end{equation}
In the combination with the probability constraint in Eq.~(\ref{eq:FourierSeries_PDF}), this relation could then lead always to some trivial built-in correlation among flow amplitudes. For instance, if anisotropic flow is quantified only with two amplitudes $v_m$ and $v_n$ and if the RHS in Eq.~(\ref{eq:ParsevalIdentity}) would be constrained to some constant value, we would then have:
\begin{equation}
v_m^2 + v_n^2 = \mathrm{const}\,,
\label{eq:exampleFromParseval}
\end{equation}
i.e. amplitudes $v_m$ and $v_n$ would be always trivially \textit{anticorrelated}, due to Parseval identity (on the other hand, \textit{correlated} amplitudes can be simply generated with the different relative signature in an analogous expression, i.e. $v_m^2 - v_n^2 = \mathrm{const}$). However, that is not the case, since the knowledge of integral of a function has a priori no relation with the knowledge of integral of that function squared. That means that even though in each high-energy nuclear collision the condition $\int_{0}^{2\pi}f(\varphi)\,d\varphi = 1$ is strictly satisfied, $\int_{0}^{2\pi}|f(\varphi)|^2\,d\varphi$ can nevertheless take any value. From another perspective, even if Parseval identity in Eq.~(\ref{eq:ParsevalIdentity}) would enforce some correlation among flow amplitudes $v_m^2$ and $v_n^2$, it would not be able to explain why then some amplitudes would be correlated, and some anticorrelated, as was observed experimentally for $v_2^2$ and $v_4^2$, and $v_2^2$ and $v_3^2$, respectively~\cite{ALICE:2016kpq}. We conclude that there is no trivial built-in correlation among different flow amplitudes which is originating solely from the fact that flow amplitudes are the degrees of freedom in the Fourier series expansion, to which also the probabilistic interpretation was attached via Eq.~(\ref{eq:FourierSeries_PDF}).


\end{document}